\documentclass[twocolumn,showpacs,superscriptaddress]{revtex4-1} 
\usepackage{graphicx,amsfonts,amsmath,subfigure}
\usepackage[normalem]{ulem}
\usepackage{color}
\usepackage{comment}

\newcommand{\bra}[1]{\ensuremath{\langle #1|}}
\newcommand{\ket}[1]{\ensuremath{|#1 \rangle}}

\newcommand{\eref}[1]{Eq.(\ref{#1})}

\newcommand{\jl}[1]{{\color{black}{#1}}}

\begin{document}

\title{Exact Simulation of Pigment-Protein Complexes Unveils \\ Vibronic Renormalization of Electronic Parameters in Ultrafast Spectroscopy}

\author{F. Caycedo-Soler}
\affiliation{Institute of Theoretical Physics and IQST, Ulm University, Albert-Einstein-Allee 11, 89081 Ulm, Germany}

\author{A. Mattioni}\email{Present address: Department of Chemistry, School of Natural
Sciences, The University of Manchester, Oxford Road,
Manchester, M13 9PL, United Kingdom}

\affiliation{Institute of Theoretical Physics and IQST, Ulm University, Albert-Einstein-Allee 11, 89081 Ulm, Germany}
\author{J. Lim}
\affiliation{Institute of Theoretical Physics and IQST, Ulm University, Albert-Einstein-Allee 11, 89081 Ulm, Germany}

\author{T. Renger}
\affiliation{Institute of Theoretical Physics, Department of Theoretical Biophysics, Johannes Kepler University Linz, Altenberger Str. 69, 4040 Linz, Austria}

\author{S. F. Huelga}\email{susana.huelga@uni-ulm.de}
\affiliation{Institute of Theoretical Physics and IQST, Ulm University, Albert-Einstein-Allee 11, 89081 Ulm, Germany}

\author{M. B. Plenio}\email{martin.plenio@uni-ulm.de}
\affiliation{Institute of Theoretical Physics and IQST, Ulm University, Albert-Einstein-Allee 11, 89081 Ulm, Germany}

\begin{abstract}{The primary steps of photosynthesis rely on the generation, transport and trapping of excitons in pigment-protein complexes (PPCs). Generically, PPCs possess highly structured vibrational spectra, combining many discrete intra-pigment modes and a quasi-continuous of protein modes, with vibrational and electronic couplings of comparable strength. The intricacy of the resulting vibronic dynamics poses significant challenges in establishing a quantitative connection between spectroscopic data and underlying microscopic models. Here we show how to address this challenge using numerically exact simulation methods by considering two model systems, namely the water-soluble chlorophyll-binding protein of cauliflower and the special pair of bacterial reaction centers. We demonstrate that the inclusion of the full multi-mode vibronic dynamics in numerical calculations of linear spectra leads to systematic and quantitatively significant corrections to electronic parameter estimation. These multi-mode vibronic effects are shown to be relevant in the longstanding discussion regarding the origin of long-lived oscillations in multidimensional nonlinear spectra.}
\end{abstract}
\maketitle

Light-harvesting (LH) antennas and photo-chemical reaction centers (RC) provide the elementary building blocks of the photosynthetic
apparatus of plants, algae and bacteria \cite{blankenship2014molecular}. Primarily these molecular aggregates consist of absorbing molecules
(pigments) complexed with specific proteins to form a PPC. Despite its fundamental importance to biology, the dynamical characterization of these complexes
to a degree that can reproduce all reported spectroscopic data in a single microscopic model remains an outstanding challenge.

Reduced models of excitonic dynamics subject to purely thermal fluctuations can achieve reasonable agreement with linear optical spectra
\cite{Monshouwer1997,Trinkunas_PRL2001,Jordanides_2001JPCB,Hu_2002,Renger_2004PRL,Renger_2005BioJ,vanGrondelle_review,Renger_2009PhRes}.
The quantitative explanation of all relevant aspects of multi-dimensional nonlinear spectroscopy though, requires a more
detailed model of the system-environment interaction that takes into account the full complexity of the environmental
structure \cite{jumper2018vibronic}. Indeed, spectroscopic studies of PPCs at low temperatures \cite{Freiberg_2009JCP, Freiberg_2011JCP,Pieper_2011JPCB,Piper_2018JPCB} reveal the presence of vibrational environments that consist of a broad spectrum of low-frequency protein modes with room
temperature energy scales, and several tens of discrete high-frequency modes that originate mainly from intra-pigment
dynamics \cite{Small_JPCB2001,Freiberg_2009JCP,Freiberg_2011JCP}. Nonlinear optical experiments on monomer pigments
in solution at both 77 K \cite{Ogilvie_2018JPClett,bukarte2020revealing} and room temperature \cite{Harel2018JPCL,Collini_ChemPhys2019} as
well as first-principles calculations \cite{Rive13,Bennet_PNAS2018} further corroborate the underdamped
nature of intra-pigment vibrational modes with picosecond lifetimes.

Recently, a range of vibronic models in which pigments are subject to the combined influence of a broad unstructured bosonic
environment and a small number of vibrational modes with frequencies in the vicinity of excitonic transitions have been formulated
\cite{Prior_PRL2010,chin2010noise,WomickMoran2011,Chin_2013NPhys,Lovett_2014PRL,Dijkstra_2015JPClett,Novelli_JPCLett2015,Nazir_JCP2016,Maly_2016CPC,
Caycedo_JPClett2018}. In this picture, vibrational lifetime borrowing can lead to long-lasting oscillatory dynamics of coherences between excitonic states and observations of long-lasting oscillatory features in multi-dimensional spectroscopy \cite{Engel_Nature2006,Flemming_2007Science,Engel2010,Hildner2013,Romero_NPhys2014,Ogilvie_NChem2014} have been attributed
to this effect \cite{Christensson_JPCB2012,butkus2012molecular,Plenio_JCP2013,tiwari2013electronic,Chenu_2013SciRep,huelga2013vibrations}. Notwithstanding, the identification of a universally accepted origin of these long-lived oscillations remains a subject of active discussion \cite{Engel2010,Miller_PNAS2017,Donatas_NatChem2018,SMLimPRL2019,cao2020quantum}.

An important obstacle that prevents the conclusive resolution of this debate is the fact that the
interpretation of spectroscopic data and their underpinning dynamical features can be influenced significantly
by the specific choice of electronic and vibrational parameters that enter the PPC models. We will demonstrate that by accounting for the full environmental spectral density, involving more than $50$ intra-pigment modes per site in addition to a broad background, the presence of high-frequency long-lived vibrational
modes  {can lead} to quantitatively significant modification of the calculated linear spectra of PPCs and consequently the estimated
values of electronic parameters to recover a best fit with actual measurements. These corrections do not appear when considering
only selected resonant modes and go well beyond predictions that can be obtained by using conventional line shape theory
\cite{Novo04,Abra10,Lewi13,Jankowiak_2019JPCB}.

\begin{figure}
\includegraphics[width=0.46\textwidth]{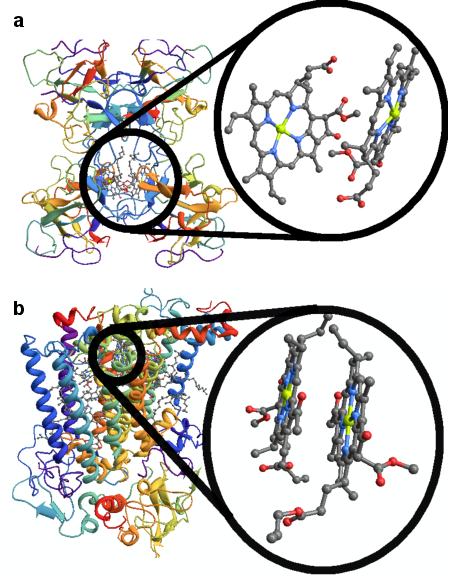}
\caption{Photosynthetic pigment-protein complexes. (a) Molecular structure of water-soluble chlorophyll-binding protein from cauliflower,
a natural dimeric PPC, with Chl{\it b} homodimer shown in detail. (b) Molecular structure of bacterial reaction center from purple
bacterium {\it Rb. Sphaeroides} with a (hetero)-dimeric unit of special pair highlighted. Site energies and couplings for the relevant
pigments are obtained from models that combine the crystal structure together with a comparison of calculated and measured spectra
\cite{adolphs2006proteins}.}
\label{fig1}
\end{figure}

To present our results, we provide an analytical theory of renormalisation effects due to multi-mode vibronic mixing
in model excitonic systems of two prototypical PPCs, namely the water-soluble chlorophyll-binding protein (WSCP) of cauliflower
and the special pair (SP) of bacterial reaction centers, depicted in Fig.~\ref{fig1}. By considering realistic environmental spectral densities, we corroborate
our predictions using two independent numerically exact methods (the temperature-dependent time evolving density matrix
using orthogonal polynomials algorithm, T-TEDOPA \cite{Prior_PRL2010,Chin_JMP2010,tamascelli2018nonperturbative,Tamascelli_2019PRL},
and the hierarchical equations of motion, HEOM \cite{Tanimura_JPSJ1989}). We show that the hybridisation of electronic
and vibrational degrees of freedom requires a significant renormalisation of electronic couplings.
Importantly, this renormalisation of electronic parameters, in turn, is shown to have a
significant impact on the dynamics of excitonic coherences, notably the lifetimes of their oscillatory dynamics.\\

{\bf Results}

{\bf Electronic and vibronic couplings of PPCs}. Absorption spectra of PPCs are determined by the electronic energy-level
structure of pigments, their mutual electronic interactions and the coupling of the resulting excitons to vibrational
degrees of freedom of the pigment's environment. In the following we will restrict our analysis to the ${\rm Q}_y$ transition between electronic ground and first excited states of the pigments which suffices for the evaluation of the low-energy part of absorption
spectra and is relevant for photosynthetic energy transfer \cite{blankenship2014molecular}. For the dimeric WSCP and SP, the electronic Hamiltonian is then described by (see Supplementary Note 1)
\begin{equation}
    H_{e}=\sum_{i=1}^{2}\varepsilon_{i}|\varepsilon_{i}\rangle\langle \varepsilon_{i}| +V(|\varepsilon_{1} \rangle\langle \varepsilon_{2}|+|\varepsilon_{2}\rangle\langle \varepsilon_{1}|)\, .
\end{equation}
Here $\ket{\varepsilon_{i}}$ denotes the singly excited state of site $i$ with on-site energy $\varepsilon_{i}$ that is in the visible (WSCP) or in the near infrared spectrum (SP). The on-site energies depend on their local environment and therefore suffer from static disorder inducing ensemble dephasing that will be included in our numerical treatment. The electronic coupling $V$ leads to delocalised electronic eigenstates (excitons), $H_{e}\ket{E_\pm}=E_\pm\ket{E_\pm}$, and an excitonic splitting $\Delta=E_{+}-E_{-}=\sqrt{4V^{2}+(\varepsilon_1-\varepsilon_2)^{2}}$. In WSCP, the mean site energies are identical, $\langle \varepsilon_{1}\rangle=\langle \varepsilon_{2}\rangle$, due to the symmetry of molecular structure, while in SP, the mean site energies are different as pigments are surrounded by nonidentical local protein environments. Another difference concerns the electronic coupling strength, which is stronger in SP due to electron exchange giving rise to short-range Dexter type contributions~\cite{WarshelJACS1987,Renger_2009JPCB}.

The exciton dynamics of PPCs is driven by vibrational modes that induce fluctuations in the transition energies $\varepsilon_i$ of pigments.
The full electronic-vibrational interaction, induced by $N$ vibrational modes per site, is described by the Hamiltonian $H = H_e + H_v + H_{e-v}$ where
\begin{eqnarray}
    H_v &=& \sum_{i=1}^{2} \sum_{k=1}^{N} \omega_k b_{i,k}^{\dagger}b_{i,k},\\
    H_{e-v} &=& \sum_{i=1}^{2} |\varepsilon_i\rangle\langle\varepsilon_i|\sum_{k=1}^{N}\omega_k\sqrt{s_k}(b_{i,k} + b_{i,k}^{\dagger}).
\end{eqnarray}
Here the annihilation (creation) operators $b_{i,k}$ ($b_{i,k}^{\dagger}$) describe a local vibrational mode of frequency
$\omega_k$ coupled to site $i$ with a strength quantified by the Huang-Rhys (HR) factor $s_k$. For an environment initially
in a thermal state, the ensuing dynamics is fully determined by the environmental spectral density $J(\omega) = \sum_k
\omega_k^2 s_k \delta(\omega-\omega_k)$ whose structure needs to be determined experimentally or theoretically.\\

\begin{figure*}
\includegraphics[width=\textwidth]{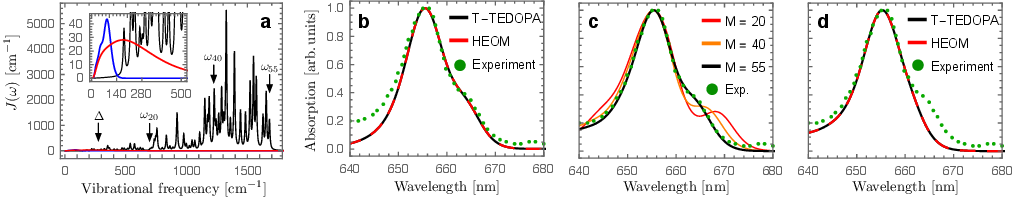}
\caption{Absorption spectra of WSCP. (a) Experimentally estimated spectral density of WSCP, consisting of 55 intra-pigment modes $J_{h}(\omega)$~\cite{Pieper_2011JPCB} and low-frequency protein modes $J_{l}^{\rm WSCP}(\omega)$~\cite{Jankowiak_2013JPCB}, shown in black and blue, respectively. Experimentally estimated spectral density $J_{l}^{\rm B777}(\omega)$ of B777 complexes is shown in red~\cite{Renger_2015JCP}. The position of the excitonic splitting $\Delta= 280\,{\rm cm}^{-1}$ obtained for an electronic coupling $V=140\,{\rm cm}^{-1}$ is indicated by a black arrow. The 20th, 40th and 55th lowest vibrational frequencies of the intra-pigment modes are marked by black arrows with $\omega_{20}$, $\omega_{40}$ and $\omega_{55}$, respectively. (b) Experimental absorption spectrum of WSCP at $77\,{\rm K}$, shown in green dots, and numerical results obtained by T-TEDOPA and HEOM, shown in black solid and red dashed lines, respectively,
for $V=69\,{\rm cm}^{-1}$ and  $J_{l}^{\rm B777}(\omega)~$\cite{Renger_2015JCP}. (c) For $V=140\,{\rm cm}^{-1}$ and $J_{l}^{\rm WSCP}(\omega)+J_{h}(\omega)$, T-TEDOPA and HEOM results can reproduce the experimental absorption spectrum, as shown in black. Numerically exact absorption spectra for the $M\in\{20,40,55\}$ lowest frequency intra-pigment modes are displayed where $M=55$ represents the full experimentally estimated spectral density. (d) For $V=69\,{\rm cm}^{-1}$ and $J_{l}^{\rm WSCP}(\omega)+J_{h}(\omega)$, T-TEDOPA and HEOM results cannot reproduce the experimental absorption spectra. See Supplementary Note 5 for details of the other molecular parameters used in these simulations. We note that the maximum amplitudes of simulated absorption spectra at 656\,nm are normalised to unity for a comparison with experimental absorption line shape.}
\label{fig2_r}
\end{figure*}

{\bf Structure of the environmental spectral density}. Generally, in PPCs the spectral density $J(\omega)$ consists of a
broad background and multiple sharp peaks distributed across a broad range of frequencies. These can be determined by
fluorescence line-narrowing (FLN) and hole burning experiments which reveal that the environmental spectral densities of WSCP and SP
consist of low-frequency broad features originating from protein motions, and $55$ intra-pigment modes resulting in multiple narrow peaks in the high-frequency part of the spectrum. The contribution of the protein modes of WSCP may be described by log-normal distribution functions of the form $J_{l}^{\rm WSCP}(\omega) =
\sum_{m} (\omega c_m/\sigma_m) \,\exp(-[\ln(\omega/\Omega_m)]^2/2\sigma_m^2)$, which provides a satisfactory description of the low-energy part of experimentally measured FLN spectra of WSCP~\cite{Jankowiak_2013JPCB}. Alternatively, the protein motions of WSCP have been modelled by the following functional form: $J_{l}^{\rm B777}(\omega)=\frac{S}{s_1+s_2}\sum_{i=1}^{2}\frac{s_i}{7! 2 \omega_i^4}\omega^5 e^{-(\omega/\omega_i)^{1/2}}$ that has been extracted from FLN spectra of B777 photosynthetic complexes \cite{Renger_2002JCP} and considered in the simulations of WSCP \cite{Renger_2015JCP}. Every underdamped
intra-pigment mode contributes a Lorentzian of width $\gamma_k \sim 1\,{\rm ps}^{-1}$, resulting
in $J(\omega)=J_l(\omega)+J_h(\omega)$ where
\begin{equation}
    J_{h}(\omega) = \sum_{k=1}^{55} \frac{4 \omega_k s_k \gamma_k (\omega_{k}^2+\gamma_{k}^2)\omega}{\pi((\omega+\omega_k)^{2}+\gamma_{k}^2)((\omega-\omega_k)^{2}+\gamma_{k}^2)},
    \label{SpectralDensity}
\end{equation}
and the reorganisation energy of the high-frequency modes is given by $\lambda_h=\int_{0}^{\infty}d\omega J_{h}(\omega)/\omega=\sum_{k=1}^{55}
\omega_k s_k$. The reorganization energy of the 55 intra-pigment modes of WSCP \cite{Pieper_2011JPCB} (SP \cite{Small_JPCB2001}) is $660\,{\rm cm}^{-1}$ ($379\,{\rm cm}^{-1}$), which is several times larger than that of quasi-continuous protein spectrum \cite{Jankowiak_2013JPCB,Jankowiak_2015JCP} and quasi-resonant intra-pigment modes with $\omega_k\approx \Delta$ (see Supplementary Note 5). The presence of underdamped vibrational modes can lead to long-lived correlations between electronic and vibrational degrees of freedom that make the rigorous numerical treatment of the ensuing vibronic dynamics very costly. In non-perturbative HEOM simulations, where experimentally or theoretically estimated spectral densities are fitted by the sum of Drude-Lorentz peaks \cite{Kramer_2012JPCL,Bennet_PNAS2018}, the simulation cost of a dimeric system exceeds several hundreds of terabytes when 55 intra-pigment modes are considered per site (see Supplementary Note 4) and, therefore, is infeasible with current computer architectures. In this work, we employ T-TEDOPA method where an experimentally estimated vibrational spectral density is mapped to a one-dimensional chain of quantum harmonic oscillators whose complexity is unaffected by the number of long-lived intra-pigment modes in the spectral density. We also employ optimised HEOM method where simulation parameters are determined by fitting the bath correlation function of highly structured environments for a finite time window corresponding to the line width of experimentally measured absorption spectra. These two methods enable one to consider the full environmental structures of WSCP and SP with a moderate simulation cost of the order of a few gigabytes or less (see Supplementary Note 3 and 4). In addition, numerically exact results obtained by these two independent methods coincide, demonstrating the high accuracy and reliability of our simulated data (see Supplementary Note 6).

{\bf WSCP homodimer}. The electronic parameters of PPCs have been estimated based on a comparison of experimentally measured spectroscopic data with approximate theoretical results where environmental structures are coarse-grained or vibronic couplings are treated perturbatively. Based on a coarse-grained spectral density $J_{l}^{\rm B777}(\omega)$, shown in red in Fig.~\ref{fig2_r}a, a best fit to the experimental absorption spectra of WSCP homodimers implies an electronic coupling strength estimate of $V\approx 70\,{\rm cm}^{-1}$~\cite{Renger_2015JCP}, as shown in red in Fig.~\ref{fig2_r}b. Such an electronic coupling results in an excitonic splitting $\Delta\approx 2V \approx 140\,{\rm cm}^{-1}$ which is consistent with the experimentally observed energy-gap between two absorption peaks  {at 656\,nm and 662\,nm, respectively. Since all the high frequency intra-pigment modes are neglected in the coarse-grained spectral density and the energy-gap between absorption peaks is smaller than the vibrational frequencies of the intra-pigment modes ($\Delta < \omega_k$), the estimated value could be interpreted as the effective coupling $V_{00}$ between $\ket{\varepsilon_1,0}$ and $\ket{\varepsilon_2,0}$ where $\ket{0}$ denotes the common vibrational ground states of the intra-pigment modes in the electronic excited state manifold. As shown in Fig.~\ref{fig3_r}a, the transition dipole strength between $\ket{g,0}$ and $\ket{\varepsilon_i,0}$ (0-0 transition) of a mononer is reduced by a factor of $\exp(-\sum_{k} s_k/2)$, as the total transition dipole strength of the monomer is redistributed to 0-1 transitions between $\ket{g,0}$ and $\ket{\varepsilon_i,1_k}$ where only the $k$-th mode is singly excited (see Supplementary Note 10). As a result, the effective coupling between 0-0 transitions, shown in Fig.~\ref{fig3_r}b, is reduced to $V_{00}=V\exp(-\sum_k s_k)$ depending on the HR factors $s_k$ of the intra-pigment modes. This implies that $V_{00}\approx 70\,{\rm cm}^{-1}$ corresponds to a bare electronic coupling $V=V_{00}\exp(\sum_{k=1}^{55}s_k)\approx 2V_{00}\approx 140\,{\rm cm}^{-1}$ under the full environmental spectral density $J_{l}^{\rm WSCP}(\omega)+J_{h}(\omega)$, including the 55 intra-pigment modes shown in black in Fig.~\ref{fig2_r}a. The renormalised electronic coupling $V\approx 140\,{\rm cm}^{-1}$ yields a best fit to experimentally measured absorption spectra, as shown in black in Fig.~\ref{fig2_r}c, when all the $M=55$ intra-pigment modes are considered in simulations. The energy-gap between absorption peaks is gradually reduced from excitonic splitting $\Delta\approx 2V\approx 280\,{\rm cm}^{-1}$ to $\Delta'\approx 2V_{00}\approx 140\,{\rm cm}^{-1}$, as the number $M$ of the lowest-frequency intra-pigment modes considered in simulations is increased from 20 via 40 to 55 (see Fig.~\ref{fig2_r}a and c). The electronic coupling $V\approx 70\,{\rm cm}^{-1}$ estimated based on the coarse-grained low-frequency spectral density cannot reproduce the experimental results when the full spectral density is considered in simulations, as shown in Fig.~\ref{fig2_r}d. The energy-gap between absorption peaks shown in Fig.~\ref{fig2_r}c and d can be quantitatively well described by the splitting of 0-0 transitions, $2V_{00}=2V\exp(-\sum_{k=1}^{M}s_k)$, implying that the effective couplings $V_{01}$ between 0-0 and 0-1 transitions, schematically shown in Fig.~\ref{fig3_r}b, is not strong enough to modify the energy-gap between low-energy absorption peaks of WSCP. However, the weak $V_{01}$ couplings can redistribute the transition dipole strength from 0-0 to 0-1 transitions and significantly modify the high-energy part of absorption spectra, which cannot be described by conventional line shape theory (see Supplementary Note 10).

\begin{figure}
\includegraphics[width=0.46\textwidth]{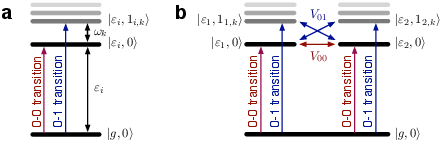}
\caption{Vibronic energy-levels in site basis. (a) Energy-level structure of monomer with 0-0 and 0-1 transitions highlighted in red and blue, respectively. (b) Energy-level structure of dimer with $V_{00}$ and $V_{01}$ representing the effective coupling between 0-0 transitions and the interaction between 0-0 and 0-1 transitions, respectively.}
\label{fig3_r}
\end{figure}

\begin{figure}
\includegraphics[width=0.46\textwidth]{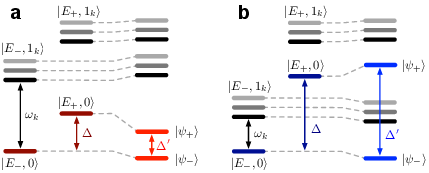}
\caption{Vibronic energy-levels in exciton basis. (a,b) Effect of multi-mode vibronic mixing on vibronic energy-level structure when excitonic splitting $\Delta$ is smaller (larger) than vibrational frequencies $\omega_k$ of intra-pigment modes, leading to reduction (increment) of the energy gap $\Delta'$ between vibronic eigenstates.}
\label{fig4}
\end{figure}

\

{\bf Multi-mode vibronic mixing in exciton basis}. In contrast to WSCP, the bare excitonic splitting of SP is of the order of the typical vibrational frequencies 
of the intra-pigment modes and the resulting redistribution of oscillator strengths and shifts of optical lines
are much more difficult to predict. To qualitatively  estimate these effects, we consider second-order perturbation theory starting
from the full Hamiltonian $H=H_{e}+H_{v}+H_{e-v}$ in the single-exciton manifold. In that case, the vibronic mixing is induced by the relative motion of the intra-pigment modes with identical frequency $\omega_k$, described by $b_k = (b_{1,k} - b_{2,k})/\sqrt{2}$, as the center of mass motion, described by $B_{k}=(b_{1,k} + b_{2,k})/\sqrt{2}$, merely induces the homogeneous broadening of absorption line shapes without affecting exciton dynamics (see Supplementary Note 1). Hence, we can discard the center-of-mass part of the total Hamiltonian to find $H=H_0 + H_I$ where
\begin{equation}
    H_0 = H_e + H_v + \cos(\theta)\ \sigma_z \sum_{k=1}^{55} \omega_k \sqrt{s_k/2}(b_k+b_k^\dagger),
    \label{eq:H0_JL}
\end{equation}
with $H_v=\sum_{k}\omega_k b_{k}^{\dagger}b_k$, and
\begin{equation}
    H_I = -\sin(\theta)\ \sigma_x \sum_{k=1}^{55} \omega_k \sqrt{s_k/2} (b_k+b_k^\dagger).
    \label{relative}
\end{equation}
Here $\theta=\tan^{-1}[2V/(\varepsilon_1-\varepsilon_2)]$, while $\sigma_x = |E_+\rangle\langle E_-| + |E_-\rangle\langle E_+|$ and $\sigma_z = |E_+\rangle\langle E_+|
- |E_-\rangle\langle E_-|$ are the Pauli matrices in the exciton basis. The Hamiltonian $H_0$ is diagonalised
by the polaron transformation in the exciton basis, $U=\ket{E_+}\bra{E_+}D_\theta + \ket{E_-}\bra{E_-}D_\theta^\dagger$ with $D_\theta
=\exp[\cos(\theta)\sum_k\sqrt{s_k/2}(b_k^\dagger-b_k)]$. For typical HR factors of PPCs, of the order of
$s_k\lesssim 0.01$, the vibronic mixing is dominated by contributions from the single vibrational excitation subspace where
it leads to eigenstates of $H$ of the form
\begin{equation}
	\ket{\psi_\pm}=a_{\pm,0}\ket{E_\pm,0}+\sum_{k=1}^{55}a_{\mp,1_k}\ket{E_\mp,1_k},
\end{equation}
with $\ket{0}$ and $\ket{1_k}$ representing vibrational states where all the intra-pigment modes are in their ground states or only one mode described by $b_{k}$ is singly excited.
In second-order perturbation theory, these vibronic eigenstates $\ket{\psi_\pm}$ have energies
\begin{equation}
    E'_\pm = E_\pm \pm \alpha \frac{2V^2}{\Delta^2} \sum_{k=1}^{55} \frac{s_k \omega_k^2}{\Delta\mp\omega_k},
    \label{energies}
\end{equation}
and the purely excitonic splitting  {$\Delta=E_+-E_-$} is shifted to a vibronic splitting
\begin{equation}
    \Delta'=E_{+}'-E_{-}'= \Delta\left(1 + \alpha \frac{4V^2}{\Delta^2} \sum_{k=1}^{55} \frac{s_k\omega_k^2}{\Delta^2-\omega_k^2}\right),
    \label{levelshift}
\end{equation}
where $\alpha = \exp(-2\cos^2(\theta)\ \sum_{k=1}^{55}s_{k})$. These energetic corrections are in complete analogy
to the well-known light shifts in atomic physics. The sign of these energy shifts is determined by the difference in
excitonic splitting and vibrational frequency, $\Delta-\omega_k$. We note that the vibronic energy renormalization can also be described in the regular electronic-vibrational basis without the polaron transformation using second order perturbation theory (see Supplementary Note 2).

For an excitonic splitting that is smaller than
the vibrational frequencies, $\Delta\lesssim \omega_k$, the energy-gap $\Delta'$ between vibronic eigenstates
$\ket{\psi_+}$ and $\ket{\psi_-}$ is reduced compared to the bare excitonic splitting $\Delta$ (see Fig.~\ref{fig4}a). This is in line with our numerically exact simulations of WSCP where the bare excitonic splitting $\Delta\approx 2V$ is reduced to $\Delta'\approx 2V_{00}\approx V$. It is notable that for PPCs consisting of chlorophylls or
bacteriochlorophylls, the HR factors of the intra-pigment modes are of the order of $s_k\approx 0.01$, independent of the vibrational frequencies $\omega_k$. In
case the excitonic splitting is significantly smaller than the vibrational frequencies of the intra-pigment modes, the detuning between them is well approximated by $\Delta_k=\omega_k - \Delta \approx \omega_k$, thus exhibiting the same scaling in $\omega_k$ as the electronic-vibrational coupling, $g_k=\omega_k\sqrt{s_k}$ . This implies that the coupling of higher-frequency modes increases with the detuning $\Delta_k$ so that they cannot simply be ignored on the basis of being off-resonant.

When the excitonic splitting is larger than the vibrational frequencies, $\Delta\gtrsim \omega_k$, the situation is reversed (see Fig.~\ref{fig4}b), resulting in an increased vibronic splitting $\Delta'$ compared to
the bare excitonic splitting $\Delta$. This case cannot be described by the splitting of 0-0 transitions, since the effective coupling $V_{00}=V\exp(-\sum_k s_k)$ is smaller in magnitude than a bare electronic coupling $V$ for arbitrary HR factors defined by $s_k\ge 0$. This implies that the mixing of 0-0 and 0-1 transitions can result in two absorption peaks with an energy gap $\Delta'$ being larger than the bare excitonic splitting $\Delta$.

\

\begin{figure*}
\includegraphics[width=\textwidth]{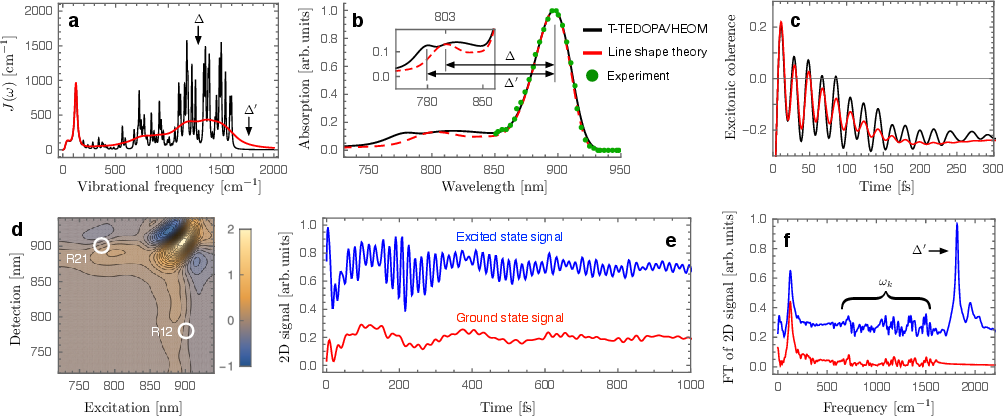}
\caption{Absorption and 2D electronic spectra of SP. (a) Experimentally estimated  spectral density of the SP \cite{Jankowiak_2015JCP,Small_JPCB2001} is shown in black for an intra-pigment mode vibrational damping rate $\gamma_{k}=(1\,{\rm ps})^{-1}$. Coarse-grained version for $\gamma_{k}=(50\,{\rm fs})^{-1}$ is shown in red and the excitonic and vibronic splittings, $\Delta\approx 1290\,{\rm cm}^{-1}$ and $\Delta'\approx 1800\,{\rm cm}^{-1}$, are highlighted. (b) Experimental absorption spectrum of the bacterial reaction center at $5\,{\rm K}$, shown in green dots, and numerically exact absorption line shape, obtained by TEDOPA and HEOM, of the SP, shown in black. Approximate absorption spectrum of the SP computed by second-order cumulant expansion is shown in red where the energy-gap between absorption peaks at $803\,{\rm nm}$ and $897\,{\rm nm}$ is approximately $\Delta\approx 1300\,{\rm cm}^{-1}$. (c) Excitonic coherence dynamics for the experimentally estimated and coarse-grained environmental structures, shown in black and red, respectively when only site 1 is initially excited. (d) Rephasing 2D spectra of the SP at waiting time $T=0$. (e,f) 2D signals at a cross-peak R12, marked in (d), and corresponding Fourier transformation where ground and excited state signals are shown in red and blue, respectively. Note that excited state signals are dominated by vibronic coherence $|\psi_+\rangle\langle \psi_-|$, leading to 2D oscillations with frequency $\Delta'\approx 1800\,{\rm cm}^{-1}$. The transient of the other cross-peak R21 is provided in Supplementary Note 8 and all molecular parameters used in these simulations are given in Supplementary Note 5.}
\label{fig5}
\end{figure*}

{\bf Special pair in bacterial reaction center}. The photosynthetic reaction center which drives exciton dissociation into free charges consists of the SP and four additional pigments~\cite{Renger_2006JPC}. The SP is a strongly coupled dimeric unit with an electronic coupling estimated to be $V=625\,{\rm cm}^{-1}$, a difference in mean site energies of $\langle \varepsilon_1-\varepsilon_2 \rangle=315\,{\rm cm}^{-1}$ and consequently a bare excitonic splitting of $\Delta\approx 1290\,{\rm cm}^{-1}$.  These electronic parameters have been estimated based on a best fit to absorption, linear dichroism and hole burning spectra of bacterial reaction centers using conventional line shape theory~\cite{Jankowiak_2019JPCB}. In what follows, we neglect the order of magnitude weaker electronic coupling of the SP to the four additional pigments and do not aim to reproduce experimentally measured absorption spectra of the whole bacterial reaction centers and re-estimate electronic parameters. Rather we concentrate on the effect of multi-mode vibronic mixing on the SP and its consequences regarding the nature and lifetimes of excitonic coherence and long-lived oscillatory signals in 2D electronic spectra.

While in WSCP the excitonic splitting is far detuned from high-frequency modes, the situation is markedly different
for the SP. Here the environmental spectral density contains high-frequency intra-pigment
modes both above and below the bare excitonic gap, as shown in black in Fig.~\ref{fig5}a. The smaller
frequency differences between vibrational modes and excitonic splitting and the varying sign of their detuning
makes the effect of multimode mixing harder to predict analytically. Indeed, the perturbation procedure for obtaining
\eref{levelshift} will be inaccurate for a larger number of modes. The vibronic splitting can be estimated beyond the perturbation theory by numerically diagonalising the Hamiltonian $H=H_0+H_I$
in Eq.(\ref{eq:H0_JL}-\ref{relative}), leading to $\Delta'\approx 1744\,{\rm cm}^{-1}$ (see Supplementary Note 7). This estimate
is in line with numerically exact simulated results where the energy-gap between absorption peaks is approximately
$1710\,{\rm cm}^{-1}$ (see $780\,{\rm nm}$ and $900\,{\rm nm}$ peaks in Fig.~\ref{fig5}b, corresponding to $\ket{\psi_+}$
and $\ket{\psi_-}$, respectively) and the oscillatory dynamics of excitonic coherence is dominated by $1755\,{\rm cm}^{-1}$
frequency component (see Fig.~\ref{fig5}c). We note that the difference between excitonic and vibronic splittings is significant,
of the order of $\Delta'-\Delta\approx 465\,{\rm cm}^{-1}$, and this shift cannot be described by conventional line shape
theory where multi-mode vibronic mixing is ignored and as a result the energy-gap between absorption peaks is reduced to the excitonic splitting (see the inset in Fig.~\ref{fig5}b).

\

{\bf Long-lived multi-mode vibronic coherence}. The considerable size of the multi-mode mixing effects on excitonic
energy gaps suggest a possibly significant influence on coherent excitonic
dynamics. The coarse-grained spectral density shown in red in Fig.~\ref{fig5}a, which corresponds to a vibrational lifetime of
$\gamma_{k} =(50\,{\rm fs})^{-1}$, yields short-lived oscillatory dynamics of excitonic coherence $\rho_{\pm}(t)
=\bra{E_-}\hat{\rho}_e(t)\ket{E_+}$ with $\hat{\rho}_e(t)$ denoting reduced electronic density matrix (see red line
in Fig.~\ref{fig5}c). Even if a few intra-pigment modes near-resonant with excitonic splitting are selected to be
weakly damped, $\gamma_{k}=(1\,{\rm ps})^{-1}$, the vibronic mixing with the large number of remaining strongly-damped
modes, $\gamma_{k}=(50\,{\rm fs})^{-1}$, suppresses the lifetime of excitonic coherences, making the resulting dynamics
essentially identical to that where all the modes are strongly damped (see Supplementary Note 7 for detailed analysis of 
multi-mode vibronic mixing). In sharp contrast, when the picosecond lifetime of
actual intra-pigment modes is considered, $\gamma_{k}=(1\,{\rm ps})^{-1}$, the excitonic coherence dynamics is dominated
by long-lived oscillations with frequency $\Delta' \approx 1755\,{\rm cm}^{-1}$, associated with the vibronic coherence
between $\ket{\psi_+}$ and $\ket{\psi_-}$ states (see black line in Fig.~\ref{fig5}c).

In 2D electronic spectroscopy, the third order nonlinear optical response of molecular systems is measured by using a sequence of femtosecond pulses with controlled time delays \cite{SMBrixner,SMJonas}. As is the case of pump probe experiments \cite{SMMukamel}, electronically excited state populations and coherences can be created by a pair of pump pulses, and the molecular dynamics in the electronic excited state manifold can be monitored by controlling the time delay $T$ between pump and probe. The additional time delay between two pump pulses enables one to monitor the molecular dynamics as a function of excitation and detection wavelengths for each waiting time $T$. The optical transitions induced by the pump pulses can also create vibrational coherences in the electronic ground state manifold, making it challenging to extract the information about coherent electronic dynamics from multidimensional spectroscopic data \cite{SMLimPRL2019}.

Our numerically exact simulations of the SP demonstrate that long-lived oscillatory signals in 2D electronic spectra can originate from purely vibrational coherences or from vibronic coherences induced by multi-mode mixing. The latter have been ignored in previous numerical studies which considered only
a few intra-pigment modes quasi-resonant with excitonic splitting and neglected all the modes that are far detuned from
excitonic transitions as they were deemed to have a negligible effect \cite{nalbach2011exciton}. However, the correct
assessment of the nature of oscillatory 2D signals requires the computation of 2D spectra under the influence of the full
 spectral density. In order to make such computation feasible, in Supplementary Note 8, we provide an approximate
master equation for vibronic dynamics, which takes into account multi-mode mixing effects and quantitatively reproduces
numerically exact absorption line shape of the SP. Fig.~\ref{fig5}d shows the resulting rephasing 2D spectra at waiting
time $T=0$ in the presence of inhomogeneous broadening. The 2D 
lineshape, shown as a function of excitation and detection {wavelengths}, is dominated by a diagonal peak excited and 
detected at $900\,{\rm nm}$ which coincides with the position of the main absorption peak (see Fig.~\ref{fig5}b). To 
investigate the excited state coherence between vibronic eigenstates $|\psi_+\rangle$ and $|\psi_-\rangle$, which induce 
the absorption peaks at $780$ and $900\,{\rm nm}$, respectively, we focus on a cross-peak R12 marked in Fig.~\ref{fig5}d. Fig.~\ref{fig5}e shows the 
transient of the cross-peak as a function of the waiting time $T$ where the oscillatory 2D signals originating from electronic ground state manifold, shown in 
red, are comparable to those of excited state signals, shown in blue. The ground state signals consist of multiple frequency 
components below $1600\,{\rm cm}^{-1}$, corresponding to the vibrational frequencies $\omega_k$ of underdamped intra-pigment 
modes, as shown in Fig.~\ref{fig5}f. It is important to note that the excited state signals include a long-lived oscillatory 
component with frequency $\sim 1800\,{\rm cm}^{-1}$, which is not present in the ground state signals and cannot originate 
from purely vibrational effects as they exceed the high-frequency cut-off of the environmental spectral density 
(see Fig.~\ref{fig5}a). This component must therefore originate from long-lived vibronic coherence due to multi-mode mixing. The long-lived oscillations at $\Delta'\approx 1800\,{\rm cm}^{-1}$ frequency cannot be described by coarse-grained environment models where only a few intra-pigment modes near-resonant with the excitonic splitting $\Delta\approx 1300\,{\rm cm}^{-1}$ are weakly damped ($\gamma_k=(1\,{\rm ps})^{-1}$), while all the other intra-pigment modes are strongly damped ($\gamma_k=(50\,{\rm fs})^{-1}$) or neglected ($s_k=0$) in 2D simulations (see Supplementary Note 8).
Our results demonstrate that while some oscillatory components in 2D spectra can originate from purely vibrational motions, 
long-lived 2D oscillations can also be the result of a strong vibronic mixing of excitons with a large number of underdamped 
intra-pigment modes.

\

{\bf Discussion}

Employing numerically exact methods and an analytical theory, we have investigated exciton-vibrational 
dynamics under the complete vibrational spectrum that has been estimated in earlier experiments. We considered
two paradigmatic regimes. The first regime, represented by an excitonic dimer in WSCP, is characterized by an excitonic 
splitting that is smaller than vibrational frequencies of intra-pigment modes. In this case, one main effect 
of vibronic coupling to the intra-pigment modes is a reduction of the dipole strength of 0-0 transitions
of monomers and of their effective coupling strength $V_{00}$ that determines the splitting between absorption 
peaks in the low-energy spectrum. A second important effect concerns the modulation of the vibrational sideband of optical transitions by a vibronic mixing between 0-0 and 0-1 transitions. Although the vibronic mixing is not strong enough to modulate the low-energy part of absorption spectra of WSCP, it can induce a notable dipole strength redistribution between 0-0 and 0-1 transitions, which cannot be described by approximate theories where the vibronic mixing is ignored.

In the second regime, represented by the SP of the photosynthetic reaction center of purple bacteria, the 
excitonic splitting is located in the middle of the high frequency part of the intra-pigment vibrational 
spectrum. In this case, the splitting between main absorption peaks can be even larger than the bare excitonic 
splitting, due to multi-mode vibronic mixing effects. This regime is found to be particularly suitable 
for the discovery of new long-lived quantum coherences in photosynthesis. We found that the coherence time 
of excitonic dynamics is not simply governed by the lifetime of quasi-resonant intra-pigment modes. Rather 
it is determined by the lifetimes of individual intra-pigment modes involved in a multi-mode vibronic mixing. 
This implies that approximate theoretical models based on coarse-graining of the high frequency part of
the vibrational environments \cite{Bennet_PNAS2018} may underestimate the lifetime of excitonic coherences 
and could be inappropriate to analyse quantum coherences observed in nonlinear experiments on photosynthetic 
systems. In addition, our results demonstrate that even if the frequency $\Delta'$ of oscillatory 2D signals is not well matched to one of the vibrational frequencies $\omega_k$ of intra-pigment modes, the long-lived 2D oscillations can be vibronic in origin, rather than being purely electronic, as is the case of the SP where $\omega_k \lesssim 1600\,{\rm cm}^{-1}<\Delta'\approx 1800\,{\rm cm}^{-1}$. This implies that the origin of long-lived oscillatory 2D signals cannot be identified based only on a comparison of the frequency spectrum of nonlinear signals with the vibrational frequency spectrum of underdamped modes. Hence, we contend that previously ignored multi-mode vibronic effects must be included in the 
interpretation of nonlinear spectroscopic signals before the current debate regarding the presence and nature 
of long-lived quantum coherences in pigment-protein complexes can be settled conclusively.

Our results suggest the possibility that the energy transfer dynamics between electronic states, such as excitons and charge-transfer states, could be governed by the multi-mode nature of the total vibrational environments, rather than a few vibrational modes quasi-resonant with electronic energy-gaps (see Supplementary Note 9).
The generality of the methods employed here also suggest that our results have a broad scope and can be of relevance
in a wide variety of scenarios involving strong hybridization of electronic and vibrational degrees of freedom,
such as recent observations of nonadiabatic dynamics in cavity polaritonics \cite{vergauwe2019modification,lather2019cavity}.
We expect that renormalization effects considered here may open an entirely new toolbox for vibrational reservoir
engineering with possible applications in information technologies and polaritonic chemistry.

{\bf Acknowledgments}. F.C.-S., A.M., J.L., S.F.H. and M.B.P. acknowledge financial support by the ERC Synergy grants BioQ and HyperQ, and support by the state of Baden-W\"urttemberg through bwHPC and the German Research Foundation (DFG) through grant no INST 40/575-1 FUGG (JUSTUS 2 cluster). A.M. acknowledges financial support by an IQST PhD fellowship. T.R. acknowledges financial support by the Austrian Science Fund (FWF): P 33155-NBL.




%

\begin{widetext}

\clearpage

\renewcommand{\figurename}{Supplementary Figure}
\renewcommand{\tablename}{Supplementary Table}
\renewcommand{\thetable}{\arabic{table}}
\renewcommand{\thesection}{Supplementary Note \arabic{section}}

\newcommand{\ad}{\ensuremath{a^\dagger}}
\def\bea{\begin{eqnarray}}
\def\eea{\end{eqnarray}}

\setcounter{figure}{0}
\setcounter{page}{1}
\setcounter{table}{0}

\thispagestyle{empty}
\begin{center}
{\bf\large Supplementary Information: \vspace{5mm} \\ Exact Simulation of Pigment-Protein Complexes Unveils \\ Vibronic Renormalization of Electronic Parameters in Ultrafast Spectroscopy}
\end{center}

\


\vspace{2cm}


{\bf\large \noindent Supplementary Figures:}
\\

\begin{figure*}[b]
\includegraphics[width=1\textwidth]{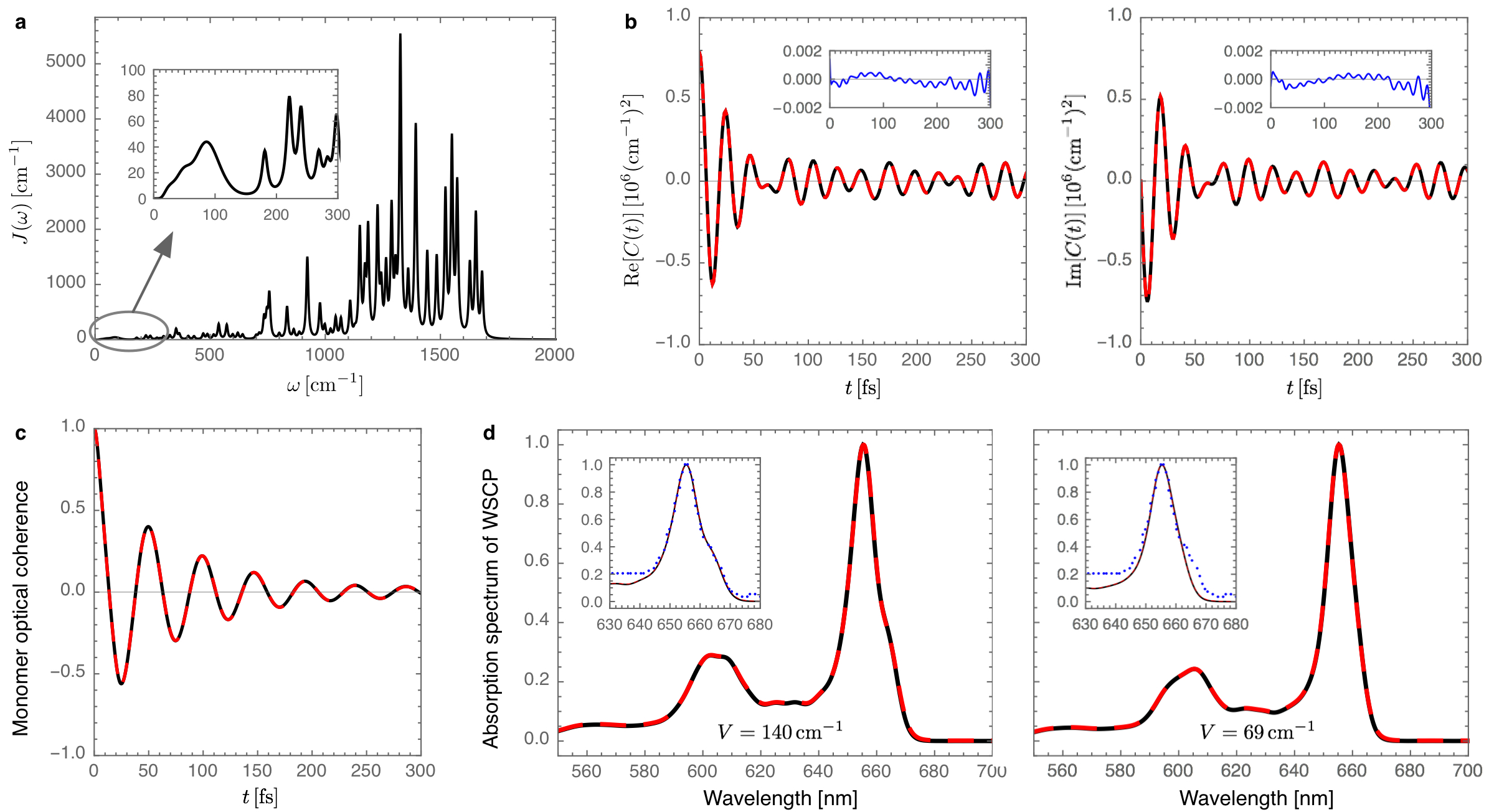}
\caption{({\bf a}) The experimentally estimated phonon spectral density $J(\omega)$ of WSCP consists of three log-normal functions at low frequencies
$\omega\lesssim 150\,{\rm cm}^{-1}$ and $55$ high-frequency underdamped modes modelled by Lorentzian functions with a vibrational damping rate of
$(1\,{\rm ps})^{-1}$. The inset shows the low energy part of the spectral density in detail. ({\bf b}) The real and imaginary parts of the
bath correlation function (BCF) $C(t)$ at $T=77\,{\rm K}$, shown in black, are fitted up to $300\,{\rm fs}$ with the sum of 13 exponentials as shown in red dashed
lines. The insets show the difference between target BCF and the fitting function, demonstrating that the fitting error is three orders of magnitude
smaller than the amplitudes of the target BCF. ({\bf c}) The dynamics of the real part of the optical coherence of a two-level monomer in the absence of static
disorder. The simulated HEOM results, shown in red, are well matched to the analytical solution shown in black: the site energy of the monomer is taken
to be zero for better visibility. The imaginary part of the coherence computed by HEOM is also well matched to the analytical solution (not shown here).
({\bf d}) The absorption spectra of WSCP homodimers with $V=140\,{\rm cm}^{-1}$ or $V=69\,{\rm cm}^{-1}$. The simulated HEOM results shown in red are well matched to the TEDOPA results shown in black. The simulated optical coherence of the dimer system decays within $300\,{\rm fs}$ mainly due to the static disorder (not shown here), implying that the fitting of the BCF up to $300\,{\rm fs}$ is sufficient for HEOM simulations. Experimentally measured absorption spectrum of WSCP is shown in blue as a guide for the eye. For $V=140\,{\rm cm}^{-1}$ ($V=69\,{\rm cm}^{-1}$), the mean site energies $\langle\varepsilon_1\rangle = \langle\varepsilon_2\rangle$ are taken to be 631.5 nm (629.8 nm) and the standard deviation of the Gaussian static disorder is $80\,{\rm cm}^{-1}$ ($72\,{\rm cm}^{-1}$). The angle between transition dipole moments of monomers is taken to be $39^{\circ}$.}
	\label{Fig_HEOM1}
\end{figure*}

\begin{figure*}[t]
	\includegraphics[width=1\textwidth]{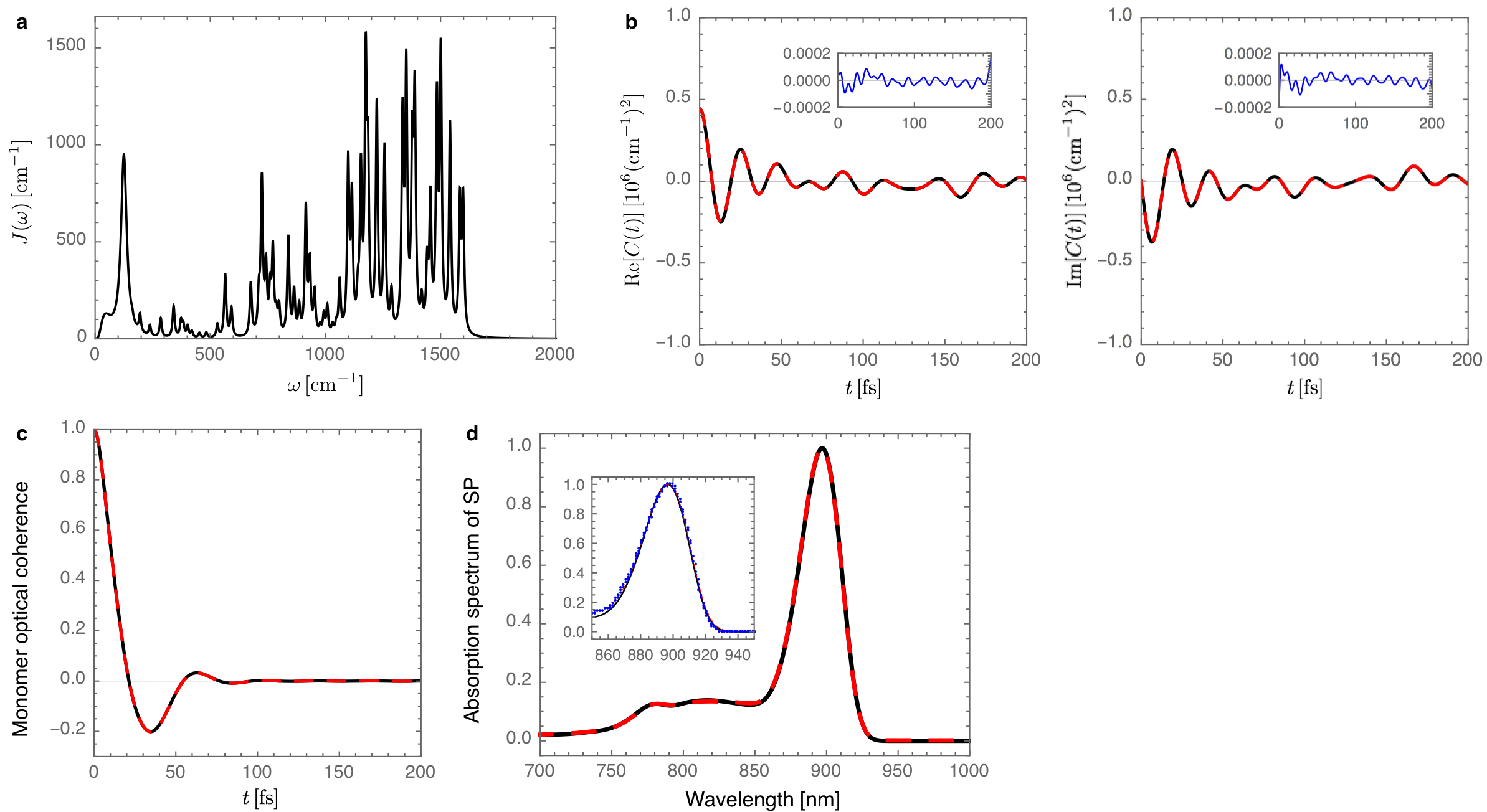}
	\caption{({\bf a}) Experimentally estimated phonon spectral density $J(\omega)$ of the special pair in bacterial reaction centers. ({\bf b}) The bath correlation function (BCF) at $T=5\,{\rm K}$, denoted by $C(t)$, is shown in black, which is fitted up to $200\,{\rm fs}$ with the sum of 16 exponentials as shown in red dashed lines. The difference between target BCF and fitting function is shown in blue, demonstrating that the fitting error is three orders of magnitude smaller than the amplitudes of the target BCF. ({\bf c}) The dynamics of the real part of the optical coherence of a two-level monomer in the absence of static disorder. The simulated HEOM results shown in red are well matched to the analytical solution shown in black: the site energy of the monomer is taken to be zero for better visibility. ({\bf d}) The absorption spectra of the special pair. The simulated HEOM results shown in red are well matched to the TEDOPA results shown in black. Experimentally measured absorption spectra of bacterial reaction centers are shown in blue as a guide for the eye. The mean site energies of sites 1 and 2 are taken to be 814.2 nm and 835.7 nm, respectively, with the detuning of $\langle\varepsilon_1-\varepsilon_2\rangle=315\,{\rm cm}^{-1}$. The standard deviation of local static disorder is taken to be $105\,{\rm cm}^{-1}$. The angle between transition dipole moments of monomers is taken to be $143^{\circ}$.}
	\label{Fig_HEOM2}
\end{figure*}

\begin{figure*}
	\includegraphics[width=0.8\textwidth]{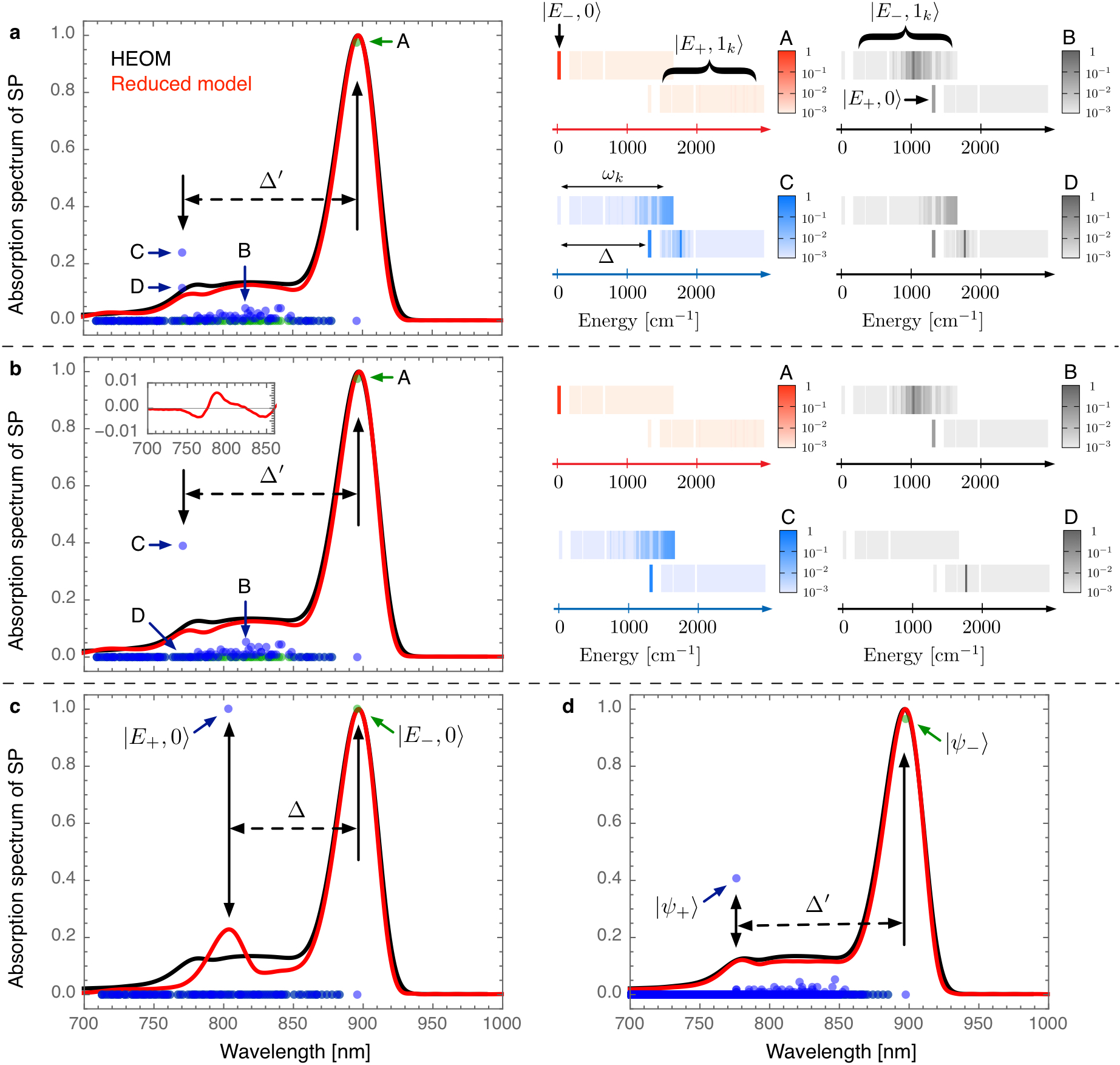}
	\caption{({\bf a}) Comparison of absorption line shapes of the special pair in bacterial reaction centers computed by HEOM (black) and reduced vibronic model (red). For each vibronic eigenstate of $\tilde{H}$, its transition energy from global ground state $\ket{g,0}$ and the fidelity with a vibrationally cold exciton state $|E_-,0\rangle$ ($|E_+,0\rangle$) are shown as a green (blue) dot. For vibronic eigenstates A-D, the fidelity with vibrationally cold $|E_\pm,0\rangle$ and vibrationally hot $|E_\pm,1_k\rangle$ states is displayed to visualize the presence of multi-mode vibronic mixing. ({\bf b}) Reduced model results without higher-order interaction terms. The difference between approximate absorption line shapes obtained by reduced models with and without the higher-order couplings is shown in the inset. ({\bf c}) Reduced model results without off-diagonal vibronic couplings. ({\bf d}) Reduced model results where double vibrational excitations are considered: in ({\bf a})-({\bf c}), single vibrational excitation subspace is considered.}
	\label{Fig_RM_abs}
\end{figure*}

\begin{figure*}
	\includegraphics[width=0.95\textwidth]{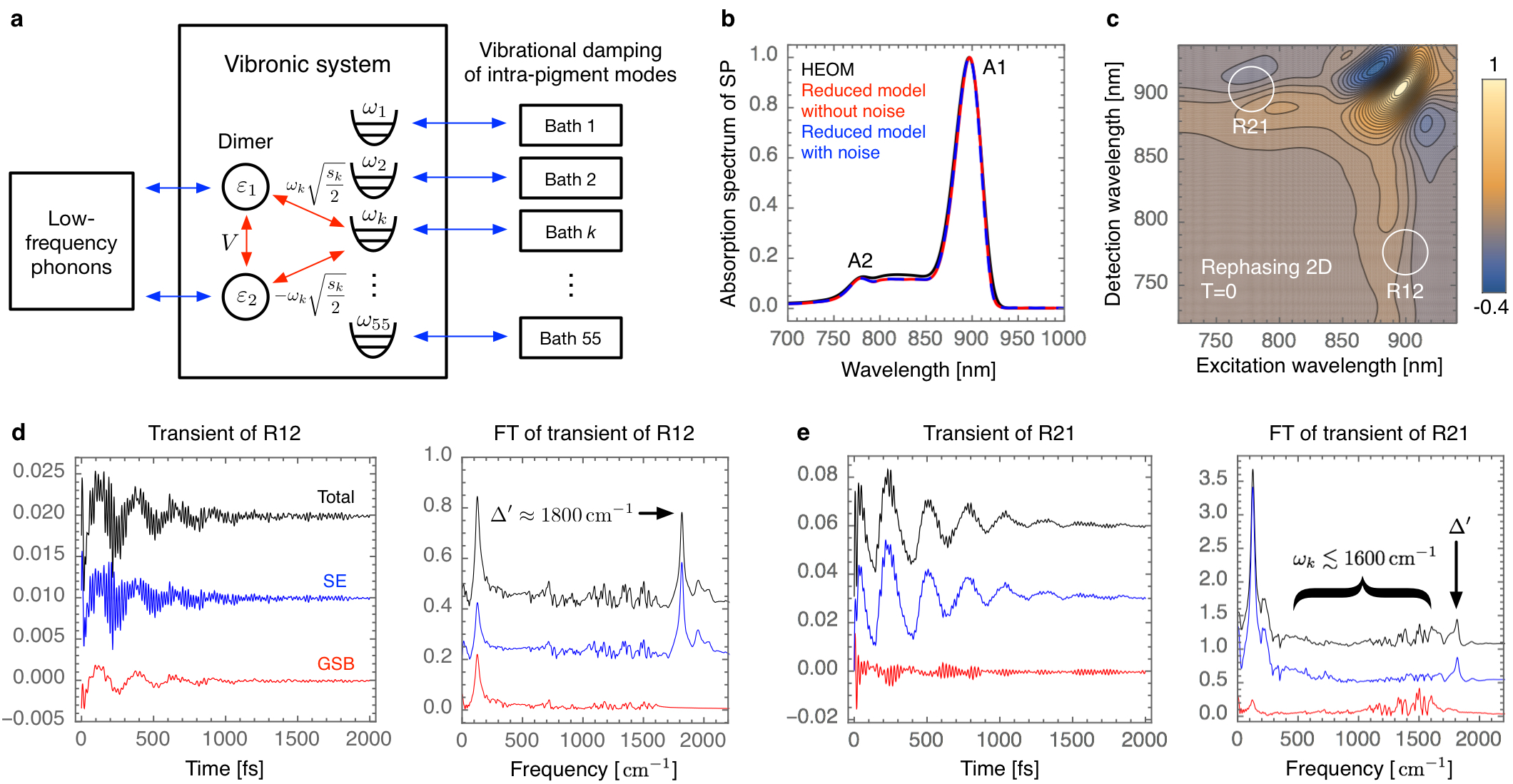}
	\caption{({\bf a}) Schematic representation of a vibronic model in the presence of noise. The vibronic system is characterized by the interaction between electronic states of a dimer and 55 oscillators associated with the relative motions of intra-pigment modes. The coupling between electronic states and low-frequency phonons induces electronic dephasing. In addition, each oscillator is coupled to an independent secondary bath, inducing vibrational damping. ({\bf b}) Absorption spectra of SP computed by HEOM and reduced vibronic models without and with noise are shown in black, red and blue, respectively, where double vibrational excitation subspace is considered. ({\bf c}) Rephasing 2D spectra at waiting time $T=0$. ({\bf d}) 2D signals at a cross-peak R12, marked in ({\bf c}), and corresponding Fourier transformation where total (GSB+SE), ground state (GSB) and excited state (SE) signals are shown in black, red and blue, respectively. Note that excited state signals as well as total 2D signals, namely the sum of ground and excited state signals, are dominated by vibronic coherence $|\psi_+\rangle\langle \psi_-|$, leading to 2D oscillations at $\Delta'\approx 1800\,{\rm cm}^{-1}$. ({\bf e}) 2D signals at a cross-peak R21, marked in ({\bf c}). Note that excited state signals are dominated by 2D oscillations at $\Delta'\approx 1800\,{\rm cm}^{-1}$, while ground state coherences are dominated by multiple components with frequencies $\omega_k\lesssim 1600\,{\rm cm}^{-1}$, which are lower than $\Delta'\approx 1800\,{\rm cm}^{-1}$. The offset values of transients and FTs are shifted for better visibility.}
	\label{Fig_RM_noise_abs_2D}
\end{figure*}

\begin{figure*}
	\includegraphics[width=1\textwidth]{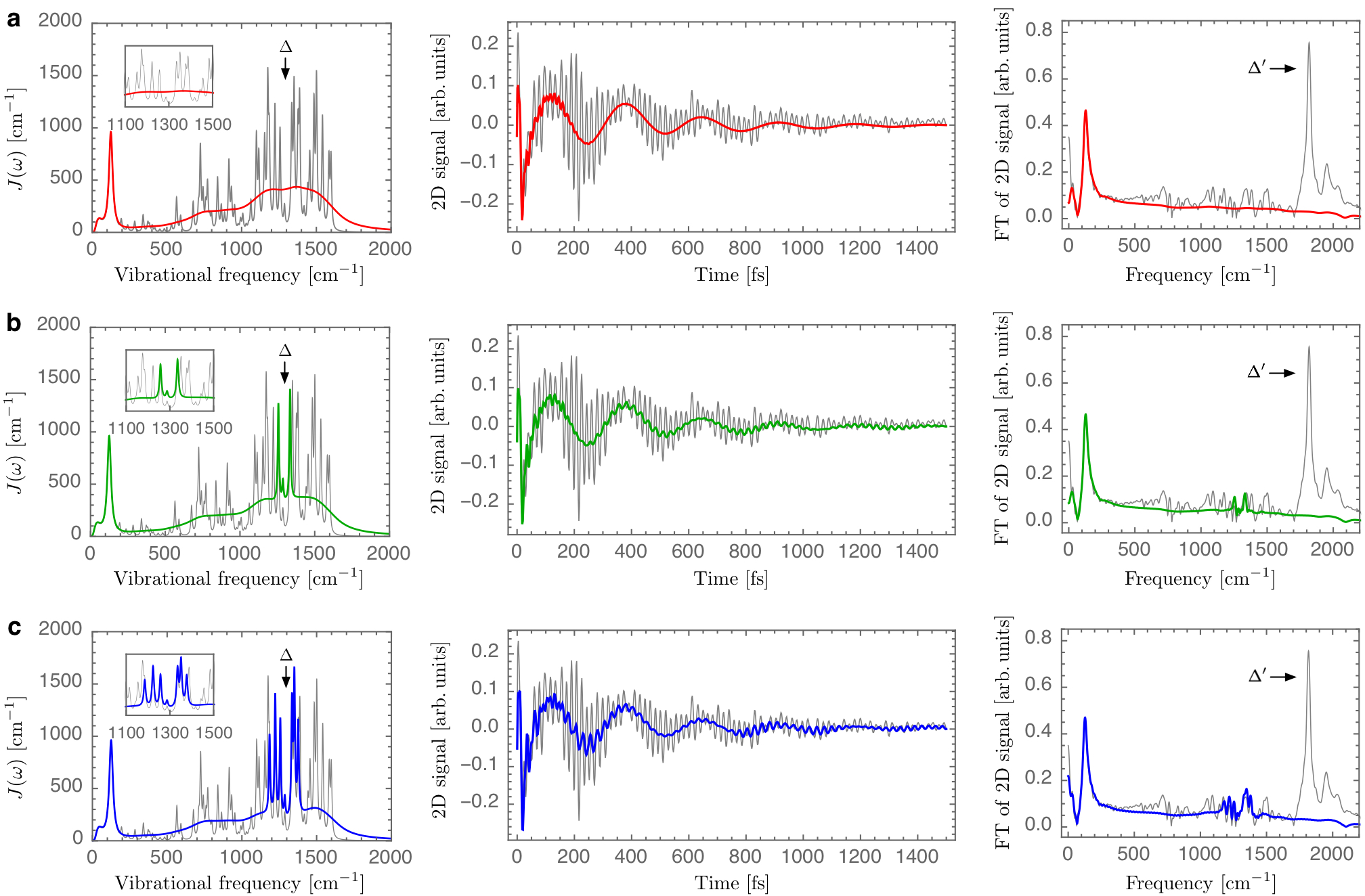}
	\caption{(\jl{{\bf a}) Fully coarse-grained spectral density is shown in red where all the 55 intra-pigment modes are modelled by broad Lorentzian spectral densities with a uniform width of $\gamma_k=(50\,{\rm fs})^{-1}$. Experimentally estimated phonon spectral density of SP is shown in grey where $\gamma_k=(1\,{\rm ps})^{-1}$ for all the 55 modes. Transient of a cross-peak R12 (SE component only, see Supplementary Figure~\ref{Fig_RM_noise_abs_2D}c and d) and its frequency spectrum, computed based on the fully coarse-grained spectral density, are shown in red where long-lived oscillations at $\Delta'\approx 1800\,{\rm cm}^{-1}$ frequency are absent. ({\bf b}) Partially coarse-grained spectral density is shown in green where the three intra-pigment modes whose frequencies are near-resonant with the excitonic splitting $\Delta\approx 1300\,{\rm cm}^{-1}$ of the SP are modelled by narrow Lorentzian spectral densities with a width of $\gamma_k=(1\,{\rm ps})^{-1}$, while all the other intra-pigment modes are modelled by the broad Lorentzian functions with $\gamma_k=(50\,{\rm fs})^{-1}$. The transient of the cross-peak R12 includes long-lived oscillations at the vibrational frequencies $\omega_k\approx \Delta$ of the near-resonant modes, as shown in green, but long-lived $\Delta'\approx 1800\,{\rm cm}^{-1}$ oscillations are absent. ({\bf c}) Similar features of 2D spectra are obtained when seven near-resonant intra-pigment modes are modelled by the narrow Lorentzian spectral densities, as shown in blue. We note that in 2D simulations, static disorder is considered where the standard deviation of a Gaussian distribution of the excitonic splitting is taken to be $\sqrt{2}\sigma\approx 150\,{\rm cm}^{-1}$.}}
	\label{Fig_SA}
\end{figure*}

\begin{figure*}
	\includegraphics[width=1\textwidth]{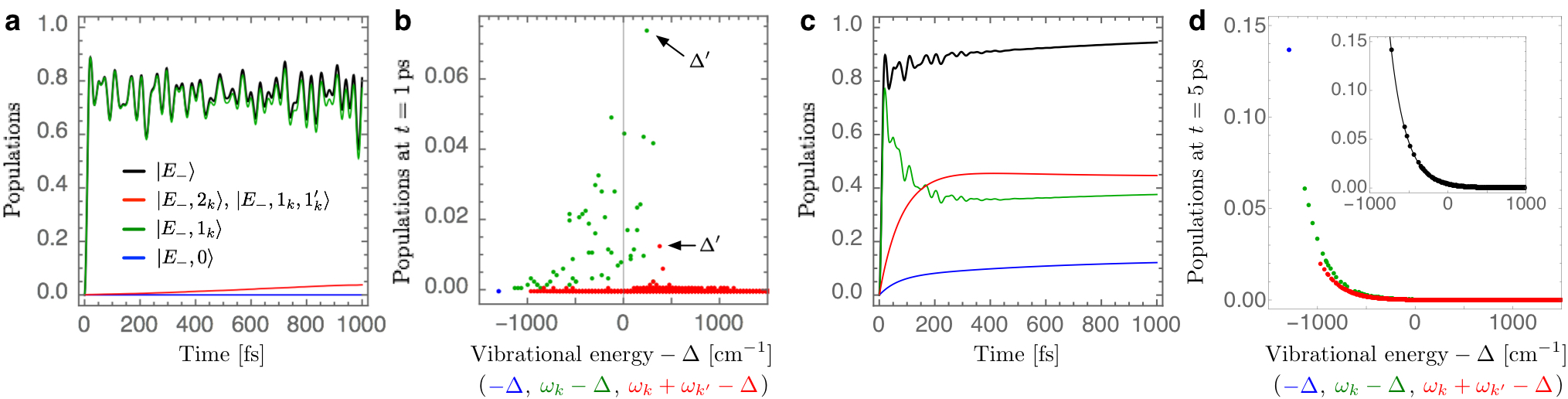}
	\caption{\jl{({\bf a}) Population dynamics of the exciton states of SP coupled to undamped 55 intra-pigment modes (relative motion modes). The vibronic coupling to low-frequency phonon environments is ignored, so that the population dynamics is coherent. The total populations of vibrationally cold ($\ket{E_-,0}$), singly excited ($\ket{E_-,1_k}$), and doubly excited states ($\ket{E_-,2_k}$ and $\ket{E_-,1_k,1_{k'}}$) of a lower-energy exciton state $\ket{E_-}$ are shown in blue, green, and red, respectively, and the total population of the lower-energy exciton states is shown in black. ({\bf b}) The populations of $\ket{E_-,0}$, $\ket{E_-,1_k}$, $\ket{E_-,2_k}$ and $\ket{E_-,1_k,1_{k'}}$ at time $t=1\,{\rm ps}$ are shown in black, green, red dots, respectively, as a function of the difference in vibrational energy and excitonic splitting $\Delta$. ({\bf c}) The population dynamics in the presence of the Lindblad noise induced by vibrational damping of the intra-pigment modes and the vibronic couplings to the low-frequency phonon environments at room temperature ($T=293\,{\rm K}$). ({\bf d}) The populations of $\ket{E_-,0}$, $\ket{E_-,1_k}$, $\ket{E_-,2_k}$ and $\ket{E_-,1_k,1_{k'}}$ at time $t=5\,{\rm ps}$ in the presence of the noise. In the inset, the populations of vibronic eigenstates at time $t=5\,{\rm ps}$ are shown in black dots as a function of the difference in vibronic energies and excitonic splitting, which are well matched to the Boltzmann distribution of a thermal state at room temperature, shown in black line.}}
	\label{Fig_SB}
\end{figure*}

\begin{figure*}
	\includegraphics[width=1\textwidth]{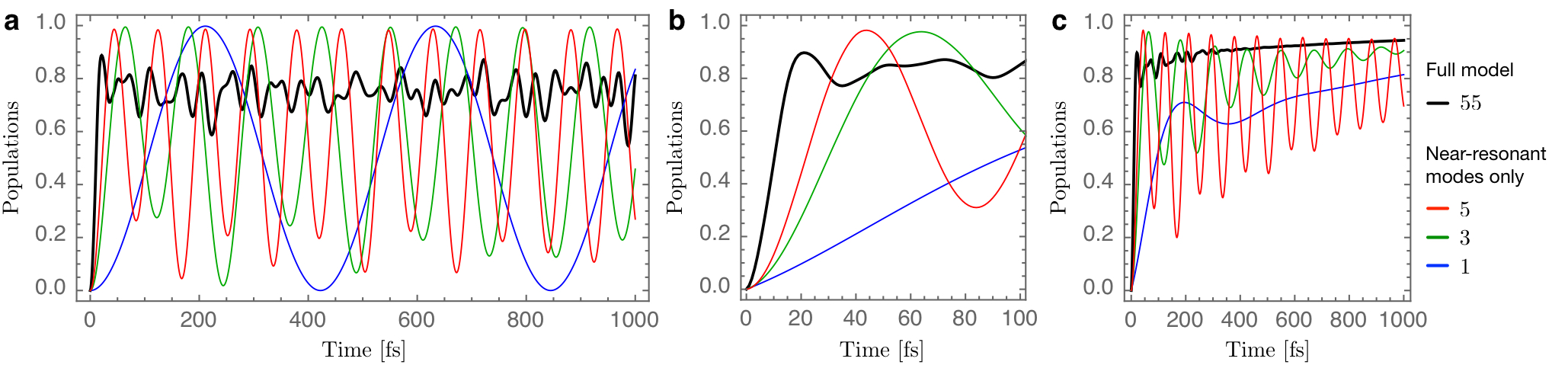}
	\caption{\jl{({\bf a}) Population dynamics of a lower-energy exciton state $\ket{E_-}$ of the SP when excitons are coupled to 55 intra-pigment modes (see black line) or coupled only to $M=1,3,5$ intra-pigment modes near-resonant with the excitonic splitting $\Delta\approx 1300\,{\rm cm}^{-1}$ of the SP (see blue, green, red lines, respectively). Here the Lindblad noise induced by vibrational damping of the intra-pigment modes and the vibronic couplings to low-frequency phonon environments at room temperature is neglected. ({\bf b,c}) The population dynamics of $\ket{E_-}$ in the presence of the Lindblad noise.}}
	\label{Fig_SC}
\end{figure*}

\clearpage

\begin{figure}
\includegraphics[width=0.9\textwidth]{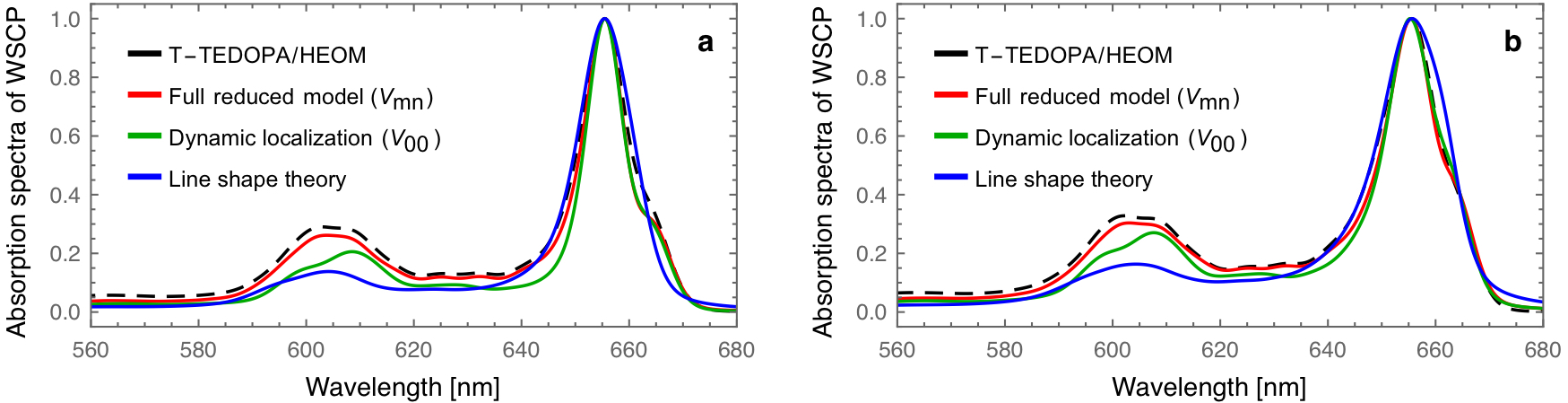}
\caption{({\bf a},{\bf b}) For different spectral densities of WSCP, ($J_{l}^{\rm WSCP}(\omega)+J_{h}(\omega),J_{l}^{\rm B777}(\omega)+J_{h}(\omega)$), numerically exact absorption spectra of WSCP are shown in black, which are well matched to approximate results obtained by a full reduced vibronic model taking into account all $V_{mn}$ couplings, shown in red. Approximate absorption spectra obtained by a reduced model where all the $V_{mn}$ couplings are ignored except for $V_{00}$ (dynamic localization) are shown in green. Approximate results computed by the conventional line shape theory are shown in blue where the full environmental spectral density $J(\omega)$ including 55 intrapigment vibrational modes is considered.}
\label{fig4}
\end{figure}

\begin{figure*}[t]
\includegraphics[width=1\textwidth]{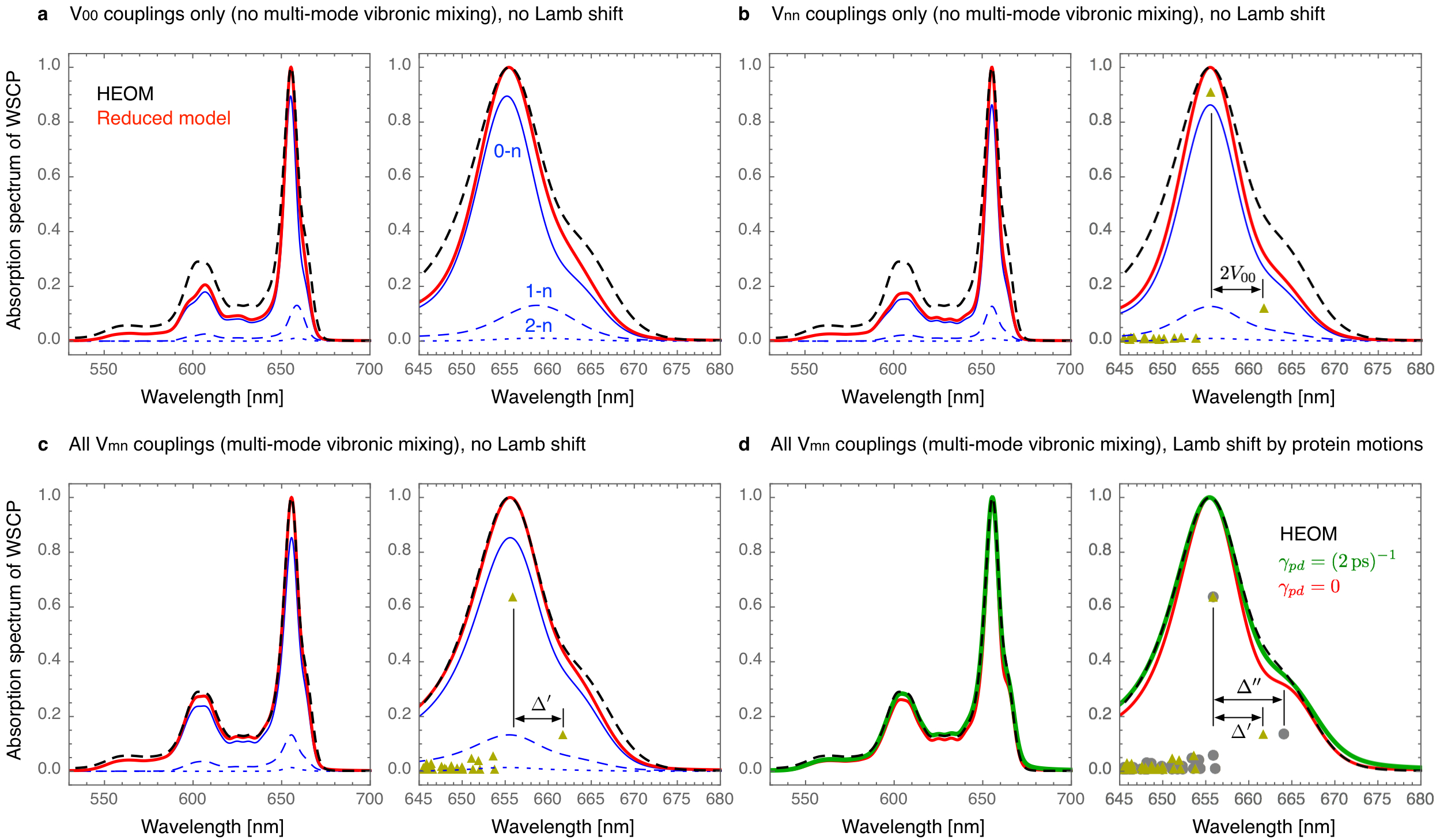}
\caption{({\bf a}) Numerically exact absorption spectra of WSCP at $T=77\,{\rm K}$ shown in a black dashed line, and approximate absorption line shape computed by numerical diagonalisation of vibronic Hamiltonian $H_0'+V_{00}(\ket{\varepsilon_1,0}\bra{\varepsilon_2,0}+h.c.)$, shown in a red solid line. The total approximate absorption line shape is a sum of 0-$n$, 1-$n$ and 2-$n$ transitions shown in blue solid, blue dashed and blue dotted lines, respectively. ({\bf b}) Approximate absorption line shape obtained by numerical diagonalisation of $H_0'+H_{I}^{(nn)}$ where $H_{I}^{(nn)}$ contains the couplings between local electronic excitations with identical vibrational states. The transition dipole strengths of 0-$n$ transitions are displayed in yellow triangles as a function of transition energies. ({\bf c}) Approximate absorption spectra based on the full Hamiltonian $H_0'+H_I'$ including the couplings amongst 0-0 and 0-1 transitions. ({\bf d}) Approximate absorption spectra based on $H_0'+H_I'+H_{LS}$ including Lamb shift $H_{LS}$. Approximate results in the absence and presence of an empirical pure dephasing rate, $\gamma_{pd}=0$ and $\gamma_{pd}=(2\,{\rm ps})^{-1}$, are shown in red and green solid lines, respectively. The transition dipole strengths and energies of 0-$n$ transitions in the presence of the Lamb shift are shown in grey dots.}
	\label{Fig_Renger_diagonalisation}
\end{figure*}

\clearpage

{\bf\large \noindent Supplementary Tables:}
\\

\begin{table}[h!]
\caption{Vibrational frequencies $\omega_k$ and Huang-Rhys factors $s_k$ of 55 intra-pigment modes of WSCP consisting of Chl{\it b} pigments~\cite{Pieper_2011JPCB}.}
\label{Table_WSCP}
\begin{ruledtabular}
\begin{tabular}{lllllllllll}
$k$ & 1 & 2 & 3 & 4 & 5 & 6 & 7 & 8 & 9 & 10 \\
\hline
$\omega_k\,[{\rm cm}^{-1}]$ & 181 & 221 & 240 & 269 & 283 & 298 & 325 & 352 & 366 & 405 \\
$s_k$ & 0.0173 & 0.0246 & 0.0182 & 0.0064 & 0.0036 & 0.0104 & 0.0112 & 0.0249 & 0.0112 & 0.0061 \\
\hline\hline
$k$ & 11 & 12 & 13 & 14 & 15 & 16 & 17 & 18 & 19 & 20 \\
\hline
$\omega_k\,[{\rm cm}^{-1}]$ & 430 & 470 & 488 & 515 & 537 & 572 & 598 & 620 & 641 & 700 \\
$s_k$ & 0.0050 & 0.0075 & 0.0061 & 0.0045 & 0.0157 & 0.0132 & 0.0036 & 0.0047 & 0.0033 & 0.0019 \\
\hline\hline
$k$ & 21 & 22 & 23 & 24 & 25 & 26 & 27 & 28 & 29 & 30 \\
\hline
$\omega_k\,[{\rm cm}^{-1}]$ & 713 & 734 & 746 & 757 & 800 & 834 & 863 & 887 & 922 & 977 \\
$s_k$ & 0.0025 & 0.0107 & 0.0112 & 0.0229 & 0.0022 & 0.0140 & 0.0033 & 0.0019 & 0.0291 & 0.0110 \\
\hline\hline
$k$ & 31 & 32 & 33 & 34 & 35 & 36 & 37 & 38 & 39 & 40 \\
\hline
$\omega_k\,[{\rm cm}^{-1}]$ & 998 & 1023 & 1045 & 1068 & 1108 & 1128 & 1150 & 1172 & 1186 & 1227 \\
$s_k$ & 0.0036 & 0.0022 & 0.0056 & 0.0050 & 0.0087 & 0.0011 & 0.0244 & 0.0121 & 0.0226 & 0.0249 \\
\hline\hline
$k$ & 41 & 42 & 43 & 44 & 45 & 46 & 47 & 48 & 49 & 50 \\
\hline
$\omega_k\,[{\rm cm}^{-1}]$ & 1243 & 1264 & 1288 & 1305 & 1326 & 1360 & 1393 & 1443 & 1484 & 1522 \\
$s_k$ & 0.0090 & 0.0126 & 0.0224 & 0.0093 & 0.0509 & 0.0093 & 0.0328 & 0.0121 & 0.0107 & 0.0185 \\
\hline\hline
$k$ & 51 & 52 & 53 & 54 & 55 &  &  &  &  &  \\
\hline
$\omega_k\,[{\rm cm}^{-1}]$ & 1550 & 1573 & 1628 & 1654 & 1681 &  &  &  &  &  \\
$s_k$ & 0.0241 & 0.0182 & 0.0081 & 0.0135 & 0.0067 &  &  &  &  &  \\
\end{tabular}
\end{ruledtabular}
\end{table}

\begin{table}[h!]
\caption{Vibrational frequencies $\omega_k$ and Huang-Rhys factors $s_k$ of 55 intra-pigment modes of BChl{\it a} pigments~\cite{Small_JPCB2001}.}
\label{Table_SP}
\begin{ruledtabular}
\begin{tabular}{lllllllllll}
$k$ & 1 & 2 & 3 & 4 & 5 & 6 & 7 & 8 & 9 & 10 \\
\hline
$\omega_k\,[{\rm cm}^{-1}]$ & 161 & 195 & 238 & 285 & 341 & 373 & 383 & 402 & 420 & 453 \\
$s_k$ & 0.0150 & 0.0400 & 0.0160 & 0.0200 & 0.0230 & 0.0100 & 0.0070 & 0.0060 & 0.0030 & 0.0020 \\
\hline\hline
$k$ & 11 & 12 & 13 & 14 & 15 & 16 & 17 & 18 & 19 & 20 \\
\hline
$\omega_k\,[{\rm cm}^{-1}]$ & 483 & 531 & 565 & 592 & 676 & 711 & 724 & 742 & 760 & 772 \\
$s_k$ & 0.0020 & 0.0040 & 0.0170 & 0.0070 & 0.0100 & 0.0060 & 0.0250 & 0.0100 & 0.0060 & 0.0120 \\
\hline\hline
$k$ & 21 & 22 & 23 & 24 & 25 & 26 & 27 & 28 & 29 & 30 \\
\hline
$\omega_k\,[{\rm cm}^{-1}]$ & 787 & 799 & 839 & 864 & 886 & 915 & 932 & 953 & 977 & 993 \\
$s_k$ & 0.0026 & 0.0036 & 0.0120 & 0.0050 & 0.0030 & 0.0130 & 0.0068 & 0.0040 & 0.0007 & 0.0017 \\
\hline\hline
$k$ & 31 & 32 & 33 & 34 & 35 & 36 & 37 & 38 & 39 & 40 \\
\hline
$\omega_k\,[{\rm cm}^{-1}]$ & 1008 &1031 & 1047 & 1062 & 1099 & 1115 & 1141 & 1154 & 1175 & 1185 \\
$s_k$ & 0.0024 & 0.0007 & 0.0007 & 0.0040 & 0.0120 & 0.0090 & 0.0020 & 0.0100 & 0.0160 & 0.0090 \\
\hline\hline
$k$ & 41 & 42 & 43 & 44 & 45 & 46 & 47 & 48 & 49 & 50 \\
\hline
$\omega_k\,[{\rm cm}^{-1}]$ & 1223 & 1257 & 1287 & 1335 & 1351 & 1377 & 1388 & 1418 & 1442 & 1456 \\
$s_k$ & 0.0130 & 0.0100 & 0.0020 & 0.0100 & 0.0120 & 0.0076 & 0.0100 & 0.0012 & 0.0026 & 0.0052 \\
\hline\hline
$k$ & 51 & 52 & 53 & 54 & 55 &  &  &  &  &  \\
\hline
$\omega_k\,[{\rm cm}^{-1}]$ & 1484 & 1501 & 1541 & 1584 & 1598 &  &  &  &  &  \\
$s_k$ & 0.0087 & 0.0104 & 0.0075 & 0.0044 & 0.0044 &  &  &  &  &  \\
\end{tabular}
\end{ruledtabular}
\end{table}

\clearpage

{\bf\large \noindent Supplementary Notes:}
\\

\section{V\MakeLowercase{ibronic} H\MakeLowercase{amiltonian} \MakeLowercase{and} P\MakeLowercase{olaron Tranformation}}

%
Here we present the vibronic Hamiltonian that describes the interaction between excitons and
vibrational modes in photosynthetic complexes in the exciton basis. This serves to define off-diagonal
vibronic couplings that play an essential role in the multi-mode vibronic mixing discussed in the main text.

For simplicity, we consider an excitonically coupled dimer where the electronic excitation of each site is
coupled to a local vibrational environment. The total Hamiltonian is modelled by $H=H_{e}+H_{v}+H_{e-v}$.
The electronic Hamiltonian $H_{e}$ is characterized by on-site energies $\varepsilon_i$ and inter-site electronic
coupling $V$
\begin{align}
	H_{e}=\sum_{i=1}^{2}\varepsilon_i \ket{\varepsilon_i}\bra{\varepsilon_i} +
    V(\ket{\varepsilon_1}\bra{\varepsilon_2}+\ket{\varepsilon_2}\bra{\varepsilon_1}).
\end{align}
The vibrational environments are modelled by quantum harmonic oscillators
\begin{align}
	H_{v}=\sum_{i=1}^{2}\sum_k\omega_k b_{i,k}^{\dagger} b_{i,k},
\end{align}
where $b_{i,k}$ describes a vibrational mode with frequency $\omega_k$, which is locally coupled to site $i$.
The coupling of the electronic degrees of freedom to their respective vibrational environment is described
by
\begin{align}
	H_{e-v}=\sum_{i=1}^{2}\ket{\varepsilon_i}\bra{\varepsilon_i}\sum_{k}\omega_{k}\sqrt{s_{k}}(b_{i,k}+b_{i,k}^{\dagger}),
\end{align}
where the vibronic coupling strength is quantified by a Huang-Rhys factor $s_{k}$. To identify the origin of the
multi-mode vibronic mixing, we define $B_{k}=\frac{1}{\sqrt{2}}(b_{1,k}+b_{2,k})$ and $b_{k}=\frac{1}{\sqrt{2}}(b_{1,k}-b_{2,k})$,
describing the center-of-mass and relative motions, respectively, of the local vibrational modes $b_{1,k}$ and $b_{2,k}$ with
identical frequency $\omega_k$~\cite{SMLimNJP2014}. $B_k$ and $b_k$ satisfy the bosonic commutation relations and the total Hamiltonian can
be expressed as $H=H_{e}+H_{c}+H_{r}$, where
\begin{align}
    H_{c}&=\sum_{k}\omega_{k}B_{k}^{\dagger}B_{k}+\openone^{(\varepsilon)}\sum_{k}\omega_{k}\sqrt{s_{k}/2}(B_{k}+B_{k}^{\dagger}),\\
	H_{r}&=\sum_{k}\omega_{k}b_{k}^{\dagger}b_{k}+\sigma_{z}^{(\varepsilon)}\sum_{k}\omega_{k}\sqrt{s_{k}/2}(b_{k}+b_{k}^{\dagger}),
\end{align}
with $\openone^{(\varepsilon)} = \ket{\varepsilon_1}\bra{\varepsilon_1}+\ket{\varepsilon_2}\bra{\varepsilon_2}$ and $\sigma_{z}^{(\varepsilon)}=\ket{\varepsilon_1}\bra{\varepsilon_1}-\ket{\varepsilon_2}\bra{\varepsilon_2}$.
Note that the center-of-mass modes $B_k$ are coupled to both electronic excited states $\ket{\varepsilon_1}$ and
$\ket{\varepsilon_2}$ with the same coupling strength and phase. This implies that the coupling to the
center-of-mass motion, described by $H_c$, does not affect electronic dynamics within the single excitation
manifold. Instead, it induces dephasing of optical coherences between electronic ground and excited states,
which broadens the absorption line shapes. The contribution of the center-of-mass motion to absorption line
width can be taken into account analytically, as described in \ref{section_HEOM}.

The vibronic coupling to the relative motion of vibrational modes, described by $H_r$, affects electronic dynamics
in the single excitation sector. Here we describe $H_r$ in terms of exciton states $\ket{E_{\pm}}$ that
diagonalize the electronic Hamiltonian $H_{e}=E_{+}\ket{E_+}\bra{E_+}+E_{-}\ket{E_-}\bra{E_-}$ and are parameterized by
\begin{align}
	\ket{E_+}&=\cos(\theta/2)\ket{\varepsilon_1}+\sin(\theta/2)\ket{\varepsilon_2},\\
	\ket{E_-}&=-\sin(\theta/2)\ket{\varepsilon_1}+\cos(\theta/2)\ket{\varepsilon_2},
\end{align}
where $\tan(\theta)=2V/(\varepsilon_1-\varepsilon_2)$ and $\Delta=E_{+}-E_{-}=\sqrt{(\varepsilon_1-\varepsilon_2)^2+4 V^2}$ denoting an excitonic splitting.
In the exciton basis, $H_r$ is expressed as
\begin{align}
	H_{r}=\sum_{k}\omega_{k}b_{k}^{\dagger}b_{k}+\sum_{i,j}A_{ij}\ket{E_i}\bra{E_j}\sum_{k}\omega_{k}\sqrt{s_k/2}(b_{k}^{\dagger}+b_{k}),
\end{align}
where $A_{ij}=\langle E_{i}|\varepsilon_{1}\rangle\langle \varepsilon_{1}|E_{j}\rangle-\langle E_{i}|\varepsilon_{2}\rangle\langle
\varepsilon_{2}|E_{j}\rangle$, leading to $A_{++}=-A_{--}=\cos(\theta)$ and $A_{+-}=A_{-+}=-\sin(\theta)$. We call the terms proportional
to $\ket{E_i}\bra{E_i}$ diagonal vibronic couplings, and the other terms  proportional to $\ket{E_i}\bra{E_j}$ with $i\neq j$ off-diagonal
vibronic couplings.

We note that the diagonal vibronic couplings describe the shift of the equilibrium positions of the vibrational modes $b_k$ depending
on electronic states. This can be seen more clearly in a displaced vibrational basis. We decompose the total Hamiltonian into two parts
$H=H_0+H_{I}$ where $H_0$ includes the diagonal vibronic couplings
\begin{align}
	H_{0}&=H_{e}+H_{c}+\sum_{k}\omega_{k}b_{k}^{\dagger}b_{k}+\cos(\theta)\sigma_z\sum_{k}\omega_{k}\sqrt{s_k/2}(b_{k}^{\dagger}+b_{k}),
\end{align}
with $\sigma_z=\ket{E_+}\bra{E_+}-\ket{E_-}\bra{E_-}$, while $H_I$ consists of the off-diagonal vibronic couplings
\begin{align}
	H_{I}=-\sin(\theta)\sigma_x\sum_{k}\omega_{k}\sqrt{s_k/2}(b_{k}^{\dagger}+b_{k}).
\end{align}
with $\sigma_x = \ket{E_+}\bra{E_-}+\ket{E_-}\bra{E_+}$. By applying a unitary operator $U=|g\rangle\langle g|+|E_+\rangle\langle E_+|D_{\theta}+|E_-\rangle\langle E_-| D_{\theta}^{\dagger}$ with
$D_{\theta}=\exp(\cos(\theta)\sum_{k}\sqrt{\frac{s_k}{2}}(b_{k}^{\dagger}-b_{k}))$ to the total Hamiltonian $H$ (the polaron transformation),
it can be shown that $H_{0}$ is diagonalised as the diagonal vibronic couplings are reduced to the global energy-level shift of exciton states
which is proportional to the reorganization energy $\sum_{k}\omega_{k}s_{k}$
\begin{align}
	UH_{0}U^{\dagger}&=H_{e}+H_{c}+\sum_{k}\omega_{k}b_{k}^{\dagger}b_{k}-\sum_{i}\ket{E_i}\bra{E_i}\cos^{2}(\theta)\sum_{k}\omega_k\frac{s_k}{2},\nonumber
\end{align}
while $H_I$ describes the vibronic mixing between exciton states mediated by vibrational modes $b_k$
\begin{align}
	UH_{I}U^{\dagger}&=-\sin(\theta)\ket{E_+}\bra{E_-}D_{\theta}\sum_{k}\omega_{k}\sqrt{s_k/2}(b_{k}^{\dagger}+b_{k})D_{\theta}+h.c.,
\end{align}
where $h.c.$ stands for Hermitian conjugate. These off-diagonal couplings are treated perturbatively in the second order cumulant expansion
technique~\cite{SMMukamel}, which has been employed to compute optical responses of multi-site systems approximately. To fully take
into account the off-diagonal vibronic couplings in simulations, one needs to employ numerically exact methods, such as TEDOPA and HEOM
discussed in \ref{section_TEDOPA} and \ref{section_HEOM}, respectively.

\section{P\MakeLowercase{erturbative} d\MakeLowercase{escription of} m\MakeLowercase{ulti}-m\MakeLowercase{ode} v\MakeLowercase{ibronic} s\MakeLowercase{plitting without} p\MakeLowercase{olaron} t\MakeLowercase{ransformation}}


\jl{In the main text, we show how the shift of an excitonic splitting $\Delta$ to a vibronic splitting $\Delta'$ induced by a multi-mode vibronic mixing is described by second order perturbation theory in the polaron basis. Here we show that a qualitatively similar vibronic energy renormalization can be derived in the regular basis without the polaron transformation by using the second order perturbation theory.}

\jl{As shown in \ref{section_vibronic_H_polaron_transformation}, the exciton states are mixed with the relative motions of the local vibrational modes and the center-of-mass motions of the local modes do not modulate the energy-gap between exciton states. Therefore we decompose the Hamiltonian of the excitons coupled to the relative motion modes $b_k$ into two parts $H_0=\sum_{i}E_{i}\ket{E_i}\bra{E_i}+\sum_{k}\omega_k b_{k}^{\dagger}b_{k}$ and $H_I=\sigma_{z}^{(\varepsilon)}\sum_{k}\omega_k\sqrt{s_k/2}(b_k+b_{k}^{\dagger})$ with $\sigma_{z}^{(\varepsilon)}=\ket{\varepsilon_1}\bra{\varepsilon_1}-\ket{\varepsilon_2}\bra{\varepsilon_2}$ where $H_I$ is considered a perturbation term. The eigenstates of $H_0$ include $\ket{E_+,0}$, $\ket{E_+,1_k}$, $\ket{E_-,0}$, $\ket{E_-,1_k}$ where $\ket{0}$ denotes a global vibrational ground state where all the modes are in their vibrational ground states, and $\ket{1_k}$ represents a singly excited vibrational state where only the $k$-th mode is singly excited and all the other modes are in their vibrational ground states. Using the second order perturbation theory, it can be shown that the excitonic splitting $\Delta=E_+-E_-$ between $\ket{E_+,0}$ and $\ket{E_-,0}$ is shifted to a vibronic splitting in the form
\begin{equation}
	\Delta'=\Delta \left( 1 + \frac{4 V^2}{\Delta^2}\sum_{k}\frac{s_k \omega_{k}^2}{\Delta^2 - \omega_{k}^{2}}\right),
\end{equation}
which is qualitatively similar to the vibronic splitting estimated in the polaron basis in the sense that $\Delta'<\Delta$ if $\Delta<\omega_k$, and $\Delta'>\Delta$ if $\Delta>\omega_k$ due to the dependence of $\Delta'$ on $\Delta^2 - \omega_{k}^{2}$. Note that when the total Huang-Rhys factor $\sum_{k}s_k$ with $s_k\ge 0$ is sufficiently small, namely in the weak vibronic coupling limit, the vibronic splittings $\Delta'$ estimated in the regular and polaron bases coincide, as $\alpha=\exp(-2\cos^{2}(\theta)\sum_{k}s_k)\rightarrow 1$ (see Eq.~(9) in the main text).}


\section{TEDOPA}\label{section_TEDOPA}

The goal of this work is the numerical evaluation of absorption spectra in the presence of a structured environment
as well as energetic and orientational disorder of the sample. To this end we will employ algorithms with controlled numerical error
that can be reduced to any desired degree. The two methods that we choose are the time-evolving density matrix with
orthogonal polynomials algorithm (TEDOPA) \cite{Prior_PRL2010,Chin_JMP2010,Tamascelli_2019PRL}, to be described in
this section, and a variation of the hierarchical equations of motion (HEOM) that we will describe in the following
\ref{section_HEOM}.

The original formulation of TEDOPA for finite temperature environments requires the time consuming preparation of
the thermal state of the environment and subsequent time evolution by means of matrix product operators (MPO)
\cite{Chin_2013NPhys} as opposed to the zero temperature case for which the initial state of the environment is trivial and
the system can be described by the more efficient matrix product state (MPS) formalism \cite{Prior_PRL2010,Chin_JMP2010}.
Recently, however, it was demonstrated that there is an exact analytical mapping of the finite temperature case to the zero temperature case
which allows for the direct computation of system observables \cite{tamascelli2018nonperturbative,Tamascelli_2019PRL}. This approach, termed T-TEDOPA,
leads to a major reduction in computational cost and, crucially, provides opportunities for further simplifications in the computations of absorption spectra that lead to additional significant reductions in computational cost by several orders of magnitude.

The calculation of the absorption spectrum requires the Fourier-Laplace transform of the correlation function $D(t)$ of
the electronic transition dipole operator $\mu=\sum_i^N (\boldsymbol{\mu}_i\cdot{\bf e}) \ket{\varepsilon_i}\bra{g} + h.c.$ averaged
over orientational and energetic disorder \cite{Renger_2002JCP} where $\ket{\varepsilon_i}$ denotes the local excitation at site $i$, $\ket{g}$ the global electronic ground state, $\boldsymbol{\mu}_i$ the dipole moment of site $i$ and ${\bf e}$ the direction of polarisation of the electric field inducing absorption. This dipole-dipole correlation function is given by
\bea
    \langle D(t) \rangle_{{\bf e},\varepsilon_i}=\langle \mbox{Tr}[ \mu e^{-iHt}\mu\rho_{th}e^{iHt}]\rangle_{{\bf e},
    \varepsilon_i},
\eea
where $\rho_{th} = e^{-\beta H}/ \mbox{Tr}[e^{-\beta H}]$. Even well beyond physiological temperatures $\rho_{th}$ is
exceedingly well approximated by $\rho_{th} = |g\rangle\langle g| \otimes e^{-\beta H_v}/\mbox{Tr}[e^{-\beta H_v}]$ as 
the exciton energy is typically found to be in the eV range. With this assumption, the T-TEDOPA approach obtains identically 
the same dynamics of electronic observables under a mapping that takes the initial state to $|g\rangle\langle g| \otimes 
|vac\rangle\langle vac|$, where $|vac\rangle$ denotes the vacuum state of the environment for which all vibrational modes 
are in their ground state and, simultaneously, takes the environmental spectral density from the original $J(\omega)$ on 
the interval  $[0,\infty]$ to
\begin{equation}
    J_{\beta}(\omega) = \frac{1}{2} sign(\omega)J(|\omega|) \left[ 1 + \coth\left( \frac{\beta\omega}{2}\right)\right],
\end{equation}
which is now defined on the entire real axis $[-\infty,\infty]$. Then the dipole-dipole correlation function is reduced to
\bea
    \langle D(t) \rangle_{{\bf e},\varepsilon_i} &=& \langle \mbox{Tr}[ \mu e^{-iH_{\beta}t}\mu\, |g\rangle\langle g|
    \otimes |vac\rangle\langle vac|e^{iH_{\beta}t}]\rangle_{{\bf e},\varepsilon_i}.
\eea
Now making use of $H|g\rangle|vac\rangle = 0$ and specialising to the case of a dimer ($N=2$) we find
\bea
    \langle D(t) \rangle_{{\bf e},\varepsilon_i} &=& \langle \mbox{Tr}[ \mu e^{-iH_{\beta}t}\mu\, |g\rangle\langle g|
    \otimes |vac\rangle\langle vac|]\rangle_{{\bf e},\varepsilon_i}\\
    &=& \langle \sum_{i,j=1}^{2} (\boldsymbol{\mu}_i\cdot{\bf e})^* (\boldsymbol{\mu}_j\cdot{\bf e}) \langle vac|\langle \varepsilon_i|e^{-iH_{\beta}t} |\varepsilon_j\rangle|vac\rangle \rangle_{{\bf e},\varepsilon_i}\\
    &=& \langle \sum_{i,j=1}^{2} (\boldsymbol{\mu}_i^*\cdot\boldsymbol{\mu}_j) \langle vac|\langle \varepsilon_i|e^{-iH_{\beta}t}
    |\varepsilon_j\rangle|vac\rangle \rangle_{\epsilon_i}. \label{eq2}
\eea
The computational effort of TEDOPA simulations typically grows stronger than linear with the simulated time $t$,
because it requires longer environmental chains and higher bond dimension of the employed matrix product states. 
Hence it proves advantageous to compute Eq.~(\ref{eq2}) via
\begin{eqnarray*}
    |\psi_i(t/2)\rangle &=& e^{-iH_{\beta}t/2}|\varepsilon_i\rangle|vac\rangle = \left(e^{iH_{\beta}t/2}|\varepsilon_i\rangle|vac\rangle\right)^*,
\end{eqnarray*}
so that
\bea
    \langle D(t) \rangle_{{\bf e},\varepsilon_i} &=& \langle
    || \sum_{i=1}^{2} \boldsymbol{\mu}_j |\psi_j(t/2)\rangle||^2 \rangle_{\epsilon_i}.
\eea
For the parameters in our work, this rearrangement of the dipole-dipole correlation function leads to more than one order of magnitude of reduction in the simulation time. The average over the energetic disorder can be simplified by two measures. First, instead of an independent
average over site energies $\varepsilon_i$, described by
\bea
	\langle D(t) \rangle_{{\bf e},\varepsilon_i}=\frac{1}{2\pi\sigma^2}\int_{-\infty}^{\infty}\int_{-\infty}^{\infty}d\varepsilon_{1}d\varepsilon_{2}\exp\left(-\frac{1}{2\sigma^2}(\varepsilon_1-\langle\varepsilon_1\rangle)^{2}-\frac{1}{2\sigma^2}(\varepsilon_2-\langle\varepsilon_2\rangle)^{2}\right)|| \sum_{i=1}^{2} \boldsymbol{\mu}_j |\psi_j(t/2)\rangle||^{2},
\eea
we compute the ensemble average over $\xi_1=\varepsilon_1-\varepsilon_2$ and $\xi_2=\frac{1}{2}(\varepsilon_1+\varepsilon_2)$
\bea
	\langle D(t) \rangle_{{\bf e},\varepsilon_i}=\frac{1}{2\pi\sigma^2}\int_{-\infty}^{\infty}\int_{-\infty}^{\infty}d\xi_{1}d\xi_{2}\exp\left(-\frac{1}{4\sigma^2}(\xi_1-\langle\xi_1\rangle)^{2}-\frac{1}{\sigma^2}(\xi_2-\langle\xi_2\rangle)^{2}\right)|| \sum_{i=1}^{2} \boldsymbol{\mu}_j |\psi_j(t/2)\rangle||^{2}.
\eea
The ensemble average over $\xi_2=\frac{1}{2}(\varepsilon_1+\varepsilon_2)$ does not require the repetition of simulations, as the global shift in site energies does not affect electronic dynamics, but simply induces a time-dependent phase factor, which can be taken into account analytically. In other words, after numerical simulation of $|| \sum_{i=1}^{2} \boldsymbol{\mu}_j |\psi_j(t/2)\rangle||^{2}_{\xi_1,\langle\xi_2\rangle}$ for $\xi_1=\varepsilon_1-\varepsilon_2$ and $\xi_2=\langle\xi_2\rangle=\frac{1}{2}(\langle\varepsilon_1\rangle+\langle\varepsilon_2\rangle)$, the average over $\xi_2$ can be computed by
\bea
	\frac{1}{\sqrt{\pi}\sigma}\int_{-\infty}^{\infty}d\xi_{2} e^{-i(\xi_2-\langle\xi_2\rangle) t_2} e^{-\frac{1}{\sigma^2}(\xi_{2}-\langle\xi_2\rangle)^{2}} || \sum_{i=1}^{2} \boldsymbol{\mu}_j |\psi_j(t/2)\rangle||^{2}_{\xi_1,\langle\xi_2\rangle}=\exp(-\frac{1}{4}\sigma^2 t^2)|| \sum_{i=1}^{2} \boldsymbol{\mu}_j |\psi_j(t/2)\rangle||^{2}_{\xi_1,\langle\xi_2\rangle},
\eea
where the Gaussian dephasing $e^{-\frac{1}{4}\sigma^2 t^2}$ leads to a broadening of absorption line shapes. The numerical average over energy difference $\xi_1=\varepsilon_1-\varepsilon_2$ can be carried out  most efficiently by means of a Gauss-Hermite interpolation of the integral over the frequency difference. Compared to a straightforward averaging with equidistant sampling points, this leads to a reduction of computational effort by almost one order of magnitude.

A further significant step in the optimisation of the T-TEDOPA simulation concerns an essential part of the
time-evolving block-decimation (TEBD) algorithm. This step selects the best MPS-approximation with fixed bond dimension
through a decimation technique that relies on the singular value decomposition (SVD). As the complexity of each
time-evolution step is dominated by the SVD we follow Refs.~\onlinecite{tamascelli2015improved,kohn2018probabilistic} and
apply a randomized version of the SVD routine
(rSVD) which trades a reduction in numerical accuracy for a significant increase in computational efficiency.
At first sight, for the relatively moderate matrix sizes in our simulations, one may not expect significant computational
gains due to the use of the rSVD. However, two specific aspects of the present problem lead to a different conclusion.
As the absorption spectrum is obtained by the Fourier transform of the correlation function $D(t)$ and is further
broadened due to the average over orientational and energetic disorder, small errors in the time evolution tend to
be averaged out. Furthermore, the uncertainty in the experimentally determined environmental spectral density makes
it unnecessary to aim for extremely high precision in the absorption spectra. A relative error of the order of
about $10^{-3}$ suffices for our application. This allows us to accept a larger tolerated error in the rSVD which
increases its speed significantly. As a result of the use of the rSVD we have been able to obtain more than one order
of magnitude reduction in CPU time without noticeably affecting the quality of the results. As a result of these
and some other minor measures, the computation of a single absorption spectrum fully averaged over energetic
and orientational disorder for the full spectral density of WSCP takes about 3 minutes on 7 processors (2\,GHz).

We note that the simulation cost of TEDOPA can be reduced further by transforming the two local baths to two effective
baths describing the center-of-mass and the relative motion of the local modes with identical vibrational frequencies
(see \ref{section_HEOM}). As the influence of the center-of-mass modes on the reduced electronic system dynamics
can be treated analytically, one can remove the center-of-mass modes from the dynamics and hence the TEDOPA simulations
and consider a dimeric electronic system coupled to a single global bath describing the relative motion modes. The application
of orthogonal polynomial algorithm \cite{Prior_PRL2010,Chin_JMP2010,Tamascelli_2019PRL} to this model leads to a single
chain of coupled harmonic oscillators where the first oscillator is also coupled to the electronic states of a dimeric
system. This enables one to reduce the simulation cost of TEDOPA compared to the case that two local baths are mapped
to two chains coupled via the electronic states of the dimer. This also implies that one can consider a trimeric electronic
system coupled to three local baths in TEDOPA simulations, as three local baths can be mapped to three global baths,
including a bath describing the center-of-mass motions which can be neglected in simulations. In this case, the remaining
two global baths, describing relative motion modes, can be mapped to two chains coupled via the electronic states of a
trimer, which can be efficiently simulated within the framework of TEDOPA.


\section{HEOM}\label{section_HEOM}


Hierarchical equations of motion (HEOM) are a numerically exact method for computing reduced system dynamics where a quantum system
is linearly coupled to a thermal environment modelled by quantum harmonic oscillators.
In the following we will explain the specifics of our HEOM implementation including the measures that we have taken to optimise
the calculation of absorption spectra.

We assume that at the initial time, system and environment are uncorrelated and the environment is in a thermal state at temperature
$T$. In this work, we consider a local bath model where each site is coupled to an independent bath modelled by an identical phonon
spectral density $J(\omega)$, characterizing system-environment coupling strengths as a function of environmental mode frequencies
$\omega$. The influence of such a harmonic bath on the reduced system dynamics is fully characterized by the bath correlation function
\begin{align}
	C(t)&=\int_{0}^{\infty}d\omega J(\omega)\left(\coth\left(\frac{\beta\omega}{2}\right)\cos(\omega t)-i\sin(\omega t)\right).
\end{align}
The parameters of HEOM simulations are determined by a multi-exponential fitting of the bath correlation function
\begin{align}
	C(t)\approx\sum_{j=1}^{N_C}(d_{j+} e^{-f_j t}+d_{j-} e^{-f_j^* t})+\sum_{k=1}^{N_R}d_k e^{-\gamma_k t},\label{eq_lim:multi_exponential_fitting}
\end{align}
where the fitting variables $d_{j+}$, $d_{j-}$, $f_j$, $d_k$ are complex-valued, while $\gamma_k$ are real-valued. The total number of
exponential functions is given by $2N_C+N_R$, which is one of the key factors determining the simulation cost of HEOM. The simulation of reduced system
dynamics within a finite time window, $0\le t\le t_{\rm max}$, requires the multi-exponential fitting of $C(t)$ within that time window,
as the fitting quality outside of the time window of interest, namely $t>t_{\rm max}$, does not affect the system dynamics for $0\le t\le
t_{\rm max}$. The fitting quality can be monitored by comparing simulated HEOM results with the analytical solutions of solvable models or
the data obtained by alternative numerically exact methods. In this work, we consider an analytically solvable monomer model, namely a
two-level system under dephasing noise. For a given set of fitting parameters,
the dynamics of the reduced system density matrix $\hat{\rho}_{\bf 0}$ can be non-perturbatively computed by using HEOM in the form
\begin{align}
	\frac{d\hat{\rho}_{\bf n}}{dt}&=-\frac{i}{\hbar}[\hat{H}_s,\hat{\rho}_{\bf n}]-\sum_{m=1}^{N_{S}}\left(\sum_{j=1}^{N_C}(n_{mj+}f_{j}+n_{mj-}f_{j}^{*})+\sum_{k=1}^{N_R}n_{mk}\gamma_{k}\right)\hat{\rho}_{\bf n}-i\sum_{m=1}^{N_{S}}\left[\hat{\sigma}_{m}^{\dagger}\hat{\sigma}_{m},\sum_{j=1}^{N_C}(\hat{\rho}_{{\bf n}^{+}_{m j+}}+\hat{\rho}_{{\bf n}^{+}_{m j-}})+\sum_{k=1}^{N_R}\hat{\rho}_{{\bf n}^{+}_{m k}}\right]\nonumber\\
	&\quad-i\sum_{m=1}^{N_{S}}\sum_{j=1}^{N_C}n_{mj+}\left(d_{j+}\hat{\sigma}_{m}^{\dagger}\hat{\sigma}_{m}\hat{\rho}_{{\bf n}^{-}_{m j+}}-d_{j-}^{*}\hat{\rho}_{{\bf n}^{-}_{m j+}}\hat{\sigma}_{m}^{\dagger}\hat{\sigma}_{m}\right)-i\sum_{m=1}^{N_{S}}\sum_{j=1}^{N_C}n_{mj-}\left(d_{j-}\hat{\sigma}_{m}^{\dagger}\hat{\sigma}_{m}\hat{\rho}_{{\bf n}^{-}_{m j-}}-d_{j+}^{*}\hat{\rho}_{{\bf n}^{-}_{m j-}}\hat{\sigma}_{m}^{\dagger}\hat{\sigma}_{m}\right)\nonumber\\
	&\quad-i\sum_{m=1}^{N_{S}}\sum_{k=1}^{N_R}n_{mk}\left(d_{k}\hat{\sigma}_{m}^{\dagger}\hat{\sigma}_{m}\hat{\rho}_{{\bf n}^{-}_{m k}}-d_{k}^{*}\hat{\rho}_{{\bf n}^{-}_{m k}}\hat{\sigma}_{m}^{\dagger}\hat{\sigma}_{m}\right),\label{eq_lim:HEOM_standard}
\end{align}
where $N_S$ denotes the number of sites of the system ($N_S=1$ for a monomer, $N_S=2$ for a dimer, such as WSCP and the special pair in
bacterial reaction centers), and $\hat{\sigma}_m$ ($\hat{\sigma}_{m}^{\dagger}$) is the annihilation (creation) operator of an electronic
excitation at site $m$. The information about the reduced system dynamics and system-environment correlations is included in the auxiliary
operators $\hat{\rho}_{\bf n}$, which are labelled by vectors ${\bf n}=(n_{mj+},\cdots,n_{mj-},\cdots,n_{mk},\cdots)$ with non-negative
integer elements. The sum of the integer elements of the vector ${\bf n}$ is denoted by $N_{\rm rank}$. The reduced system density matrix is
described by the lowest-rank operator $\hat{\rho}_{\bf 0}$ with $N_{\rm rank}=0$ where all the integer elements of ${\bf n}$ are zero. All the other higher-rank
auxiliary operators contain the information about system-environment correlations and have the same dimension as the reduced system density matrix $\hat{\rho}_{\bf 0}$ and are taken
to be null matrices at the initial time. The interaction between auxiliary operators is described by ${\bf n}^{\pm}_{m j+}$, ${\bf n}^{\pm}_{m j-}$,
${\bf n}^{\pm}_{m k}$, which are defined by ${\bf n}^{\pm}_{m j+}=(n_{mj+}\pm 1,\cdots,n_{mj-},\cdots,n_{mk},\cdots)$, ${\bf n}^{\pm}_{m j-}=
(n_{mj+},\cdots,n_{mj-}\pm 1,\cdots,n_{mk},\cdots)$, ${\bf n}^{\pm}_{m k}=(n_{mj+},\cdots,n_{mj-},\cdots,n_{mk}\pm 1,\cdots)$, which
are one-rank higher or lower than ${\bf n}$. To obtain numerically exact results, one needs to increase the number of auxiliary operators, determined by the maximum value of $N_{\rm rank}$, until the simulated system dynamics shows numerical convergence. In this work, the convergence
of HEOM simulations is achieved for the maximum $N_{\rm rank}=10$.

In absorption simulations, we compute the dynamics of optical coherence in the form $\sum_{i=1}^{N_S}\psi_{i}(t)|\varepsilon_i\rangle\langle g|$,
where $|\varepsilon_i\rangle$ denotes a singly excited state of site $i$, and $|g\rangle$ is the global electronic ground state. Note that
$\hat{\sigma}^{\dagger}_{m}\hat{\sigma}_m|g\rangle=0$ due to the absence of electronic excitations in the global ground state. This implies
that in absorption simulations, the mathematical form of HEOM can be simplified to
\begin{align}
	\frac{d\hat{\rho}_{\bf n}}{dt}&=-\frac{i}{\hbar}\hat{H}_s \hat{\rho}_{\bf n}-\sum_{m=1}^{N_{S}}\left(\sum_{j=1}^{N_C}(n_{mj+}f_{j}+n_{mj-}f_{j}^{*})+\sum_{k=1}^{N_R}n_{mk}\gamma_{k}\right) \hat{\rho}_{\bf n}-i\sum_{m=1}^{N_{S}}\hat{\sigma}_{m}^{\dagger}\hat{\sigma}_{m}\left(\sum_{j=1}^{N_C}(\hat{\rho}_{{\bf n}^{+}_{m j+}}+\hat{\rho}_{{\bf n}^{+}_{m j-}})+\sum_{k=1}^{N_R}\hat{\rho}_{{\bf n}^{+}_{m k}}\right)\nonumber\\
	&\quad-i\sum_{m=1}^{N_{S}}\sum_{j=1}^{N_C}n_{mj+}d_{j+}\hat{\sigma}_{m}^{\dagger}\hat{\sigma}_{m}\hat{\rho}_{{\bf n}^{-}_{m j+}}-i\sum_{m=1}^{N_{S}}\sum_{j=1}^{N_C}n_{mj-}d_{j-}\hat{\sigma}_{m}^{\dagger}\hat{\sigma}_{m}\hat{\rho}_{{\bf n}^{-}_{m j-}}-i\sum_{m=1}^{N_{S}}\sum_{k=1}^{N_R}n_{mk}d_{k}\hat{\sigma}_{m}^{\dagger}\hat{\sigma}_{m}\hat{\rho}_{{\bf n}^{-}_{m k}},\label{eq_lim:HEOM_absorption}
\end{align}
where the auxiliary operators are described by $\hat{\rho}_{\bf n}=\sum_{i=1}^{N_S}\psi_{{\bf n},i}(t)|\varepsilon_i\rangle\langle g|$,
similar to $N_S$-dimensional states with unnormalized amplitudes $\psi_{{\bf n},i}(t)$. We note that $\hat{H}_s$ is the electronic
Hamiltonian in this work, satisfying $\hat{H}_s|g\rangle = 0$. It is notable that in Eq.~(\ref{eq_lim:HEOM_standard}), $d_{j+}$, namely
the amplitude of $e^{-f_j t}$ considered in the multi-exponential fitting (see Eq.~(\ref{eq_lim:multi_exponential_fitting})), induces the
coupling between $\hat{\rho}_{\bf n}$ and $\hat{\rho}_{{\bf n}^{-}_{m j+}}$ (see $d_{j+}$ in Eq.~(\ref{eq_lim:HEOM_standard})) and that
between $\hat{\rho}_{\bf n}$ and $\hat{\rho}_{{\bf n}^{-}_{m j-}}$ (see $d_{j+}^{*}$ in Eq.~(\ref{eq_lim:HEOM_standard})). This implies
that even if we consider only $d_{j+} e^{-f_j t}$ in Eq.~(\ref{eq_lim:multi_exponential_fitting}) by setting $d_{j-}=0$, the auxiliary
operators are labelled by ${\bf n}=(n_{mj+},\cdots,n_{mj-},\cdots,n_{mk},\cdots)$ with both $n_{mj+}$ and $n_{mj-}$ included, as
is the case that both $d_{j+} e^{-f_j t}$ and $d_{j-} e^{-f_{j}^{*} t}$ are considered in the fitting. Hence, setting $d_{j-}=0$
does not reduce the number of auxiliary operators for a given maximum value of $N_{\rm rank}$ in general. However, in absorption simulations, as shown in Eq.~(\ref{eq_lim:HEOM_absorption}), $d_{j+}$ only induces the interaction between
$\hat{\rho}_{\bf n}$ and $\hat{\rho}_{{\bf n}^{-}_{m j+}}$, not the interaction between $\hat{\rho}_{\bf n}$ and $\hat{\rho}_{{\bf n}^{-}_{m j-}}$.
Therefore, in absorption simulations, one does not need to fit the bath correlation function with pairwise exponentials,
$d_{j+} e^{-f_j t}+d_{j-} e^{-f_j^* t}$, and can consider a sum of fully independent exponentials in the form
\begin{align}
	C(t)\approx\sum_{j=1}^{N_C}d_{j} e^{-f_j t},
\end{align}
with complex-valued $d_j$ and $f_j$. In this case, HEOM is simplified to
\begin{align}
	\frac{d\hat{\rho}_{\bf n}}{dt}&=-\frac{i}{\hbar}\hat{H}_s \hat{\rho}_{\bf n}-\sum_{m=1}^{N_{S}}\sum_{j=1}^{N_C}n_{mj}f_{j} \hat{\rho}_{\bf n}-i\sum_{m=1}^{N_{S}}\hat{\sigma}_{m}^{\dagger}\hat{\sigma}_{m}\sum_{j=1}^{N_C}\hat{\rho}_{{\bf n}^{+}_{m j}}-i\sum_{m=1}^{N_{S}}
\sum_{j=1}^{N_C}n_{mj}d_{j}\hat{\sigma}_{m}^{\dagger}\hat{\sigma}_{m}\hat{\rho}_{{\bf n}^{-}_{m j}},\label{eq:HEOM_absorption_simplified}
\end{align}
where ${\bf n}=(n_{mj},\cdots)$. This approach provides more flexibility in the multi-exponential fitting of the bath correlation function and can reduce the number of exponentials required to achieve a desired fitting quality. We note that
this approach cannot be employed when the population dynamics of electronic excited states is involved in the physical quantity of
interest, such as nonlinear optical responses.

To further reduce the simulation cost of HEOM, we map local vibrational modes to global modes including the center-of-mass modes. This enables one to perform HEOM simulations without considering the center-of-mass modes, which reduces the dimension of the vectors
${\bf n}$ labelling auxiliary operators $\hat{\rho}_{\bf n}$. As an example, we consider a dimer model where each site is coupled to a single local mode with an identical vibrational frequency $\omega$ and
a Huang-Rhys factor $s$
\begin{align}
	H=H_{e}+\sum_{i=1}^{2}\left(\omega b_{i}^{\dagger}b_{i}+\omega\sqrt{s}|\varepsilon_i\rangle\langle \varepsilon_i|\otimes(b_{i}^{\dagger}+b_{i})\right).
\end{align}
By introducing a center-of-mass mode $B=\frac{1}{\sqrt{2}}(b_{1}+b_{2})$ and a relative-motion mode $b=\frac{1}{\sqrt{2}}(b_{1}-b_{2})$, satisfying
the bosonic commutation relations, the Hamiltonian can be expressed as
\begin{align}
	H&=H_{e}+\omega (B^{\dagger}B+b^{\dagger}b)+\omega\sqrt{s}A_+\otimes(B^{\dagger}+B)+\omega\sqrt{s}A_-\otimes(b^{\dagger}+b),
\end{align}
where $A_\pm=(|\varepsilon_1\rangle\langle \varepsilon_1|\pm |\varepsilon_2\rangle\langle \varepsilon_2|)/\sqrt{2}$. It is notable that
$A_+\propto |\varepsilon_1\rangle\langle \varepsilon_1|+|\varepsilon_2\rangle\langle \varepsilon_2|$ is proportional to the identity operator
of the single electronic excitation subspace. This implies that the center-of-mass mode induces fully correlated fluctuations of the energy levels of
the two sites, without modifying the energy-level difference between them. Therefore the influence of the center-of-mass mode on the
dynamics of optical coherence can be computed in an analytical way, as is the case of the Kubo's lineshape theory for a two-level monomer.
The influence of the relative-motion mode on the system dynamics, described by $A_-$, still requires numerically exact simulations. The mapping
from local to global modes can be generalized to a multi-mode case characterized by a phonon spectral density. For a dimer
system coupled to identical local baths, the influence of the relative-motion modes on optical coherence can be computed by HEOM in the form
\begin{align}
	\frac{d\hat{\rho}_{\bf n}}{dt}&=-\frac{i}{\hbar}\hat{H}_s \hat{\rho}_{\bf n}-\sum_{j=1}^{N_C}n_{j}f_{j} \hat{\rho}_{\bf n}-i \frac{|\varepsilon_1\rangle\langle \varepsilon_1|-|\varepsilon_2\rangle\langle \varepsilon_2|}{\sqrt{2}}\sum_{j=1}^{N_C}\hat{\rho}_{{\bf n}^{+}_{j}}-i\sum_{j=1}^{N_C}n_{j}d_{j}\frac{|\varepsilon_1\rangle\langle \varepsilon_1|-|\varepsilon_2\rangle\langle \varepsilon_2|}{\sqrt{2}}\hat{\rho}_{{\bf n}^{-}_{j}},\label{eq_lim:HEOM_relative_motions}
\end{align}
where ${\bf n}=(n_{j},\cdots)$. It is notable that the dimension of ${\bf n}=(n_{j},\cdots)$ of the relative-motion model is the half of the
dimension of ${\bf n}=(n_{1j},\cdots,n_{2j},\cdots)$ of the original model with two local baths. This significantly reduces the number of auxiliary
operators for a given maximum value of $N_{\rm rank}$ and corresponding simulation cost. This approach can be generalized to a multi-site system consisting of $N_S$ sites where
${\bf n}=(n_{1j},\cdots,n_{N_{S}j},\cdots)$ is reduced to ${\bf n}=(n_{1j},\cdots,n_{N_{S}-1,j},\cdots)$ by removing a center-of-mass bath.

We note that the dynamics of the optical coherence $\hat{\rho}_{\bf 0}(t)=\sum_{i=1}^{2}\psi_{{\bf 0},i}(t)|\varepsilon_i\rangle\langle g|$
computed by Eq.~(\ref{eq_lim:HEOM_relative_motions}) only considers the dephasing induced by the relative-motion modes. The additional dephasing
caused by the center-of-mass modes can be taken into account analytically by multiplying
an additional time-dependent factor $\exp(-\frac{1}{2}G(t))$ that is independent of the electronic excited states $|\varepsilon_i\rangle$, namely $\sum_{i=1}^{2}\psi_{{\bf 0},i}(t)|\varepsilon_i
\rangle\langle g|\exp(-\frac{1}{2}G(t))$ where
\begin{align}
	G(t)=-i\lambda t+\int_{0}^{\infty}d\omega\frac{J(\omega)}{\omega^2}\left((1-\cos(\omega t))\left(1+\frac{2}{e^{\omega/k_B T}-1}\right)+i\sin(\omega t)\right),
\end{align}
with the reorganization energy $\lambda=\int_{0}^{\infty}d\omega J(\omega)/\omega$. The factor $1/2$ in $\exp(-\frac{1}{2}G(t))$, which does not
appear in the conventional lineshape theory, originates from the $1/\sqrt{2}$ factor in $A_+$ describing the coupling to the center-of-mass
modes. We note that this approach can be generalized to the simulations of nonlinear optical spectra by taking into account the contribution of the center-of-mass
modes analytically.

We would like to stress that the techniques devised to reduce the computational cost of absorption simulations by T-TEDOPA can also be applied to
HEOM. As explained in \ref{section_TEDOPA}, in T-TEDOPA the finite temperature environments can be mapped to zero temperature environments for which the initial state becomes a pure state, while the phonon spectral density becomes
temperature-dependent. This enables one to compute the dipole-dipole correlation function up to time $t$ by simulating the time evolution of a pure electronic-vibrational state
up to time $t/2$. This technique can be employed in HEOM. In absorption simulations, the initial state of the system density matrix is given by $\hat{\rho}_{\bf 0}
=\ket{g}\bra{g}$ and all the higher-rank auxiliary operators are taken to be null matrices at the initial time, namely $\hat{\rho}_{\bf n}=0$ for all ${\bf n}\neq{\bf 0}$.
When a transition dipole moment operator $\mu$ is multiplied to the left-hand side of the auxiliary operators, the lowest-rank operator becomes a linear
combination of optical coherences, $\hat{\rho}_{\bf 0}(0)=\sum_{i=1}^{N_S}\psi_{{\bf 0},i}(0)\ket{\varepsilon_i}\bra{g}$. In this case, as shown in
Eq.~(\ref{eq:HEOM_absorption_simplified}), the time evolution of the auxiliary operators is governed by a propagator that is multiplied only to the left-hand
side of $\hat{\rho}_{\bf n}$. This makes it straightforward to define a high-dimensional vector $\boldsymbol{\rho}(t)$, consisting of all the elements
$\psi_{{\bf n},i}(t)$ of the auxiliary operators $\hat{\rho}_{\bf n}(t)=\sum_{i=1}^{N_S}\psi_{{\bf n},i}(t)\ket{\varepsilon_i}\bra{g}$, whose dynamics is governed by $\frac{d}{dt}\boldsymbol{\rho}(t)=G\boldsymbol{\rho}(t)$ with a propagator $G$. The dipole-dipole correlation function is then described
by
\bea
    \langle D(t) \rangle_{{\bf e},\varepsilon_i} = \langle \boldsymbol{\rho}_{\bf n}(0)\cdot(e^{Gt}\boldsymbol{\rho}_{\bf n}(0))\rangle_{{\bf e},\varepsilon_i} = \langle
    (e^{G^{\dagger}t/2}\boldsymbol{\rho}_{\bf n}(0))\cdot(e^{Gt/2}\boldsymbol{\rho}_{\bf n}(0))\rangle_{{\bf e},\varepsilon_i},
\eea
demonstrating that one can reduce the simulation time from $t$ to $t/2$ in HEOM simulations, as is the case of T-TEDOPA. The reduced time window can
decrease the number of auxiliary operators required to obtain numerically exact absorption line shapes, as higher-rank auxiliary operators are initially
null matrices and they are populated only by the interaction with lower-rank auxiliary operators.

The computational cost of HEOM is determined by the number of the auxiliary operators, controlled by the maximum value of $N_{\rm rank}$. As detailed in \ref{section_HEOM_TEDOPA}, we consider experimentally estimated phonon spectral densities, including 55 intra-pigment modes per site, and perform HEOM simulations with the maximum $N_{\rm rank}=10$ to obtain fully converged absorption spectra for the cases that the bath correlation functions of WSCP and SP are fitted with the sum of 13 and 16 exponentials, respectively. For the maximum $N_{\rm rank}=10$ and $N_{\rm exp}=13$ (or $N_{\rm exp}=16$) exponentials, the number of the auxiliary operators is given by $(N_{\rm exp}+N_{\rm rank})!/(N_{\rm exp}!N_{\rm rank}!)$. For absorption simulations of a dimeric system, it is sufficient to consider a subspace spanned by optical coherences $\ket{\varepsilon_1}\bra{g}$ and $\ket{\varepsilon_2}\bra{g}$ for each auxiliary operator, leading to the simulation cost of 37\,MB and 170\,MB, respectively, for WSCP and SP.

When the experimentally estimated spectral densities are fitted with Drude-Lorentz peaks, one needs to consider at least two exponentials to describe each local intra-pigment mode. In such a conventional HEOM approach, a dimeric system coupled to 55 intra-pigment modes per site corresponds to a case of 220 exponentials. In this case, the simulation cost for the maximum $N_{\rm rank}=10$ becomes of the order of $10^{6}$\,TB. Even if the number of auxiliary operators is reduced for approximate HEOM simulations, the simulation cost is of the order of 1\,TB for the maximum $N_{\rm rank}=6$, for which we found that the numerical errors of simulated absorption spectra are not negligible. We note that this estimate does not include the number of exponentials required to consider low-frequency protein motions and Matsubara frequencies originating from the intra-pigment modes, implying that the computational cost of the conventional approach can be even higher than the estimate provided here.


\section{E\MakeLowercase{lectronic} p\MakeLowercase{arameters and} p\MakeLowercase{honon} s\MakeLowercase{pectral} d\MakeLowercase{ensities of} WSCP \MakeLowercase{and} SP}\label{section_params}

Here we provide a summary of the electronic parameters and phonon spectral densities of WSCP and SP considered in simulations.

The electronic parameters of WSCP have been estimated based on the phonon spectral density of B777 photosynthetic complexes~\cite{Renger_2015JCP}
\begin{equation}
	J_{l}^{\rm B777}(\omega)=\frac{S}{s_1+s_2}\sum_{i=1}^{2}\frac{s_i}{7! 2 \omega_i^4}\omega^5 e^{-(\omega/\omega_i)^{1/2}},
	\label{eq:B777_SD}
\end{equation}
where $S=0.8$, $s_1=0.8$, $s_2=0.5$, $\omega_1=0.069\,{\rm meV}$ and $\omega_2=0.24\,{\rm meV}$. By using second order cumulant expansion, it is found that the low-energy part of experimental absorption spectra of WSCP can be well reproduced when the electronic coupling between two pigments is $V=69\,{\rm cm}^{-1}$ and the angle between monomer transition dipole moments is $39^{\circ}$.

The phonon spectral density of WSCP has been estimated based on experimentally measured fluorescence line-narrowing spectra of WSCP~\cite{Pieper_2011JPCB,Jankowiak_2013JPCB}. It is found that the low-frequency part of the phonon spectral density can be well described by a sum of three log-normal distributions in the form
\begin{equation}
	J_{l}^{\rm WSCP}(\omega)=\sum_{m=1}^{3}\frac{S_m}{\sigma_m \sqrt{2\pi}}\omega\exp(-\frac{[\ln(\omega/\Omega_m)]^2}{2\sigma_m^2}),
\end{equation}
with $S_1=0.39$, $S_2=0.23$, $S_3=0.23$, $\sigma_1=0.4$, $\sigma_2=0.25$, $\sigma_3=0.2$, $\Omega_1=26\,{\rm cm}^{-1}$,
$\Omega_2=51\,{\rm cm}^{-1}$, $\Omega_3=85\,{\rm cm}^{-1}$. The high-frequency part of the phonon spectral density consists of 55 intra-pigment vibrational modes, which are modelled by Lorentzian functions in the form
\begin{equation}
    J_{h}(\omega) = \sum_{k=1}^{55} \frac{4 \omega_k s_k \gamma_k (\omega_{k}^2+\gamma_{k}^2)\omega}{\pi((\omega+\omega_k)^{2}+\gamma_{k}^2)((\omega-\omega_k)^{2}+\gamma_{k}^2)},
\end{equation}
where vibrational frequencies $\omega_k$ and Huang-Rhys factors $s_k$ are summarised in Supplementary Table \ref{Table_WSCP}. The vibrational damping rates of the intra-pigment modes are taken to be $\gamma_{k}=(1\,{\rm ps})^{-1}$. It is found that experimentally measured absorption spectra of WSCP can be reproduced by numerically exact simulations when $J_{l}^{\rm WSCP}(\omega)+J_{h}(\omega)$ and $V=140\,{\rm cm}^{-1}$ are considered with independent static disorder in site energies modelled by Gaussian distributions with the standard deviation of $80\,{\rm cm}^{-1}$ (see \ref{section_HEOM_TEDOPA}).

The electronic parameters of SP heterodimers have been estimated based on experimentally measured absorption, linear dichroism and hole burning spectra of bacterial reaction centers~\cite{Jankowiak_2019JPCB}. By using conventional line shape theory, it is found that the difference in mean site energies is $\langle\varepsilon_1-\varepsilon_2\rangle=315\,{\rm cm}^{-1}$,
electronic coupling is $V=625\,{\rm cm}^{-1}$, and the angle between transition dipole moments of monomers is $143^{\circ}$.

Experimentally estimated phonon spectral density of SP~\cite{Jankowiak_2015JCP,Small_JPCB2001} consists of a log-normal distribution function
\begin{equation}
	J_{l}^{\rm SP}(\omega)=\frac{S_{l}}{\sigma_l \sqrt{2\pi}}\omega\exp(-\frac{[\ln(\omega/\Omega_l)]^2}{2\sigma_l^2}),
\end{equation}
with $S_l=1.7$, $\sigma_{l}=0.47$, $\Omega_l=35\,{\rm cm}^{-1}$, and the special pair marker mode with vibrational
frequency $\omega_{\rm sp}=125\,{\rm cm}^{-1}$, Huang-Rhys factor $s_{\rm sp}=1.5$ and damping rate $\gamma_{\rm sp}=15\,{\rm cm}^{-1}$, and 55
intra-pigment vibrational modes of BChl{\it a} pigments with vibrational frequencies and Huang-Rhys factors
summarised in Supplementary Table \ref{Table_SP}. The damping rates of the intra-pigment modes are taken to be $\gamma_{k}=(1\,{\rm ps})^{-1}$. The special
pair marker mode and 55 intra-pigment modes are modelled by Lorentzian spectral densities. In experiments where phonon spectral densities are
estimated, the optical response may originate from the lowest-energy exciton rather than the lowest-energy pigment. To take into account this
effect approximately, we consider diagonal vibronic couplings of the lowest-energy exciton state of bacterial reaction centers, namely the
low-energy exciton $\ket{E_{-}}$ of the SP, where $\bra{E_-}H_{e-v}\ket{E_-} = \omega\sqrt{s}(\sin^{2}(\theta/2)(b_{1}+b_{1}^{\dagger})+\cos^{2}(\theta/2)(b_{2}+b_{2}^{\dagger}))$. We introduce an effective mode $b=(\sin^{4}(\theta/2)+\cos^{4}(\theta/2))^{-1/2}(\sin^{2}(\theta/2)b_{1}+\cos^{2}(\theta/2)b_{2})$, satisfying the bosonic commutation
relation $[b,b^{\dagger}]=1$, so that the diagonal vibronic couplings are expressed as $\bra{E_-}H_{e-v}\ket{E_-}=\omega\sqrt{s_{E_{-}}}(b+b^{\dagger})$
where the renormalised Huang-Rhys factor of the exciton state is given by $s_{E_{-}}=s(\sin^{4}(\theta/2)+\cos^{4}(\theta/2))$. Since experimental
data are fitted with line shape functions based on the effective Huang-Rhys factor $s_{E_{-}}$ in the estimation of phonon spectral densities, we
renormalise the Huang-Rhys factors of the local modes, $s=s_{E_{-}}/(\sin^{4}(\theta/2)+\cos^{4}(\theta/2))\approx 1.9\,s_{E_-}$ based on the
mean site energy difference $\langle\varepsilon_1-\varepsilon_2\rangle=315\,{\rm cm}^{-1}$ and electronic coupling $V=625\,{\rm cm}^{-1}$ of
the SP. Since the Huang-Rhys factors $S_l$ and $s_{\rm sp}$ of the log-normal distribution and special pair marker mode have been estimated in experiments on excitonic systems~\cite{Jankowiak_2015JCP}, we multiply the renormalisation factor $\sim 1.9$ to $S_l$ and $s_{\rm sp}$.
Since the parameters in Supplementary Table \ref{Table_SP} have been estimated in experiments on BChl{\it a} monomers~\cite{Small_JPCB2001}, we do not
renormalise the Huang-Rhys factors $s_k$ of the intra-pigment modes. It is found that the low-energy part of experimental absorption spectra of bacterial reaction centers, dominated by the optical responses of SP, can be well reproduced by numerically exact simulations when the experimentally estimated spectral density is considered with local static disorder modelled by Gaussian distributions with the standard deviation of $105\,{\rm cm}^{-1}$ (see \ref{section_HEOM_TEDOPA}).



\section{N\MakeLowercase{umerical} t\MakeLowercase{est and} c\MakeLowercase{omparison of} HEOM \MakeLowercase{and} TEDOPA \MakeLowercase{for} s\MakeLowercase{tructured} e\MakeLowercase{nvironments at} l\MakeLowercase{ow} t\MakeLowercase{emperatures}}\label{section_HEOM_TEDOPA}

In this section we provide direct evidence that both HEOM and TEDOPA are capable of achieving numerically exact results for dimeric systems in contact
with realistic, highly structured environmental spectral densities. To this end we apply both methods to WSCP and the special pair in bacterial reaction centers
and show that the fully converged results coincide.

Supplementary Figure~\ref{Fig_HEOM1} presents the HEOM and TEDOPA results of numerically exact absorption line shapes of WSCP at 77\,K. Supplementary Figure~\ref{Fig_HEOM1}a shows
the experimentally estimated spectral density of WSCP which consists of three log-normal functions at low vibrational frequencies
and 55 narrow Lorentzian functions corresponding to intra-pigment vibrational modes (see \ref{section_params} for all the parameters
that enter this spectral density). For the HEOM simulation, the real and imaginary part of the corresponding bath correlation function at $T=77\,{\rm K}$ is
fitted with the sum of 13 exponentials, as shown in Supplementary Figure~\ref{Fig_HEOM1}b. The quality of the fit is confirmed by its ability to reproduce the analytical
solution of monomer optical coherence dynamics, as shown in Supplementary Figure~\ref{Fig_HEOM1}c. Finally, Supplementary Figure~\ref{Fig_HEOM1}d shows numerically exact absorption line
shapes of WSCP obtained by independent calculations using HEOM and TEDOPA.  For two different electronic coupling strengths $V=69\,{\rm cm}^{-1}$ and $V=140\,{\rm cm}^{-1}$, we observe perfect overlap of the absorption spectra obtained by the two methods. This demonstrates the reliability of the two methods and our simulated data.

Supplementary Figure~\ref{Fig_HEOM2} presents HEOM and TEDOPA results of numerically exact absorption line shapes of the special pair in bacterial reaction centers at $T=5\,{\rm K}$ under the highly structured environmental spectral density shown in Supplementary Figure~\ref{Fig_HEOM2}a. The corresponding bath correlation
function at $T=5\,{\rm K}$, shown in black in Supplementary Figure~\ref{Fig_HEOM2}b, is well fitted with the sum of $16$ exponentials up to $t=200\,{\rm fs}$, as shown in red.
Supplementary Figure~\ref{Fig_HEOM2}c shows that the fitting quality is good enough to reproduce the analytical solution of monomer optical coherence dynamics. Supplementary Figure~\ref{Fig_HEOM2}d displays the absorption spectra of the special pair demonstrating that the results obtained by HEOM and TEDOPA are well matched. Our results demonstrate that low-temperature systems can be efficiently simulated by HEOM when the bath correlation function is numerically fitted with exponentials. This contrasts the conventional approach where  the bath correlation function is expanded as a sum of exponentials in an analytical way for some model spectral densities. A well-known example is the Ohmic spectral density with the Lorentz-Drude cutoff function where the number of exponentials required to achieve a desired fitting quality increases
as temperature decreases, due to Matsubara terms, making HEOM simulations more challenging at lower temperatures. Even in this case, the simulation cost can be significantly
reduced by replacing multiple Matsubara (exponential) terms with a few damped oscillations based on numerical fitting.

It is notable that the input parameters of HEOM simulations are determined based on the fitting of the bath correlation functions. This implies that one
needs to improve the fitting quality until simulated HEOM results show convergence, which requires the repetition of HEOM simulations and as a result
increases the overall simulation cost. This is contrary to TEDOPA where the input parameters are computed by orthogonal polynomial
algorithm, which makes it easier to control the accuracy of input parameters when compared to HEOM.


\section{R\MakeLowercase{educed} v\MakeLowercase{ibronic} m\MakeLowercase{odel for} a\MakeLowercase{bsorption} s\MakeLowercase{pectra}}\label{section_reduced_absorption}

Numerically exact absorption line shape of the special pair in bacterial reaction centers consists
of two narrow peaks centered at 780 and $900\,{\rm nm}$, respectively, and a relatively broad peak around $800$-$850\,{\rm nm}$. Here
we show that the $780\,{\rm nm}$ and $900\,{\rm nm}$ peaks originate from vibronic eigenstates $\ket{\psi_+}$ and $\ket{\psi_-}$,
resulting from the vibronic mixing of exciton states $\ket{E_+}$ and $\ket{E_-}$ with multiple intra-pigment modes, and the other vibrationally excited eigenstates contribute to the vibrational sideband in the $800$-$850\,{\rm nm}$ region.

To understand the origin of the three-peak structure, we consider the vibronic Hamiltonian where the excitons are coupled to the relative
motions of $55$ intra-pigment modes (see \ref{section_vibronic_H_polaron_transformation})
\begin{align}
	H &\approx H_{e}+\sum_{k=1}^{55}\omega_{k}b_{k}^{\dagger}b_k+\sum_{i,j}A_{ij}|E_i\rangle\langle E_j|\sum_{k=1}^{55}\omega_k \sqrt{s_k/2}(b_{k}^{\dagger}+b_{k}).\nonumber
\end{align}
For simplicity, the relative motions of low-frequency protein modes and the vibrational damping of the relative intra-pigment modes $b_k$ are not considered, which can induce pure dephasing and relaxation of vibronic eigenstates (see \ref{section_reduced_2DES}). The influence of the center-of-mass motions of the protein and intra-pigment modes on absorption line shapes is treated non-perturbatively. In more detail, absorption line shape is determined by the Fourier transformation of the dynamics of optical coherence,
${\rm Tr}\{\mu e^{-i H t}\mu\rho_{th}e^{i H t}\} \exp(-\frac{1}{2}G(t))$ where $e^{\pm i H t}$ describes the interaction between excitons and relative intra-pigment modes, while $G(t)$ takes into account the frequency shift and dephasing of the optical coherence induced by the center-of-mass motions
\begin{align}
	G(t)&=\int_{0}^{\infty}d\omega\frac{J(\omega)}{\omega^2}[(1-\cos(\omega t))(1+\frac{2}{e^{\omega/k_B T}-1})+i\sin(\omega t)]-i\lambda t,
\end{align}
where $J(\omega)$ denotes experimentally estimated phonon spectral density of SP, and $\lambda=\int_{0}^{\infty}d\omega J(\omega)/\omega$
is the reorganisation energy. To reduce the total number of vibrational excitations of the relative motion modes $b_k$ required to obtain
numerically converged absorption spectra, we consider the polaron transformation in \ref{section_vibronic_H_polaron_transformation}
where the vibronic Hamiltonian is transformed to
\begin{align}
	\tilde{H}&=H_{e}+\sum_{k=1}^{55}\omega_{k}b_{k}^{\dagger}b_{k}-\sum_{i=\pm}|E_i\rangle\langle E_i|\left(\sum_{k=1}^{55}\omega_k\frac{s_k}{2}A_{ii}^{2}\right)\label{lim_eq:tildeH_r}\\&\quad+|E_+\rangle\langle E_-|A_{+-}\sum_{k=1}^{55}\omega_{k}\sqrt{\frac{s_k}{2}}D_{k}\left(\sqrt{\frac{s_k}{2}}A_{++}\right)(b_{k}^{\dagger}+b_{k})D_{k}^{\dagger}\left(\sqrt{\frac{s_k}{2}}A_{--}\right)\prod_{k'\neq k}D_{k'}\left(\sqrt{\frac{s_{k'}}{2}}(A_{++}-A_{--})\right)+h.c.\nonumber
\end{align}
Similarly the transition dipole moment operator is transformed to
\begin{align}
	\tilde{\mu}=({\bf e}\cdot\boldsymbol{\mu}_{E_+})|E_+\rangle\langle g|\prod_{k=1}^{55}D_{k}\left(\sqrt{\frac{s_k}{2}}A_{++}\right)+({\bf e}\cdot\boldsymbol{\mu}_{E_-})|E_-\rangle\langle g|\prod_{k=1}^{55}D_{k}\left(\sqrt{\frac{s_k}{2}}A_{--}\right)+h.c.,
\end{align}
where $\boldsymbol{\mu}_{E_j}=\sum_{i=1}^{2}\boldsymbol{\mu}_{i}\langle E_j|\varepsilon_i\rangle$ represent the transition dipole moment vectors of
the exciton states $|E_+\rangle$ and $|E_-\rangle$ with $\boldsymbol{\mu}_{1,2}$ denoting the transition dipole moment vectors of monomers,
and ${\bf e}$ is a unit vector describing polarization of the electric field inducing absorption.

For isotropic samples, one needs to consider the orientational average of absorption line shapes as the relative angle between ${\bf e}$ and
$\boldsymbol{\mu}_{1,2}$ is random. In simulations, one can take into account the orientational average exactly by
considering three polarization directions ${\bf e}\in\{{\bf x},{\bf y},{\bf z}\}$ for fixed $\boldsymbol{\mu}_{1,2}$ and averaging the corresponding optical coherence dynamics
${\rm Tr}\{\mu e^{-i H t}\mu\rho_{th}e^{i H t}\}$ with uniform weighting factors. This can be proved by considering the formal representation
of the orientational average of absorption line shapes $\int dR\,{\rm Tr}\{(R\mu) e^{-i H t}(R\mu)\rho_{th}e^{i H t}\}$ with
$R\mu=\sum_{i=1}^{2}({\bf e}\cdot R\boldsymbol{\mu}_{i})|\varepsilon_i\rangle\langle g|+h.c.$ denoting the transition dipole moment operator
where monomer transition dipoles $\boldsymbol{\mu}_{i}$ are randomly rotated by $R$ in the laboratory frame. By applying the rotation matrix $R$
to ${\bf e}$ instead of $\boldsymbol{\mu}_{i}$, namely ${\bf e}\cdot (R\boldsymbol{\mu}_{i})=(R^{\dagger}{\bf e})\cdot \boldsymbol{\mu}_{i}$, one can
represent $R^{\dagger}{\bf e}=\cos(\theta){\bf z}+\sin(\theta)\cos(\phi){\bf x}+\sin(\theta)\sin(\phi){\bf y}$ in the spherical coordinate.
By substituting ${\bf e}$ parameterized by $(\theta,\phi)$ to $\int dR\,{\rm Tr}\{(R\mu) e^{-i H t}(R\mu)\rho_{th}e^{i H t}\}$ and then computing
the orientational average in the spherical coordinate, one can show that the absorption line shapes of isotropic samples can be computed
exactly by considering the three orientations mentioned above~\cite{SMLimPRL2019}. In 2D simulations, discussed in \ref{section_reduced_2DES}, the
orientational average of isotropic samples can be computed exactly by considering a finite number of orientations, similar to absorption simulations~\cite{SMLimPRL2019}. The required orientations can be determined via the concept of spherical t-design which allows
the average of any polynomial of the order $t$ over a sphere to be obtained by an average over a discrete set of specific orientations~\cite{Bajnok1991,MAKINO1999910}.

In Supplementary Figure~\ref{Fig_RM_abs}a, we consider the single vibrational excitation subspace where all the relative-motion modes are in their vibrational ground states or only one of the 55 modes is singly excited. In simulations, we average the dynamics of optical coherences ${\rm Tr}\{\tilde{\mu} e^{-i \tilde{H} t}\tilde{\mu}\rho_{th}e^{i \tilde{H} t}\}$ for 1000 random realizations of site energies, and then multiply $\exp(-\frac{1}{2}G(t))$ induced by center-of-mass modes, as the latter does not depend on static disorder. The eigenvalues and eigenstates of the Hamiltonian $\tilde{H}$ determine, respectively, the transition energies and dipole strengths of vibronic eigenstates. The eigenvalue spectrum depends on the random realizations of site energies, inducing ensemble dephasing of optical coherences. Supplementary Figure~\ref{Fig_RM_abs}a shows that the approximate absorption line shape based on the reduced model, shown in red, can quantitatively reproduce the numerically exact absorption line shape of SP, shown in black.

To understand the origin of the three-peak structure in absorption spectrum, we investigate four vibronic eigenstates in the absence of static disorder, including state A appearing at $900\,{\rm nm}$, state B at $817\,{\rm nm}$, and states C and D at $780\,{\rm nm}$, as highlighted in Supplementary Figure~\ref{Fig_RM_abs}a. The state A is well described by $|E_-,0\rangle$ where $|E_-\rangle$ is the lower-energy exciton state and $|0\rangle$ denotes the global vibrational ground state. It is notable that the vibronic mixing between $|E_-,0\rangle$ and $|E_+,1_k\rangle$, which can be induced by off-diagonal vibronic couplings, is negligible due to the large energy gap between $|E_-,0\rangle$ and $|E_+,1_k\rangle$, where $|1_k\rangle$ denotes a composite vibrational state with a single vibrational excitation in the $k$-th mode. For all the eigenstates of $\tilde{H}$, the fidelity with $|E_-,0\rangle$ is shown in green dots, demonstrating that the mixing with $|E_-,0\rangle$ is negligible for all the eigenstates except for A. The state B is one of the vibronic eigenstates contributing to the vibrational sideband in the $800$-$850\,{\rm nm}$ region. The state B is dominated by $|E_-,1_{1008\,{\rm cm}^{-1}}\rangle$ with a single vibrational excitation in the $1008\,{\rm cm}^{-1}$ mode, which is weakly mixed with quasi-resonant $|E_+,0\rangle$ and multiple $|E_-,1_k\rangle$ states mediated by off-diagonal vibronic couplings. The states C and D, on the other hand, have relatively large fidelity with $|E_+,0\rangle$, as shown in blue dots. Both the states C and D show a strong mixing between $|E_+,0\rangle$ and $|E_+,1_{453\,{\rm cm}^{-1}}\rangle$, and relatively weak mixing with multiple $|E_-,1_k\rangle$ states. It is notable that the mixing between $|E_+,0\rangle$ and $|E_+,1_{453\,{\rm cm}^{-1}}\rangle$ cannot be directly created by the off-diagonal vibronic couplings proportional to $|E_+\rangle\langle E_-|$ or $|E_-\rangle\langle E_+|$, hinting that it is due to the higher-order interactions mediated by $|E_-,1_k\rangle$ states.

To clarify the origin of the strong mixing between $|E_+,0\rangle$ and $|E_+,1_{453\,{\rm cm}^{-1}}\rangle$ in the states C and D, in Supplementary Figure~\ref{Fig_RM_abs}b we turn off the higher-order interaction terms in Eq.~(\ref{lim_eq:tildeH_r}) by neglecting the couplings proportional to $|E_+,1_k\rangle\langle E_-,1_l|$ with $k\neq l$, which induces the transition between exciton states and, at the same time, the exchange of a single vibrational excitation between different modes $k$ and $l$. Note that the absorption line shapes computed with and without the higher-order interaction terms, shown in Supplementary Figure~\ref{Fig_RM_abs}a and b, respectively, are almost identical (see the inset in Supplementary Figure~\ref{Fig_RM_abs}b), implying that the higher-order terms are weak couplings. It is also notable that when the higher-order terms are removed from the Hamiltonian $\tilde{H}$, the states A and B are almost unchanged, but the mixing between $|E_+,0\rangle$ and $|E_+,1_{453\,{\rm cm}^{-1}}\rangle$ of the states C and D is completely suppressed. These results can be rationalised as follows. In the absence of the higher-order interaction terms, the energy-level of higher-energy exciton state $|E_+,0\rangle$ is increased by $447\,{\rm cm}^{-1}$ due to the vibronic mixing with multiple $|E_-,1_k\rangle$ states, leading to the state C in Supplementary Figure~\ref{Fig_RM_abs}b, while the vibronic mixing of a vibrationally hot state $|E_+,1_{453\,{\rm cm}^{-1}}\rangle$ with $|E_-,0\rangle$ is negligible due to the relatively large energy-gap between them, leading to the state D in Supplementary Figure~\ref{Fig_RM_abs}b whose energy-level is almost identical to that of $|E_+,1_{453\,{\rm cm}^{-1}}\rangle$. This leads to a small energy-gap $\sim 6\,{\rm cm}^{-1}$ between states C and D, which can be strongly mixed even by weak higher-order couplings, as shown in Supplementary Figure~\ref{Fig_RM_abs}a. However, the energy-level shifts of the states C and D by the higher-order interaction terms are negligible due to the weak coupling strength, which makes the resulting absorption line shapes insensitive to the presence of the higher-order couplings.

Since the vibronic mixing induced by a weak coupling is likely to be sensitive to energy-level fluctuations induced by environments, such as protein motions, which are not considered in the above reduced model analysis, we performed numerically exact simulations of excitonic coherence dynamics of SP and controlled the vibrational damping rate of the $453\,{\rm cm}^{-1}$ modes in order to check if there is a strong vibronic mixing between excitons and $453\,{\rm cm}^{-1}$ modes. It is found that the lifetime of the $453\,{\rm cm}^{-1}$ modes does not affect excitonic coherence dynamics (not shown here), implying that the vibronic mixing with the $453\,{\rm cm}^{-1}$ modes is negligible under actual environments. Therefore we will ignore the higher-order interaction terms, $|E_+,1_k\rangle\langle E_-,1_l|$ with $k\neq l$, in the following approximate simulations of absorption and 2D spectra to avoid weak coupling effects.

Supplementary Figure~\ref{Fig_RM_abs}c shows absorption spectra in the absence of off-diagonal vibronic couplings. It is notable that the energy gap between absorption peaks is close to the bare excitonic splitting $\Delta\approx 1300\,{\rm cm}^{-1}$ of SP.

Supplementary Figure~\ref{Fig_RM_abs}d shows absorption line shapes computed by the reduced vibronic model within double vibrational excitation subspace, which includes doubly excited vibrational states where only the $k$-th mode is doubly excited, $\ket{2_k}$, or two different modes $k$ and $l$ are singly excited at the same time, $\ket{1_k,1_l}$. We note that the inclusion of doubly excited vibrational states in simulations does not qualitatively change absorption line shapes, but improves the quantitative agreement between approximate and numerically exact absorption line shapes, shown in red and black, respectively.


\section{R\MakeLowercase{educed} v\MakeLowercase{ibronic} m\MakeLowercase{odel for} t\MakeLowercase{wo}-d\MakeLowercase{imensional} e\MakeLowercase{lectronic} s\MakeLowercase{pectra}}\label{section_reduced_2DES}

Here we investigate the nonlinear optical response of SP based on the reduced vibronic model in \ref{section_reduced_absorption}, which can
quantitatively reproduce numerically exact absorption spectra of SP. Contrary to the dynamics of optical coherence of SP, which determines absorption
line shape and decays within $200\,{\rm fs}$, nonlinear spectroscopic techniques can measure electronic-vibrational dynamics on a picosecond time scale and therefore
the noise induced by relative vibrational motions can significantly modify simulated nonlinear spectra. Here we consider the noise originating from the relative motions of low-frequency protein modes and the vibrational damping of 55 intra-pigment modes, which is approximately
described by a Lindblad equation. The influence of the center-of-mass motions of the total vibrational environments on 2D electronic
spectra (2DES) can be taken into account analytically~\cite{SMMukamel}, similar to absorption simulations considered in \ref{section_reduced_absorption}. We compute rephasing 2D spectra of SP based
on the reduced vibronic model in order to demonstrate that oscillatory 2D signals are a mixture of $\Delta'\approx 1800\,{\rm cm}^{-1}$ frequency component
induced by vibronic coherence $\ket{\psi_{+}}\bra{\psi_{-}}$ in the electronic excited state manifold, and multiple oscillatory components with frequencies $\omega_k\lesssim 1600\,{\rm cm}^{-1}$ originating
from vibrational coherences of the 55 intra-pigment modes in the electronic ground state manifold.

As schematically shown in Supplementary Figure~\ref{Fig_RM_noise_abs_2D}a, we consider two noise-inducing processes. Firstly, the vibronic coupling between electronic states and low-frequency part $J_{l}(\omega)$ of the phonon spectral density of SP induces electronic dephasing. In a displaced vibrational basis defined by $U=|g\rangle\langle g|+|E_+\rangle\langle E_+|\prod_{k=1}^{55} D_{k}\left(\sqrt{\frac{s_k}{2}}A_{++}\right)+|E_-\rangle\langle E_-|\prod_{k=1}^{55} D_{k}\left(\sqrt{\frac{s_k}{2}}A_{--}\right)$, the electronic dephasing is described by the interaction Hamiltonian in the form
\begin{align}
	\tilde{H}_{ed}&=U H_{ed} U^{\dagger}=U\left(\sum_{i=1}^{2}(-1)^{i-1}|\varepsilon_i\rangle\langle \varepsilon_i| \right) U^{\dagger}\sum_{l} f_{l} (b_{l}+b_{l}^{\dagger}),
\end{align}
where $b_l$ represent the relative motions associated with low-frequency phonons including log-normal and marker mode (see \ref{section_params}). Secondly, the vibrational damping of the relative motions of high-frequency intra-pigment modes is described by the coupling of each mode to an independent secondary bath, modelled by
\begin{align}
	\tilde{H}_{vd}&=U H_{vd} U^{\dagger}=\sum_{k=1}^{55}(b_{k}^{\dagger}+b_{k})\sum_{l} g_{k,l} (c_{k,l}+c_{k,l}^{\dagger}),
\end{align}
where $c_{k,l}^{\dagger}$ denote the bath modes inducing the vibrational damping of the $k$-th intra-pigment mode $b_k$. The phonon spectral density $J_{vd, k}(\omega)$ of the $k$-th secondary bath, characterized by $g_{k,l}$, determines the damping rate of the $k$-th intra-pigment mode. We consider a Lorentzian spectral density centered at the vibrational frequency $\omega_k$ of the intra-pigment mode, so that the coupling strength between intra-pigment mode $b_k$ and environmental modes $c_{k,l}$ is maximized when environmental mode frequencies are resonant with $\omega_k$. The width of the Lorentzian spectral density is taken to be $(50\,{\rm fs})^{-1}$, so that the corresponding bath correlation function quickly decays compared to the picosecond timescale of mode damping. It is found that simulated results are not sensitive to the width of the Lorentzian spectral density when it is broader than $(50\,{\rm fs})^{-1}$. The amplitude of the Lorentzian spectral density is determined in such a way that the damping rate of the $k$-th mode is reduced to $(1\,{\rm ps})^{-1}$ when the mode is decoupled from electronic states, resulting in a single mode under damping induced by a secondary bath.

To construct a Lindblad equation, the interaction Hamiltonians $\tilde{H}_{ed}$ and $\tilde{H}_{vd}$ are expressed as
\begin{align}
	\tilde{H}_{ed}&=A_{ed} \sum_{l} f_{l} (b_{l}^{\dagger}+b_{l}),\\
	\tilde{H}_{vd}&=\sum_{k=1}^{55}A_{vd,k}\sum_{l} g_{k,l} (c_{k,l}^{\dagger}+c_{k,l}),
\end{align}
with $A_{ed}=U\left(\sum_{i=1}^{2}(-1)^{i-1}|\varepsilon_i\rangle\langle \varepsilon_i| \right) U^{\dagger}$ and $A_{vd,k}=b_{k}^{\dagger}+b_{k}$. We represent the system operators $A_{ed}$ and $A_{vd,k}$ in the eigenbasis of the vibronic system Hamiltonian, $\tilde{H}|\psi_{j}\rangle=\epsilon_{j}|\psi_{j}\rangle$, and define $A_{\alpha}(\omega)=\sum_{\epsilon'-\epsilon=\omega}\Pi(\epsilon)A_{\alpha}\Pi(\epsilon')$ for each system operator where $\Pi(\epsilon)$ describes the projection onto the subspace belonging to the eigenvalue $\epsilon$. The noise is described by dissipators in a Lindblad form~\cite{SMPetruccione}
\begin{align}
	\frac{d\rho(t)}{dt}&=-i[\tilde{H},\rho(t)]+\sum_{\alpha}\sum_{\omega}\gamma_{\alpha}(\omega)[A_{\alpha}(\omega)\rho(t)A_{\alpha}^{\dagger}(\omega)-\frac{1}{2}\{A_{\alpha}^{\dagger}(\omega)A_{\alpha}(\omega),\rho(t)\}],
\end{align}
where decoherence rates are determined by $\gamma_{\alpha}(\omega)=2\pi J_{\alpha}(\omega)(n(\omega)+1)$ with $n(\omega)=(\exp(\hbar\omega/k_B T)-1)^{-1}$, which is characterized by the corresponding spectral density, namely $J_{\alpha}(\omega)=J_{ed}(\omega)$ or $J_{\alpha}(\omega)=J_{vd,k}(\omega)$ for $k\in\{1,2,3,\cdots,55\}$, and temperature $T$.

Supplementary Figure~\ref{Fig_RM_noise_abs_2D}b displays absorption line shapes of SP computed by HEOM and reduced vibronic models without and with noise, shown in black, red and blue, respectively. Note that the reduced model results with noise are similar to those without noise. This implies that absorption line widths are mainly determined by inhomogeneous broadening and dephasing induced by center-of-mass modes, and the decohering processes described by the Lindblad equation are negligible within the lifetime of optical coherences, which is approximately $\sim 200\,{\rm fs}$.

In 2D electronic spectroscopy, a molecular sample is perturbed by three excitation pulses with controlled time delays, which generates third-order optical responses propagating in several phase-matched directions~\cite{SMJonas,SMBrixner}. Here we consider rephasing 2D spectra, which can be measured in a particular phase-matched direction~\cite{SMJonas,SMBrixner}. The time delay between first and second pulses enables one to resolve excitation wavelengths. The time interval between second and third pulses, called waiting times $T$, allows for monitoring electronic-vibrational dynamics of molecular systems. In simulations, we consider ground state bleaching (GSB) and stimulated emission (SE) pathways, describing electronic ground and excited state dynamics, respectively~\cite{SMMukamel}. This includes purely vibrational ground state coherences and the excited state coherences between vibronic eigenstates, which lead to oscillatory 2D signals~\cite{Plenio_JCP2013,Jonas_Nature,SMLimNC2015}.

Supplementary Figure~\ref{Fig_RM_noise_abs_2D}c shows rephasing 2D spectra at waiting time $T=0$ in the presence of inhomogeneous broadening. The 2D lineshape, shown as a function of excitation and detection wavelengths, is dominated by a diagonal peak excited and detected at both $\sim 900\,{\rm nm}$. The 2D peak location coincides with the transition energy of the main absorption peak A1 marked in Supplementary Figure~\ref{Fig_RM_noise_abs_2D}b. To investigate the excited state coherence between vibronic eigenstates $|\psi_-\rangle$ and $|\psi_+\rangle$, inducing absorption peaks A1 and A2, respectively, we consider two cross peaks R12 and R21 highlighted in Supplementary Figure~\ref{Fig_RM_noise_abs_2D}c. R12 (R21) is excited at the transition energy of A1 (A2) and then detected at that of A2 (A1).

Supplementary Figure~\ref{Fig_RM_noise_abs_2D}d shows the transient of R12 and its Fourier transformation visualizing frequencies present in the transient. It is notable that excited state SE signals, shown in blue, are dominated by oscillations at $\Delta'\approx 1800\,{\rm cm}^{-1}$ induced by the vibronic coherence between $|\psi_-\rangle$ and $|\psi_+\rangle$, which are brighter than the other vibronic eigenstates due to excitonic characters. This is contrary to the ground state GSB signals shown in red, which are dominated by low-frequency oscillations at $\sim 125\,{\rm cm}^{-1}$, close to the vibrational frequency of the marker mode. This makes the total 2D signals, namely the sum of GSB and SE signals, dominated by the excited state coherence $|\psi_+\rangle\langle \psi_-|$ and the vibrational coherences induced by the marker mode, leading to $\Delta'\approx 1800\,{\rm cm}^{-1}$ and $\omega_{\rm sp}=125\,{\rm cm}^{-1}$ oscillations, respectively.

Supplementary Figure~\ref{Fig_RM_noise_abs_2D}e shows the transient of the other cross-peak R21 and its frequency spectrum. Contrary to the case of R12, the oscillation amplitudes of ground state signals, shown in red, are comparable to those of excited state signals at R21. We note that the ground state oscillations consist of multiple components with different frequencies lower than $\sim 1600\,{\rm cm}^{-1}$, as the vibrational frequencies $\omega_k$ of underdamped intra-pigment modes considered in simulations are lower than $\sim 1600\,{\rm cm}^{-1}$ and the Huang-Rhys factors of these modes are of the order of 0.01, which are too small to induce ground state coherences with double vibrational excitation frequencies $\omega_k+\omega_{k'}$. Note that the excited state signals at R21 include $\Delta'\approx 1800\,{\rm cm}^{-1}$ oscillations, which are absent in the ground state signals.

Our 2D results demonstrate that long-lived 2D oscillations can be dominated by vibronic coherences even if the frequency of 2D oscillations is not near-resonant with the vibrational frequencies of underdamped modes, namely $\omega_k \lesssim 1600\,{\rm cm}^{-1}\ll\Delta'\approx 1800\,{\rm cm}^{-1}$. This implies that even if experimentally observed frequency of 2D oscillations is not matched to vibrational frequencies $\omega_k$ of intra-pigment modes observed in monomer experiments, there is possibility that the 2D oscillations originate from vibronic coherences induced by multi-mode mixing, rather than from purely electronic coherences~\cite{SMEngelScience2013}.

\jl{So far we have demonstrated that long-lived oscillations with $\sim 1800\,{\rm cm}^{-1}$ frequency appears in 2D spectra when the vibrational damping rates of all the 55 intra-pigment modes are taken to be $\gamma_k=(1\,{\rm ps})^{-1}$. In Supplementary Figure~\ref{Fig_SA}, we show that the long-lived oscillatory 2D signals originating from a multi-mode vibronic coherence cannot be modelled by coarse-grained spectral densities where only a few intra-pigment modes near-resonant with the excitonic splitting $\Delta\approx 1300\,{\rm cm}^{-1}$ of the SP are modelled by narrow Lorentzian spectral densities with a width of $\gamma_{k}=(1\,{\rm ps})^{-1}$, while all the other intra-pigment modes are modelled by broad Lorentzian functions with a width of $\gamma_k = (50\,{\rm fs})^{-1}$. In Supplementary Figure~\ref{Fig_SA}a, the vibrational damping rates of all the 55 intra-pigment modes are taken to be $\gamma_{k}=(50\,{\rm fs})^{-1}$, leading to a broad spectral density in the high-frequency region, as shown in red. In this case, all the high-frequency 2D oscillations are short-lived, and the $1800\,{\rm cm}^{-1}$ component is not visible in the frequency spectrum of 2D signals at the cross-peak R12. In Supplementary Figure~\ref{Fig_SA}b, the three intra-pigment modes whose frequencies are close to the excitonic splitting of the SP are modelled by the narrow Lorentzian functions, while all the other intra-pigment modes are approximated by the broad Lorentzian functions, as shown in green. It is notable that the three underdamped vibrational modes lead to long-lived oscillatory 2D signals at their vibrational frequencies $\omega_k\approx \Delta$, but the $1800\,{\rm cm}^{-1}$ component is short-lived as the lifetime of a multi-mode vibronic coherence is determined by the overall vibrational damping rate of several tens of the intra-pigment modes involved in a multi-mode vibronic mixing. Even when seven near-resonant intra-pigment modes are modelled by the narrow Lorentzian functions, the long-lived $1800\,{\rm cm}^{-1}$ oscillations do not appear in 2D spectra, as shown in blue in Supplementary Figure~\ref{Fig_SA}c. These results demonstrate that long-lived multi-mode vibronic coherences cannot be described by conventional vibronic models where only a few near-resonant modes are assumed to be weakly damped, while all the other modes, being relatively off-resonant from the excitonic splitting but participating in the multi-mode vibronic mixing, are severely coarse-grained. We note that the $1800\,{\rm cm}^{-1}$ oscillations are not observable in simulations when the off-resonant intra-pigment modes are ignored by taking the corresponding Huang-Rhys factors to be zero, as the multi-mode vibronic energy renormalization, namely $\Delta\approx 1300\,{\rm cm}^{-1}\rightarrow \Delta' \approx 1800\,{\rm cm}^{-1}$ of the SP, cannot be induced by a small number of near-resonant intra-pigment modes, as the Huang-Rhys factors of individual modes are small.}


\section{P\MakeLowercase{opulation} t\MakeLowercase{ransfer} d\MakeLowercase{ynamics}}

\jl{Here we show that the energy transfer between exciton states can be enhanced by multi-mode vibronic effects. We investigate which intra-pigment modes are excited during the transition from higher to lower energy exciton state in order to demonstrate that the intra-pigment modes whose vibrational frequencies are near-resonant with a vibronic splitting $\Delta'$, instead of a bare excitonic splitting $\Delta$, dominate the excitonic transition. We also compare the exciton transfer dynamics in the presence of highly structured vibrational environments with that of coarse-grained phonon spectral densities where only a few intra-pigment modes whose vibrational frequencies are near-resonant with the excitonic splitting $\Delta$ are considered and all the other off-resonant modes are neglected. We show that the transition from higher to lower energy exciton state is more efficient for the multi-mode vibrational environments than for the coarse-grained environments.}

\jl{In Supplementary Figure~\ref{Fig_SB}a and b, we consider a coherent vibronic energy transfer in the SP where exciton states are coupled to 55 intra-pigment vibrational modes. The noise induced by vibrational damping of the intra-pigment modes and vibronic couplings of the excitons to low-frequency phonon environments is ignored, which will be considered later. Starting from an initial state $\ket{E_+,0}$ with $\ket{E_+}$ denoting a higher-energy exciton state and $\ket{0}$ a global vibrational ground state where all the intra-pigment modes are in their vibrational ground states, we investigate the population dynamics of lower-energy exciton states $\ket{E_-,0}$, where all the intra-pigment modes are in their vibrational ground states, $\ket{E_-,1_k}$, where only the $k$-th intra-pigment mode is singly excited, and $\ket{E_-,2_k}$ and $\ket{E_-,1_k,1_{k'}}$, where only the $k$-th mode is doubly excited and two different modes $k$ and $k'$ are singly excited at the same time, respectively. In Supplementary Figure~\ref{Fig_SB}a, the total population of the lower-energy exciton states is shown in black, which increases rapidly within a sub-100\,fs time scale and then fluctuates around 0.75. The total population of $\ket{E_-,1_k}$, shown in green, is close to the total population of the lower-energy exciton states, shown in black, implying that the single-phonon transitions from $\ket{E_+,0}$ to $\ket{E_-,1_k}$ dominate the coherent vibronic energy transfer. The population of $\ket{E_-,0}$ is close to zero on a picosecond time scale, as shown in blue, but the total population of $\ket{E_-,2_k}$ and $\ket{E_-,1_k,1_{k'}}$ increases slowly, as shown in red, implying that the excitonic transition from higher- to lower-energy exciton state can take place mediated by the creation of two phonons.}

\jl{To identify which vibrational modes are excited during the excitonic transition, in Supplementary Figure~\ref{Fig_SB}b, the populations of the vibrationally cold ($\ket{E_-,0}$), singly excited ($\ket{E_-,1_k}$), and doubly excited states ($\ket{E_-,2_k}$ and $\ket{E_-,1_k,1_{k'}}$) at time $t=1\,{\rm ps}$ are shown in blue, green and red dots, respectively, as a function of their detunings from the initial state $\ket{E_+,0}$, namely $-\Delta$ for $\ket{E_-,0}$, $\omega_k-\Delta$ for $\ket{E_-,1_k}$, $\omega_k+\omega_{k'}-\Delta$ for $\ket{E_-,2_k}$ if $k=k'$ or $\ket{E_-,1_k,1_{k'}}$ otherwise. It is notable that the singly excited states $\ket{E_-,1_k}$ have comparable populations for a wide range of $\omega_k-\Delta$, as shown in green, demonstrating that several tens of the intra-pigment modes can contribute to the excitonic transition. More importantly, the populations of $\ket{E_-,1_k}$ become higher as the vibrational frequencies $\omega_k$ are closer to the vibronic splitting $\Delta'\approx \Delta+465\,{\rm cm}^{-1}$, which is blue-shifted from the bare excitonic splitting $\Delta$ due to a multi-mode vibronic mixing. This is in line with the populations of the doubly excited states $\ket{E_-,2_k}$ and $\ket{E_-,1_k,1_{k'}}$, shown in red, which are maximised when the sum of vibrational energy quanta $\omega_k+\omega_{k'}$ is near-resonant with the vibronic splitting $\Delta'$. These results demonstrate that the excitonic transition is governed by the resonance between intra-pigment modes and vibronic splitting $\Delta'$, instead of the bare excitonic splitting $\Delta$. These observations can be rationalised as follows. Since the Huang-Rhys factors of individual intra-pigment modes are small, even if a single intra-pigment mode is removed from a full vibronic model, the change in the vibronic splitting $\Delta'$ will be small. This implies that the full vibronic model can be decomposed into (i) a multi-mode vibronic model where the excitons are coupled to 54 intra-pigment modes per site and (ii) a perturbation term describing the interaction between excitons and the removed intra-pigment mode with vibrational frequency $\omega_k$, where the contribution of the perturbation term to the 54-mode vibronic model is approximately determined by $\omega_k-\Delta'$.}

\jl{So far we have considered a purely coherent multi-mode vibronic energy transfer. In the presence of the Lindblad noise induced by vibrational damping of the intra-pigment modes and the vibronic couplings to low-frequency phonon environments at room temperature, the oscillatory features of population dynamics are suppressed, as shown in Supplementary Figure~\ref{Fig_SB}c, and the populations of vibronic eigenstates are relaxed to the Boltzmann distribution of a thermal state, as shown in Supplementary Figure~\ref{Fig_SB}d.}

\jl{In Supplementary Figure~\ref{Fig_SC}a, the population dynamics of the lower-energy exciton state $\ket{E_-}$ in the absence of the Lindblad noise is shown for several phonon spectral densities, including the full model with 55 undamped intra-pigment modes, shown in black, and coarse-grained models where $M$ intra-pigment modes near-resonant with the excitonic splitting $\Delta\approx 1300\,{\rm cm}^{-1}$ of the SP are considered and all the other modes are neglected by taking the corresponding Huang-Rhys factors to be zero: the simulated results for $M=1,3,5$ are shown in blue, green, red, respectively. It is notable that when a single near-resonant mode is considered ($M=1$), the population of the lower-energy exciton state $\ket{E_-}$ oscillates between 0 and 1, due to a coherent resonant energy transfer between $\ket{E_+,0_r}$ and $\ket{E_-,1_r}$ with $\ket{0_r}$ and $\ket{1_r}$ denoting, respectively, the vibrational ground and first excited states of the near-resonant mode. Similarly, when a small number of near-resonant modes are considered ($M=3,5$), the population of the lower-energy exciton state $\ket{E_-}$ oscillates with a large amplitude and the population of $\ket{E_-}$ is efficiently transferred back to the higher-energy exciton state $\ket{E_+}$. This is contrary to the multi-mode case, shown in black, where the population of the lower-energy exciton state is increased rapidly on a sub-100\,fs time scale and then weakly oscillates around a mean value of $\sim 0.8$. This implies that multi-mode vibrational environments can induce an ultrafast excitonic transition from $\ket{E_+}$ to $\ket{E_-}$ and then suppresses the back transfer to the higher-energy exciton state $\ket{E_+}$. Even when the Lindblad noise induced by vibrational damping of the intra-pigment modes and low-frequency phonon environments at room temperature is considered in simulations, the energy transfer is more efficient for the multi-mode vibrational environments than for the coarse-grained models. Supplementary Figure~\ref{Fig_SC}b shows that the energy transfer rate from $\ket{E_+}$ to $\ket{E_-}$ at early times is enhanced as the number of the intra-pigment modes coupled to the exciton states is increased. In addition, Supplementary Figure~\ref{Fig_SC}c shows that the population of the lower-energy exciton state $\ket{E_-}$ is saturated more rapidly for the multi-mode vibrational environments, as the coarse-grained cases show large-amplitude oscillations of the population dynamics, induced by an efficient back transfer to the higher-energy exciton state $\ket{E_+}$. These results demonstrate that the energy transfer dynamics can be enhanced by multi-mode vibronic effects, and we found similar multi-mode vibronic enhancement for various electronic coupling strengths, including $V=50, 100, 300\,{\rm cm}^{-1}$ (not shown here) and $625\,{\rm cm}^{-1}$ of the SP considered in Supplementary Figure~\ref{Fig_SC}.}


\section{M\MakeLowercase{ulti}-m\MakeLowercase{ode} v\MakeLowercase{ibronic} e\MakeLowercase{ffects in} s\MakeLowercase{ite} b\MakeLowercase{asis}}

\jl{To explain the influence of multi-mode vibronic effects on absorption spectra, we consider the transition dipole moment operators of PPCs, describing the optical transition between electronic ground and excited states. Within the Franck-Condon approximation, the transition dipole moment operator of a dimeric system is written as
\begin{equation}
	\mu=\sum_{i=1}^{2}({\bf e}\cdot\boldsymbol{\mu}_{i})(\ket{\varepsilon_i}\bra{g}+\ket{g}\bra{\varepsilon_i}),
\end{equation}
where ${\bf e}$ is the unit vector in the direction of the electric field inducing absorption, $\boldsymbol{\mu}_i$ the 
transition dipole moment of site $i$, and $\ket{g}$ the global electronic ground state. The transition dipole moment 
operator describes the vertical optical transition from electronic ground to excited states, which launches the vibrational 
motions of PPCs, inducing 0-0  lines and vibrational sidebands in absorption spectra. The impact of the presence 
of vibrational modes on the strength of the 0-0 lines can be determined by decomposing the total Hamiltonian $H$
into $H = H_0+H_I$ where $H_I=V(\ket{\varepsilon_1}\bra{\varepsilon_2}+\ket{\varepsilon_2}\bra{\varepsilon_1})$ describes 
the electronic interaction between monomers, while $H_0$ determines the energy-level structure of each monomer. 
The Hamiltonian $H_0$ is diagonalised by the polaron transformation in site basis, $U=\ket{g}\bra{g}+\sum_{i=1}^{2}\ket{\varepsilon_i}\bra{\varepsilon_i}D_i$ with unitary displacement operators $D_i=\exp[\sum_{k=1}^{55}\sqrt{s_k}(b_{i,k}^{\dagger}-b_{i,k})]$, leading to
\begin{equation} H_{0}'=\sum_{i=1}^{2}(\varepsilon_i-\lambda_h)\ket{\varepsilon_i}\bra{\varepsilon_i}+\sum_{i=1}^{2}\sum_{k=1}^{55}\omega_{k}b_{i,k}^{\dagger}b_{i,k},
\end{equation}
where $\lambda_h=\sum_{k=1}^{55}\omega_k s_k$ denotes the reorganization energy of the intra-pigment modes. 
The eigenstates of $H_{0}'$ include the electronic ground (excited) states $\ket{g,0}$ ($\ket{\varepsilon_i,0}$), 
where all the intra-pigment modes are in their vibrational ground states, and $\ket{g,1_{i,k}}$ ($\ket{\varepsilon_i,1_{i,k}}$), 
where only one mode described by $b_{i,k}$ is singly excited. The transition dipole moment operator in the polaron basis, $\mu'=\sum_{i=1}^{2}({\bf e}\cdot\boldsymbol{\mu}_{i})(\ket{\varepsilon_i}\bra{g}D_i+h.c.)$, determines the transition dipole strengths between the eigenstates of $H_{0}'$. These transition strengths are
given by the Franck-Condon factors of uncoupled monomers, such as $\bra{g,0} \mu\ket{\varepsilon_i,0} = ({\bf e}\cdot\boldsymbol{\mu}_{i})\exp(-\frac{1}{2}\sum_{l=1}^{55}s_{l})$ and $\bra{g,0}\mu\ket{\varepsilon_i,1_{i,k}} = ({\bf e}\cdot\boldsymbol{\mu}_{i})\exp(-\frac{1}{2}\sum_{l=1}^{55}s_{l})\sqrt{s_{k}}$, which describe the dipole 
strengths of the 0-0 transition and the 0-1 transitions (vibrational sideband) in monomer
absorption spectrum. This implies that the dipole strength of the 
0-0 transition is reduced by the vibronic coupling to the intra-pigment modes by the factor $\exp(-\frac{1}{2}\sum_{l=1}^{55}s_l)$, 
as the total dipole strength of a monomer is redistributed to 0-$n$ transitions where $n$ vibrational excitations 
are present amongst the 55 intra-pigment modes. This is in line with the electronic interaction
Hamiltonian in the polaron basis, $H_{I}'=V\ket{\varepsilon_1}\bra{\varepsilon_2}\exp[\sum_{k=1}^{55}\sqrt{2s_k}(b_{k}^{\dagger}-b_{k})]+h.c.$ with $b_{k}=(b_{1,k}-b_{2,k})/\sqrt{2}$, that includes the effective coupling between 0-0 transitions, $V_{00}=\bra{\varepsilon_1,0}H_{I}'\ket{\varepsilon_2,0}=V\exp(-\sum_{k=1}^{55}s_k)\approx 0.5 V$ for WSCP, and the couplings $V_{01}$ between 0-0 and 0-1 transitions, such as $\bra{\varepsilon_1,0}H_{I}'\ket{\varepsilon_2,1_{2,k}}$.}

To investigate how well the numerically exact absorption spectra of WSCP can be described by the effective coupling $V_{00}$ between 0-0 transitions and how the absorption line shapes are modified by the mixing of 0-0 and 0-1 transitions, we consider a reduced vibronic model where the total Hamiltonian $H'=H_{0}'+H_{I}'$ in the polaron basis is numerically diagonalised. We consider a subspace that 
contains all states involving up to two vibrational excitations of 55 intrapigment modes. This approach enables one to compute the transition energies and dipole strengths of vibronic eigenstates of $H'$ for the cases that the vibronic interaction Hamiltonian $H_{I}'$ contains only the coupling $V_{00}$ between 0-0 transitions or includes all the vibronic couplings $V_{mn}$ between $m$-$n$ transitions with $m,n\in\{0,1,2\}$ where non-zero values of $m$ describe thermal populations of the intra-pigment modes at finite temperatures. The homogeneous broadening of the vibronic eigenstates is approximately described by a Lindblad equation where the Markovian noise is induced by low-frequency protein modes $J_{l}^{\rm WSCP}(\omega)$ and vibrational damping of the intra-pigment modes. The full reduced model including all the $V_{mn}$ couplings can quantitatively well reproduce the numerically exact absorption line shape of WSCP, as shown in red in Supplementary Figure~\ref{fig4}a. Taking into account only $V_{00}$ (dynamic localization) leaves the low-energy part of absorption spectra (above $650$\,nm) unchanged, as shown in green in Supplementary Figure~\ref{fig4}a, but the intensity of the vibrational sideband in a high-energy region is underestimated. This implies that the low-energy part of absorption spectra of WSCP is mainly determined by the effective coupling $V_{00}$ between 0-0 transitions. The deviations in the high-energy part of absorption spectra is due to the mixing of 0-0 and 0-1 transitions, which redistributes transition dipole strengths from 0-0 lines to
the vibrational sideband (see \ref{section_Lamb} for details). We note that the reduced model results are scaled by different factors to ensure that the maximum amplitudes of absorption spectra are unity at 656\,nm, for a better comparison with numerically exact results. When the full intrapigment spectral density $J(\omega)$, including 55 high-frequency modes, is considered in the conventional line shape theory (see \ref{section_lineshapetheory} for details), the vibronic suppression of the energy-gap between absorption peaks is overestimated by a Lamb shift \cite{SMPetruccione}, as shown in blue in Supplementary Figure~\ref{fig4}a, and the intensity of the vibrational sideband is significantly underestimated. We note that the Lamb shift originates from a perturbative treatment of vibronic couplings, which is not appropriate to describe the interaction between electronic states and underdamped intrapigment modes where electronic-vibrational correlations are maintained for a long time. These results demonstrate that the multimode vibronic mixing between 0-0 and 0-1 transitions is important for a reliable description of absorption line shapes, especially the intensity of a vibrational sideband, even if the vibronic mixing is not strong enough to modulate the low-energy part of absorption spectra, as shown in red and green in Supplementary Figure~\ref{fig4}a. These multimode effects induced by intrapigment modes are robust against variations of the low-frequency vibronic coupling spectrum of protein motions, as shown in Supplementary Figure~\ref{fig4}b, where $J_{l}^{\rm B777}(\omega)$ is considered instead of $J_{l}^{\rm WSCP}(\omega)$.

We note that our numerically exact absorption spectra of WSCP are well matched to the low-energy part of experimental absorption spectra above 640\,nm, but the intensity of the computed high-frequency vibrational sideband (with a main peak at 600-610\,nm) is stronger than experimental results \cite{Pieper_2011JPCB} (not shown here). This implies that the Huang-Rhys factors of the high-frequency intra-pigment modes may be overestimated due to an approximate theory considered in the analysis of experimental FLN data \cite{Pieper_2011JPCB}, which is in line with the spectral density of WSCP computed by first-principles methods \cite{RosnikJCTC2015}. A more reliable estimation of the Huang-Rhys factors requires numerically exact simulations of the FLN spectra, which is beyond the scope of this work. In addition, the vibrational sideband of ${\rm Q}_y$ transitions may be vibronically mixed with the zero-phonon lines of ${\rm Q}_x$ transitions between the electronic ground and second excited states of the pigments \cite{ReimersSR2013}, which is not considered in this work and deserves separate investigation.


\subsection{Multi-Mode Vibronic Mixing of 0-0 and 0-1 Transitions and Lamb Shift}\label{section_Lamb}

Here we show the influence of multi-mode vibronic mixing of 0-0 and 0-1 transitions of WSCP and Lamb shift on absorption line shapes of WSCP in more detail.

In reduced model simulations of absorption, we numerically diagonalize the total Hamiltonian $H'=H_0'+H_I'$ in the polaron basis within double vibrational excitation subspace and compute the energy-levels and transition dipole strengths of vibronic eigenstates. The broadening of absorption peaks is modelled by the static disorder in site energies and homogeneous broadening induced by the coupling to low-frequency protein motions and the vibrational damping of intra-pigment modes. The center-of-mass motions of the protein modes are treated non-perturbatively, while the relative motions of the protein modes are considered approximately by using a Lindblad equation (see \ref{section_reduced_2DES}). In addition, the vibrational damping of local intra-pigment modes is considered within the Lindblad formalism where each intra-pigment mode is coupled to an independent secondary bath in a thermal state, similar to the reduced model for SP (see \ref{section_reduced_2DES}). We assume that the intra-pigment modes are initially in a thermal state at temperature $T=77\,{\rm K}$.

In Supplementary Figure~\ref{Fig_Renger_diagonalisation}a, we consider only the $V_{00}$ coupling between 0-0 transitions and neglect all the other couplings in $H_I'$ (dynamic localization). Numerically exact absorption spectrum of WSCP is shown in a black dashed line, while approximate absorption line shape is shown in a red solid line. The contributions of 0-$n$, 1-$n$ and 2-$n$ transitions to the total approximate absorption spectra are shown in blue solid, blue dashed and blue dotted lines, respectively. The 1-1 transitions lead to a single peak at 659\,nm due to the absence of the interaction between $\ket{\varepsilon_1,1_{j,k}}$ and $\ket{\varepsilon_2,1_{j',k'}}$. Note that the total approximate absorption line shape shown in red is not well matched to numerically exact results shown in black.

In Supplementary Figure~\ref{Fig_Renger_diagonalisation}b, we consider the interaction between local electronic excitations when associated vibrational states are identical, such as the coupling between $\ket{\varepsilon_1,n_{j,k}}$ and $\ket{\varepsilon_2,n_{j,k}}$ with $n_{j,k}\in\{0,1,2\}$, and that between $\ket{\varepsilon_1,1_{j,k},1_{j',k'}}$ and $\ket{\varepsilon_2,1_{j,k},1_{j',k'}}$ where different modes $b_{j,k}$ and $b_{j',k'}$ are singly excited at the same time. This modifies the absorption line shape of 1-$n$ transitions, leading to two absorption peaks in zero-phonon line region above 650\,nm, similar to 0-$n$ transitions. To clarify the energy-gap between 0-0 transitions, the transition dipole strengths of all 0-$n$ transitions are shown in yellow triangles as a function of transition energies, where the splitting between 0-0 transitions is $\Delta'=2V_{00}\approx 142\,{\rm cm}^{-1}$. Note that approximate absorption spectra are still not well matched to numerically exact results.

In Supplementary Figure~\ref{Fig_Renger_diagonalisation}c, we now consider the full $H_I'$, containing the couplings between local electronic excitations with different vibrational states. This leads to a multi-mode vibronic mixing between 0-0 and 0-1 transitions, which redistributes the transition dipole strengths of the bright 0-0 transition at 655.5\,nm and relatively dark 0-1 transitions around 650\,nm (compare yellow triangles in Supplementary Figure~\ref{Fig_Renger_diagonalisation}b and c). This makes the main absorption peak at 655.5\,nm darker, and the vibrational sideband around 650\,nm brighter. As a result, approximate absorption line shape shown in red is quantitatively well matched to numerically exact absorption spectrum shown in black, although the approximate results underestimate absorption in the low-energy region above 662\,nm. This implies that the energy-gap $\Delta'\approx 134\,{\rm cm}^{-1}$ between absorption peaks in zero-phonon line region is underestimated when compared to numerically exact results.

So far we have not included the Lamb shift in reduced model simulations. Within the Lindblad formalism~\cite{SMPetruccione}, the energy-level shift induced by the interaction with environmental degrees of freedom is described by
\begin{align}
	H_{LS}&=\sum_{\omega}\sum_{\alpha}S_{\alpha}(\omega)A_{\alpha}^{\dagger}(\omega)A_{\alpha}(\omega)\\
	&=-\sum_{j}\ket{\psi_j}\bra{\psi_j}\sum_{\alpha}\lambda_{\alpha}|\langle \psi_j|A_{\alpha}|\psi_j\rangle|^2 +\sum_{\epsilon_k\neq \epsilon_j}\ket{\psi_k}\bra{\psi_k}\sum_{\alpha}S_{\alpha}(\epsilon_k-\epsilon_j)|\langle\psi_j|A_\alpha|\psi_k\rangle|^2,\label{eq:Lamb_general}
\end{align}
where $\ket{\psi_j}$ represent the vibronic eigenstates of the total Hamiltonian $UHU^{\dagger}=H_0+H_I$, satisfying $UHU^{\dagger}\ket{\psi_j}=\epsilon_{j}\ket{\psi_j}$, $A_\alpha$ noise operators describing the coupling to low-frequency protein motions or vibrational damping of intra-pigment modes, $S_{\alpha}(\Delta')={\rm P}\int_{-\infty}^{\infty}d\omega(J_{\alpha}(\omega)(n(\omega)+1)+J_{\alpha}(-\omega)n(-\omega))/(\Delta'-\omega)$ and $\lambda_{\alpha}=\int_{0}^{\infty}d\omega J_{\alpha}(\omega)/\omega$ with $J_{\alpha}(\omega)$ denoting the environmental spectral density describing the noise process associated with $A_{\alpha}$ (see \ref{section_reduced_2DES}). Supplementary Figure~\ref{Fig_Renger_diagonalisation}d shows that when the Lamb shift is included in simulations, the energy-gap $\Delta'\approx 134\,{\rm cm}^{-1}$ renormalised by the intra-pigment modes is increased to $\Delta''\approx 188\,{\rm cm}^{-1}$. The difference between approximate and numerically exact absorption line shapes can be further reduced by introducing a small empirical pure dephasing rate $\gamma_{pd}=(2\,{\rm ps})^{-1}$, as shown in a green solid line. Within the Lindblad formalism, the phonon spectral density of protein motions of WSCP does not induce a pure dephasing process, as log-normal distribution functions are super-Ohmic, which may underestimate the pure dephasing effect in numerically exact simulations.

These results demonstrate that the interaction between 0-0 transitions is significantly suppressed by a local vibronic coupling to high-frequency intra-pigment modes, and a multi-mode vibronic mixing of 0-0 and 0-1 transitions can modify absorption line shapes even if the mixing is not strong enough to change the energy-gap between absorption peaks originating from 0-0 transitions. This is similar to the properties of bare exciton states where the energy-gap between excitons is close to the difference in site energies, $|\varepsilon_2-\varepsilon_1|$, when the electronic coupling $V$ between monomers is sufficiently weak, but such a weak coupling can induce notable redistribution of oscillator strengths, leading to large difference in transition dipole strengths of exciton states~\cite{Renger_1996JPC}. The Lamb shift induced by low-frequency protein motions can increase or decrease the energy-gap between absorption peaks, similar to the multi-mode vibronic mixing induced by the intra-pigment modes, hinting that characterisation of the environmental structures of photosynthetic pigment-protein complexes is essential to understand the optical responses of these systems. We note that in photosynthetic systems, pigments may be coupled to non-identical vibrational environments, characterised by different spectral densities. Even in this case, one can generalise our approach and show that the effective coupling between 0-0 transitions of sites $i$ and $j$ is given by $V_{ij}\exp(-\frac{1}{2}\sum_{k}(s_{i,k}+s_{j,k}))$ where $V_{ij}$ denotes the bare electronic coupling between sites $i$ and $j$ of a multi-chromophoric system, and $s_{i,k}$ the Huang-Rhys factors of the intra-pigment modes locally coupled to site $i$.


\subsection{Line Shape Theory}\label{section_lineshapetheory}

Here we consider conventional line shape theory based on second order cumulant expansion where absorption line shape of a dimer is approximately described by
\begin{align}
	&\sum_{j=\pm}|\boldsymbol{\mu}_{E_j}|^{2}\int_{-\infty}^{\infty}dt \exp[i(\omega-E_{j}-E_{LS,j})t+G_{j}(t)-G_{j}(0)-(\gamma_{j\rightarrow k}/2+\gamma_{pd})|t|],\nonumber
\end{align}
where $\boldsymbol{\mu}_{E_j}$ and $E_{j}$ denote the transition dipole moment vector and energy-level, respectively, of an exciton state $\ket{E_{j}}$, defined by $H_{e}\ket{E_{j}}=E_{j}\ket{E_j}$, and $G_{j}(t)=\sum_{i=1}^{2}|\langle \varepsilon_i|E_j\rangle|^{4}G(t)$ with $G(t)$ characterised by the full spectral density $J(\omega)$ of WSCP, including both low-frequency protein and high-frequency intra-pigment modes
\begin{align}
	G(t)&=\int_{0}^{\infty}d\omega \,\omega^{-2}[J(\omega)(n(\omega)+1)e^{-i\omega t}+J(\omega)n(\omega)e^{i\omega t}].\nonumber
\end{align}
Here $\gamma_{pd}$ is an empirical pure dephasing rate, while $\gamma_{j\rightarrow k}$ represents an incoherent exciton population transfer rate from $\ket{E_j}$ to $\ket{E_k}$
\begin{align}
	\gamma_{\pm\rightarrow \mp}&=2\pi \sum_{i=1}^{2}|\langle E_{+}|\varepsilon_i\rangle\langle\varepsilon_i|E_{-}\rangle|^{2}(J(\pm\Delta)(n(\pm\Delta)+1)+J(\mp\Delta)n(\mp\Delta)),\nonumber
\end{align}
with $J(\omega)=0$ for $\omega<0$, $\Delta=E_+ - E_->0$ and $n(\omega)=(\exp(\omega/k_B T)-1)^{-1}$. The Lamb shift $E_{LS,j}$ describes the energy-level shift of an exciton state $\ket{E_j}$ due to the interaction with the full vibrational environments
\begin{equation}
	E_{LS,\pm}=-\sum_{i=1}^{2}|\langle\varepsilon_i|E_{\pm}\rangle|^{4}\lambda+\sum_{i=1}^{2}|\langle E_{+}|\varepsilon_i\rangle\langle\varepsilon_i|E_{-}\rangle|^{2}\,{\rm P}\int_{-\infty}^{\infty}d\omega\,\frac{J(\omega)(n(\omega)+1)+J(-\omega)n(-\omega)}{\pm\Delta-\omega},
\end{equation}
with ${\rm P}$ denoting the Cauchy principal value and $\lambda=\int_{0}^{\infty}d\omega J(\omega)/\omega$, which is a special case of Eq.~(\ref{eq:Lamb_general}). When the full phonon spectral density of WSCP is considered, the energy-gap between excitons is reduced from a bare excitonic splitting $\Delta\approx 2V=280\,{\rm cm}^{-1}$ to $\Delta'\approx 102\,{\rm cm}^{-1}$ by the Lamb shift. This implies that the Lamb shift dominated by high-frequency intra-pigment modes suppresses the energy-gap between exciton states, although the Markov approximation considered in the derivation of the Lamb shift is not appropriate to describe the long-lived correlations between excitons and underdamped intra-pigment modes. It is notable that the suppression of the excitonic splitting is over-estimated by the conventional line shape theory, which cannot reproduce numerically exact absorption spectra of WSCP (see Supplementary Figure~\ref{fig4}).



%

\end{widetext}

\end{document}